\documentclass[12pt]{article}
\usepackage{graphicx}

 \newcommand{\vbf}{\mbox{\boldmath $v$}}
 
 \newcommand{\imp}{\mbox{\boldmath $p$}}
 \newcommand{\Ebf}{\mbox{\boldmath $E$}}
 \newcommand{\kbf}{\mbox{\boldmath $k$}}
 
 \newcommand{\Hbf}{\mbox{\boldmath $H$}}
 
 \newcommand{\zbf}{\mbox{\boldmath $z$}}

 \newcommand{\nabf}{\mbox{\boldmath $\nabla$}}
 \newcommand{\nablabf}{\mbox{\boldmath $\nabla$}}
 
 \newcommand{\Sbf}{\mbox{\boldmath $S$}}
 \newcommand{\Abf}{\mbox{\boldmath $A$}}
 
 \newcommand{\e}{{\rm e}}

 \newcommand{\pa}{\partial}

 \newcommand{\text}{\rm}
 
 \newcommand{\drm}{{\rm d}}
 \newcommand{\grm}{{\rm g}}

 \newcommand{\Ncal}{{\cal N}}
 \newcommand{\ug}{ \; = \; }

 \newcommand{\ga}{\gamma}

 \newcommand{\infi}{\infty}

 \newcommand{\la}{\lambda}

 \newcommand{\kr}{k_{\rho}}

 \newcommand{\circrm}{{\rm circ}}
 \newcommand{\sinc}{{\rm sinc}}

 \newcommand{\bb}{\begin{equation}}
 \newcommand{\ee}{\end{equation}}
 \newcommand{\bc}{\begin{center}}
 \newcommand{\ec}{\end{center}}
 \newcommand{\bega}{\begin{eqnarray}}
 \newcommand{\ega}{\end{eqnarray}}
 \newcommand{\begae}{\begin{eqnarray*}}
 \newcommand{\egae}{\end{eqnarray*}}

 \newcommand{\h}{\hspace*{4ex}}
 \newcommand{\dis}{\displaystyle}

 \newcommand{\rr}{\rho}
 \newcommand{\Aove}{{\overline A}}
 
 \newcommand{\om}{\omega}

 \newcommand{\cent}{\centerline}
 \newcommand{\vs}{\vspace*}

\begin{document}
\baselineskip 0.5cm

\begin{center}

{\large {\bf NON-DIFFRACTING WAVES: \ A NEW INTRODUCTION}$^{\: (\dag)}$}
\footnotetext{$^{\: (\dag)}$  Work
partially supported by FAPESP, CAPES and CNPq (Brazil), and by INFN
(Italy). \ E-mail addresses: \ recami@mi.infn.it, mzamboni@decom.fee.unicamp.br; \
\ hugo@decom.fee.unicamp.br; \ leo@sc.usp.br}

\end{center}

\vs{5mm}

\cent{ Erasmo Recami$^{\, 1,2,3}$, M.Zamboni-Rached$^{\, 4,1}$, H.E.Herna'andez-Figueroa$^{\, 1}$, and L.A.Ambr\'osio$^{\, 5}$ }

\vs{0.2 cm}

\cent{$^{1}$ {\em DECOM, FEEC, Universidade Estadual de Campinas
(UNICAMP), Campinas, SP, Brazil}}
\cent{$^{2}$ {\em Facolt\`a di Ingegneria, Universit\`a statale di
Bergamo, Bergamo, Italy.}}
\cent{$^{3}$ {\em INFN---Sezione di Milano, Milan, Italy.}}
\cent{$^{4}$ {\em Photonics Group, Electrical \& Computer Engineering, University of Toronto, Canada.}}
\cent{$^{5}$ {\em SEL, EESC, University of Sao Paulo (USP), Sao Carlos, SP, Brazil.}}

\vs{0.5 cm}

\

{\small{{{\em PACS nos.: 42.25.Bs; 42.25.Fx; 41.20.Jb; 46.40.Cd. \ \  OCIS codes: 999.9999;  260.1960;  070.7545; 070.0070 200.0200;  050.1120; 070.1060;  280.0280;  050.1755. \ \ \ Keywords: wave equations; wave propagation; localized waves; non-diffracting waves (NDW); Bessel beams; Optics; Microwaves; Acoustics; totally forward NDW pulses; finite apertures; finite energies; functional expression for any NDW pulses; truncated waves; analytic description of truncated beams; subluminal NDWs; subluminal electromagnetic bulletts; subsonic acoustic bulletts; stationary solutions; zero-speed envelopes; "frozen waves" (FW); experimental production of FWs; Special relativity; Non-Restricted special Relativity; Lorentz transformations; electromagnetic X-shaped waves; acoustic X-shaped supersonic waves.}}}}

\vs{0.5 cm}

\section{A General Introduction}

\

\subsection{A prologue}

In this work, which essentially deals with {\em exact} solutions to the wave equations, we begin by introducing the topic
of Non-Diffracting Waves (NDW), including some brief historical remarks, and by a simple
definition of NDWs: Afterwards we present some recollections ---besides of ordinary waves (gaussian
beams, gaussian pulses)--- of the simplest non-diffracting waves (Bessel beams, X-shaped pulses,...). More
details can be found in the first two (introductory) chapters of the volume
on {\em Localized Waves} published\cite{Livro} by J.Wiley (Hoboken, NJ,
USA) in 2008. \ In Section 2 we go on to show how to eliminate {\em any} backward-traveling
components (also known as non-causal components), first in the case of ideal NDW pulses,
and then, in Section 3, for realistic, finite-energy NDW pulses. In particular, in
subsection 3.1 we forward a general functional expression for any totally-forward non-diffracting
pulses. \ Then, in Section 4 an efficient method is set forth for the {\em analytic} description of
{\em truncated} beams, a byproduct of its being the elimination of any need of lengthy numerical
calculations. \ In Section 5 we explore the not less interesting question of the {\em subluminal} NDWs, or bullets,
in terms of two different methods, the second one being introduced since it allows the
analytic description of $v=0$ NDWs, that is, of non-diffracting waves with a static envelope
(``Frozen Waves", FW), in terms of continuous Bessel beam superpositions. \ The production of such
Frozen Waves (which, indeed, have been experimentally generated in recent time for Optics) is
theoretically developed in Section 6 also for the case of absorbing media. \ Section 7 discusses
the role of Special Relativity and of Lorentz transformations, which is relevant for the
physical comprehension of the whole issue of NDWs. \ In Section 8 we present further analytic
solutions to the wave equations, with use of higher-order Bessel beams (namely, non-axially
symmetric solutions). \ Next Section, 9, deals in detail with an application of NDWs
to Biomedical Optics, by having recourse to the generalized Lorenz-Mie theory. \ In Section 10 we
exploit the important fact that ``soliton-like" solutions can be found also in the rather
different case of the ordinary, linear {\em Schroedinger equation} ---which is not a properly said
wave-equation--- within standard Quantum Mechanics; by constructing, for instance, also a general exact
non-diffracting solution for such equation.  These ``localized" solutions to the Schroedinger equation
may a priori be of help for a better understanding, say, of de Broglie's approach and of the particle-wave
duality. \ Some complementary issues are just mentioned in the last Section. \ The present work also constitutes
a part of a much longer Review in preparation.

\

\h Let us now start by recalling that diffraction and dispersion are known since long to be phenomena
limiting the applications of beams or pulses.

\h Diffraction is always present, affecting any waves that
propagate in two or three-dimensional media.
Pulses and beams are constituted by waves traveling
along different directions, which produces a gradual {\em spatial}
broadening.  This effect is a limiting factor
whenever a pulse is needed which maintains its transverse
localization, like, e.g., in free space communications, image forming,
optical lithography, electromagnetic tweezers, etcetera.

\h Dispersion acts on pulses propagating in material media,
causing mainly a {\em temporal} broadening: An effect due to
the variation of the refraction index with the frequency, so that
each spectral component of the pulse possesses a different
phase-velocity. This entails a gradual temporal widening, which
constitutes a limiting factor when a pulse is needed which
maintains its {\em time} width, like, e.g., in communication
systems.

\h It has been important, therefore, to develop techniques able to
reduce those phenomena. \ The {\em Non-diffracting Waves\/},
known also as Localized Waves, are indeed able to
resist diffraction for a long distance. \ Today, non-diffracting waves are
well-established both theoretically and experimentally, and are
having innovative applications not only in vacuum, but also in
material (linear or non-linear) media, showing to be able to
resist also dispersion. \ As we were mentioning, their potential
applications are being intensively explored, always with
surprising results, in fields like Acoustics, Microwaves, Optics,
and are promising also in Mechanics, Geophysics\cite{Sanya}, and even
Elementary particle physics\cite{schX} and Gravitational Waves. One
interesting acoustic application has been already obtained in high-resolution ultra-sound
scanning of moving organs in the human body. \ We shall see that
NDWs are suitable superpositions of Bessel beams. \ And worth noticing is
that peculiar superposition of Bessel beams can be used to obtain ``static"
non-diffracting wave fields, with high transverse localization, and whose longitudinal
intensity pattern can assume any desired shape within a chosen
interval $0 \leq z \leq L$ of the propagation axis: Such waves with a
{\em static} envelope \cite{FWart2,MichelOE2,MichelOE1,Livro,Prego}, that we called ``Frozen Waves" (FW), have been
experimentally produced in recent times in the case of Optics, as reported
elsewhere also in the Book whose introductory chapter (Chap.1) is constituted by this paper. Those Frozen Waves
promise to have very important
applications (even in the field of medicine, and of tumor curing\cite{brevetto})

\h To confine ourselves to electromagnetism, let us recall again the
present-day studies on electromagnetic
tweezers, optical (or acoustic) scalpels, optical guiding of atoms
or (charged or neutral) corpuscles, optical lithography, optical (or acoustic)
images, communications in free space, remote optical alignment, optical acceleration
of charged corpuscles, and so on.

\subsection{Preliminary, and historical, remarks}

As we were saying, ordinary beams and pulses are superpositions of plane waves which travel in different directions;
this causes diffraction,
and consequently an increasing spatial broadening of the waves during propagation. Incidentally, we are here considering only
propagating, that is, {\em non-evanescent,} waves.

\h Surprisingly, solutions to the wave equations however exist, which represent in homogeneous media beams and
pulses able to resist the effects of diffraction for long distances. Such solutions are called Non-diffracting Waves (NDW),
or Localized Waves (LW); even if a better name would be ``Limited-diffractions Waves"\cite{Lu1,Lu2}.

\h The theory of NDWs allows compensating also for effects like dispersion and attenuation. Indeed, in dispersing
homogeneous media, it is possible to construct pulses that simultaneously resist the effects {\em of diffraction and of surface

dispersion.}  And, in absorbing homogeneous media, it is also possible to construct beams that resist the
simultaneous effects {\em of diffraction and of attenuation.}

\h For earlier reviews about NDWs, we can refer the reader, for instance, to the first two chapters
of the book\cite{Livro}, as well as Ref.\cite{AIEP}, and references therein.  There, the reader
will find general and formal (simple) introductions to NDWs, with more details on the separate cases
of beams and of pulses; as well as on the rather different
characteristics of the Bessel and of the NDWs waves, with respect to
(w.r.t) the Gaussian ones. The important properties of
the former w.r.t. the latter ones can find application, as well-known and as stressed therein, in all fields in which
an essential role is played by a wave-equation (like electromagnetism, optics, acoustics,
seismology, geophysics, and also gravitation, elementary particle physics, etc.).

\

{\em Here, let us only insert the following, quite brief historical information:}

\

The non-diffracting
solutions to the wave equations (scalar, vectorial, spinorial,...) are in fashion,
both in theory and in experiment, since a couple of decades.  Rather well-known are
the ones with luminal or superluminal peak-velocity\cite{Livro}: Like the
so-called X-shaped waves(see \cite{Lu1,MRH,IEEE} and refs. therein), which are
supersonic in Acoustics\cite{Lu2}, and superluminal in Electromagnetism (see\cite{PhysicaA};
see also\cite{PhysicaA2} and \cite{PhysicaA3}).

\h By Bateman\cite{Bateman} and later on Courant \&
Hilbert\cite{Courant}, it had been already recognized that {\em luminal} NDWs
exist, which are solutions to the wave equations. \ After the subsequent early works,
already quoted by us, a great deal of results\cite{sub} has been published on NDWs, from both the theoretical and
the experimental point of view: Initially, taking only free space into account; later on,
considering more complex media which exibit effects such as dispersion (see, e.g., \cite{disp1,disp2,disp3}),
nonlinearity\cite{Conti}, anisotropy\cite{anis1,anis2,anis3},
losses\cite{MichelOE2}, and so on. Extensions of this type have been carried out along with
the development, for instance, of efficient methods for obtaining non-diffracting
beams and pulses in the subluminal, luminal and superluminal regimes, thus allowing easier experimental verifications.

\h Indeed, in recent years, some attention\cite{sub,5,6,7,8,9,10,11,12,13} started to be paid to the
(more ``orthodox") {\em subluminal\/} NDWs too.  It should be stressed that, in any case,
the interest of the NDWs resides not in their peak-velocity\cite{Cherenkov,RepToSeshadri,Utkin}, but in the
circumstance that they propagate in a homogeneous linear medium without distortion ---and in a self-reconstructing
way\cite{14,15,MichelOE2}--- (apart from local variations: In the sense that their square magnitude keeps its
shape during propagation, while local variations are shown only by its real, or imaginary, part).

\

\h  In the past, however, much attention was not even paid
to Brittingham's 1983 paper\cite{Brittingham}, wherein he
obtained {\em pulse-type} solutions to the Maxwell equations,
which propagated in free space as a new kind of $c$-speed ``solitons".  That
lack of attention was partially due to the fact that Brittingham had been
able neither to get finite-energy expressions for his ``wavelets",
nor to make suggestions about their practical production.  Two years later,
however, Sezginer\cite{Sezginer} was able to obtain
quasi-nondiffracting luminal pulses endowed with a finite energy:  Finite-energy
pulses are known to travel no longer undistorted for an infinite distance,
but nevertheless propagate without deformation for a long field-depth,
much larger than the one achieved by ordinary pulses like the gaussian ones:
Cf., e.g., Refs.\cite{16,17,18,19,Durnin1,Durnin2,Tesi1,Tesi2,Mathieu,Introd,Capitulo,PIER98} and refs. therein.

\

\h An interesting problem, indeed, was that of
investigating what it would happen to the ideal {\em Bessel beam
solution when truncated} by a finite transverse aperture. In 1987 a heuristical answer came, after the
quoted series of pioneering papers\cite{16,17,18,19}, from the known
experiment by Durnin et al.\cite{Durnin1,Durnin2}, when it was shown that a
realistic Bessel beam, passing through a finite aperture, is able to travel keeping its
transverse intensity shape approximately unchanged all along a {\em large} ``depth of field".

\h In any case, only after 1985 a general theory of NDWs started to be extensively
developed\cite{Ziolk1,Ziolk2,Ziolk3,Lu1,Donnelly,Ziolk4,PhysicaA,Esposito,Friberg,SB1,birdseye,MRH,BandS,MRH2,Kiselev1,Kiselev2},
both in the case of beams and in the case of pulses. For reviews, see for
instance the Refs.\cite{IEEE,birdseye,Introd,Capitulo,Tesi2,PIER98,AIEP,Livro} and citations therein.
 \ For the propagation of NDWs in bounded regions (like {\em wave-guides\/}),
see Refs.\cite{Coaxial,MRF,MFR} and refs therein. \ For the focussing of
NDWs, see e.g. Refs.\cite{MSR,focusing,Livro} and
quotations therein. \ For recourse to chirped optical X-type waves to obtain
pulses capable of recovering their spatial shape both transversally and longitudinally, see
e.g. Refs.\cite{chirped,Livro} and references therein. \ Not less important, as to the construction of
general NDWs propagating in
{\em dispersive} media, see, besides the quoted \cite{disp1,disp2,disp3}, also
Refs.\cite{39,40,41}; while, for {\em lossy} media, cf. both Ref.\cite{MichelOE2} and
refs. therein, and this Chapter. \ Al last, for finite-energy, or truncated, solutions
see, e.g., Refs.\cite{MRB,JosaMlast,Ziolk4,Lu4,MRBb,MRBc}, as well as this Chapter.

\h By now, the NDWs have been experimentally
produced\cite{Lu2,PSaari,Ranfagni,PSaariB,PSaari2}, and are being applied in
fields ranging from ultrasound scanning\cite{LuBiomedical,LuImaging,8,11} to
optics (for the production, e.g., of new type of
tweezers\cite{brevetto,Curtis,MichelOE2,bbi30,rfr35}). All those works have demonstrated that
non-diffracting pulses can travel with any arbitrary peak-velocities $v$,
that is, with speed $v$ in the range $0<v<\infty$.

\

\h {\bf Let us introduce at this point a first mathematical definition of NDWs:}

\

Diffraction, as a spatial transverse spreading, cannot occur in the simple case of 1 space dimension. Actually,
the 1D wave equation
\bb (\pa^2_z - 1/c^2 \pa^2_t)\psi(z,t) = 0   \label{eq1}   \ee   

admits the general solution \ $\psi = f(z-ct) + g(z+ct)$, \ quantities $f$ and $g$ being arbitrary (differentiable)
functions; and, for instance, a solution of the type \ $\psi(z-ct)$ \ travels rigidly along the positive
$z$-direction with speed $c$. Let us here recall, and stress, that, if a wave depends on $t$ and $z$ only through
the quantity $z-Vt$, it will be seen as moving without any distortion with the speed $V$: \ See, e.g.,
Ref.\cite{IEEE} and references therein.

Passing on to the 3D case, when the wave equation reads
\bb (\nabla_{\perp}^2 + \pa^2_z - 1/c^2\pa^2_t)\psi(\mathbf{r}_{\perp},z,t) \ug 0 \; ,   \label{eq2}  \ee   

quantity \ $\nabla_{\perp}^2$ \ being the transverse Laplacian, and \ $\mathbf{r}_{\perp}$ \ the transverse position
vector [so that $\mathbf{r} = \mathbf{r}_{\perp} + z \mathbf{k}$], it is natural to look for possible
solutions of the type
\bb \psi(\mathbf{r}_{\perp}, z-Vt) \,\, , \label{lw1}\ee                                  

which would correspond to waves rigidly propagating along $z$ with speed $V$, whatever be the value of $V$ (cf.
Refs.\cite{Livro,IEEE}). \ To check the mentioned possibility, let us go back to Eq.(\ref{eq2}). It is simple to show, then, that
an acceptable solution of the type (\ref{lw1}) has just to satisfy the equation
\bb (\nabla_{\perp}^2 + (1-V^2/c^2)\pa^2_{\zeta})\psi(\mathbf{r}_{\perp},\zeta) \ug 0 \,\, ,
\label{eo2} \ee                                                                              

where  \ $\zeta \; \equiv \; z-Vt$.  [Let us explicitly recall\cite{Introd} that the shape of any solutions that depend
on $z$ and on $t$  only through the quantity $z-Vt$ will always appear the same to
an observer traveling along $z$ with the speed $V$, whatever it be
(subluminal, luminal or Superluminal) the value of $V$: That is,
such a solution will propagate rigidly with speed $V$].

\h One can therefore realize that:

(i) when $V=c$, \ equation (\ref{eo2}) becomes elliptic: Namely, it becomes a Laplace equation in the transverse
variables, that prevents the free-space solution from being transversally localizable. In
other words, these solutions are plane waves, or plane wave pulses, with scars practical interest.

(ii) when $V<c$, \ equation (\ref{eo2}) is still elliptic: More specifically, it is a Laplace equation in the variables
$(x,y,\zeta\sqrt{1-V^2/c^2})$, so that the free-space solutions cannot admit any local maxima or
minima. \  No solutions of physical interest are obtainable.

(iii) when $V>c$, however, equation (\ref{eo2})  is hyperbolic, and it becomes possible to obtain
non-diffracting solutions of the type \ $\psi(\mathbf{r}_{\perp},z-Vt)$, both for beams and for pulses.

\h The latter simple and interesting result shows that, when basing ourselves on Eq.(\ref{eo2}), the solutions
that can propagate rigidly (that is, without any spatial modifications) are those corresponding to $V>c$.
In the case of beams, $V$
is merely the phase velocity; but in the case of pulses it is the peak velocity (sometimes identified
with the group-velocity). Incidentally, it is known that, when one superposes waves whose phase-velocity
does not depend on their frequency, such a phase-velocity becomes\footnote{Let us here recall that the group velocity
can be written as \ $\vbf_\grm = \nablabf_k \omega = {\partial \omega} / {\partial k_z \; \zbf}$ \ only when $k_x$ and
$k_y$ remain (almost) constant in the considered superposition, as it happens e.g. in the case of metallic guides.}
the actual peak-velocity\cite{MRF,Introd,Majorana}.

\h Many interesting solutions of this kind exist\cite{Livro,Lu1,Lu2,Ziolk4,PhysicaA}, and some of them
will be mentioned in this Chapter, and in the related 2014 Book. \ From the historical point of view, let us repeat that such
solutions to the wave equations (and, in particular, to the Maxwell equations, under weak hypotheses)
were theoretically predicted long time ago\cite{BarutMR,Stratton,Courant,Bateman}, mathematically constructed
in more recent times\cite{Lu1,PhysicaA,SupCh}, and soon after experimentally
produced\cite{Lu2,PSaari,Ranfagni,PSaari2}.

\h However, {\em it is rather restrictive to define a NDW as a solution of the type} (\ref{lw1}), {\em with} $V>c$.
Actually, also {\em subluminal} NDW solutions to the wave equations exist\cite{sub}, and they too are rather
interesting, as we shall discuss below.

\subsection{Definition of Non-diffracting Wave (NDW)}

Therefore, it is convenient to formulate a {\em more comprehensive} definition, wherefrom to derive a much ampler set of solutions
(super-luminal, luminal, or sub-luminal) capable of withstanding diffraction: Both for infinite distances, in the ideal
case (of infinite energy), and for large but finite distances, in the realistic case (of finite energy).

\h Let us start by formulating an adequate definition of an {\em ideal} Non-diffracting Wave.

\h Let us consider a linear and homogeneous wave equation in free
space. In cylindrical coordinates $(\rho,\phi,z)$ and using a
Fourier-Bessel expansion, its general solution $\psi(\rho,\phi,z,t)$
can be expressed, when disregarding evanescent waves, as
\bb \Psi(\rho,\phi,z,t) \ug
\sum_{n=-\infty}^{\infty}\left[\int_{0}^{\infty}d\kr\,\int_{-\infi}^{\infty}dk_z\,\int_{-\infty}^{\infty}d\om\,\kr
A_n^{'}(\kr,k_z,\om) J_n(\kr \rho)e^{ik_z z}e^{-i\om t}e^{i n
\phi} \right] \label{S2geral1} \ee                                                

with
\bb A_n^{'}(\kr,k_z,\om) \ug A_n(k_z,\om)\,\delta\left[\kr^2 -
\left(\frac{\om^2}{c^2} - k_z^2 \right)\right]
\label{S2sepc1} \; , \ee                                           

the $A_n(k_z,\om)$ being arbitrary functions, and $\delta(.)$ the
Dirac delta function. It is important to emphasize that the $J_n(\kr \rho)$ are $n$-order {\em Bessel functions.}
For simplicity, many authors often confined themselves to the zero-order Bessel functions $J_0(.)$.

\h An ideal NDW is a wave that must be capable of maintaining indefinitely its spatial form (except for local
variations) while propagating. This property may be mathematically expressed, when assuming propagation in the
$z$ direction, as follows:
\bb \Psi(\rho,\phi,z,t) \ug \Psi \left( \rho,\phi,z + \Delta z_0,t +
\frac{\Delta z_0}{V} \right) \; , \label{S2def} \ee                   

where $\Delta z_0$ is a chosen length, and $V$ is the
pulse-peak velocity, with $0\leq V \leq \infty$. \ Then, by using Eq.(\ref{S2geral1}) into Eq.(\ref{S2def}), and taking
account of Eq.(\ref{S2sepc1}), one can show\cite{MRH2,Livro,MRH} that any non-diffracting solution can be written

\bb \begin{array}{clcr} \Psi(\rho,\phi,z,t) \ug &
\dis{\sum_{n=-\infty}^{\infty}\,\sum_{m=-\infty}^{\infty}}\left[\dis{\int_{-\infty}^{\infty}\drm\om\,\int_{-\om/c}^{\om/c}\drm
k_z
A_{nm}(k_z,\om)}\right. \\

\\

& \times \left.J_n\left(\rho\sqrt{\dis{\frac{\om^2}{c^2} - k_z^2}}
\right)e^{ik_z z}e^{-i\om t}e^{i n \phi} \right]                                
\end{array}\label{S2geral2} \ee

with
\bb A_{nm}(k_z,\om) \ug S_{nm}(\om)\delta\left(\om -
(V k_z + b_m) \right)\label{Anm} \; , \ee                                       

it being $b_m = 2m\pi V/\Delta z_0$ (with $m$ an integer number too, of course), while quantity $S_{nm}(\om)$
is an arbitrary {\em frequency spectrum.} \ Notice that, due to Eq.(\ref{Anm}), each term in the
double sum (\ref{S2geral2}), namely in its expression within
square brackets, is a truly non-diffracting wave (beam or pulse);
and their sum (\ref{S2geral2}) is just the most general form
representing {\em an ideal NDW} according to definition (\ref{S2def}).

\h One should also notice that (\ref{S2geral2}) {\em is nothing but a
superposition of Bessel beams} with a specific ``space-time coupling",
characterized by {\em linear relationships}
between their angular frequency $\om$ and their longitudinal wave number $k_z$.

\h Concerning such a superposition, the Bessel beams with $\omega/k_z > 0$
($\omega/k_z < 0$) propagate in the positive (negative) $z$ direction. As
we wish to obtain NDWs propagating in the positive $z$ direction, it is not desirable
the presence of ``backward" Bessel beams, i.e. of
``backward components" ---often called {\em non-causal}, since they should be {\em entering}
the antenna or generator---.
The problems with the backward-moving components, that so frequently plague the
localized waves, can be however overcome by appropriate choices of the spectrum
(\ref{Anm}), which can totally eliminate those components, or
minimize their contribution, in superposition (\ref{S2geral2}). \ Let us notice that
often only positive velues of $\omega$ are considered ($0 \leq \omega \leq \infty$).

\h Another important point refers to the energy\cite{Sezginer,PIER98,Hillion1,MRH} of the
NDWs.  It is well known that any ideal
NDW, i.e., any field with the spectrum (\ref{Anm}), possesses
infinite energy. However finite-energy NDWs can be constructed by
concentrating the spectrum $A_{nm}(k_z,\om)$ in the surrounding of
a straight line of the type $\om = V k_z + b_m$ instead
of collapsing it exactly over that line\cite{MRH,MRH2}. In such a case, the NDWs
get a finite energy, but, as we knw, are endowed with {\em finite} field depths:
i.e., they maintain their spatial forms for long (but not
infinite) distances.

\h Despite the fact that expression (\ref{S2geral2}), with
$A_{nm}(k_z,\om)$ given by (\ref{Anm}), does represent ideal
non-diffracting waves, it is difficult to use it for obtaining
analytical solutions, especially when having the task of
eliminating the backward components.  This difficulty becomes even
worse in the case of finite-energy NDWs. We shall come back to this point
in Section 2.

\subsection{First examples}

Before going on, let us be more concrete. \ First of all, let us notice that
Eq.(\ref{S2geral1}), for $n=0$ and on
integrating over $k_z$, reduces to the less general ---but still quite
useful--- solution
\bb \psi(\rho,z,t) \ug
\dis{\int_{-\infi}^{\infi} \, \int_{0}^{\om/ck_\rr} \,
J_0(k_\rr\,\rho)\,e^{i\sqrt{\om^2/c^2
\,-\,k_\rr^2}\,\,z}\,e^{-i\om t}\,S(k_\rr,\om)\,\drm
k_\rr\,\drm\om} \label{sg} \ee                                     

where $S(k_\rr,\om)$ is now the chosen spectral function, with only $k_z>0$ (and
we still disregard evenescent waves). We are using the standard relation
\bb \frac{\om^2}{c^2} \ug k_\rr^2 + k_z^2  \; . \label{c1} \ee         

\h From the integral solution (\ref{sg}) one can get in particular, for instance,
the ({\bf non-localized}) gaussian beams and pulses, to which we
shall refer for illustrating the differences of the NDWs with respect to them.

\

{\em The Gaussian Beam} --- A very common (non-localized) beam is
the gaussian beam\cite{Molone}, corresponding to the spectrum
\bb S(k_\rr,\om) \ug 2a^2\,e^{-a^2 k_\rr^2}\,\delta(\om - \om_0) \; .
\label{eg} \ee

In Eq.(\ref{eg}), $a$ is a positive constant, which will be shown
to depend on the transverse aperture of the initial pulse.

\h The integral solution (\ref{sg}), with spectral function (\ref{eg}),
can be regarded as a {\em superposition} of plane waves: \ Namely, {\em of plane waves
propagating in all directions (always with $k_z \geq 0$), the most
intense ones being those directed along (positive) $z$\/} [especially when $\Delta k_\rho \equiv 1/a << \omega_0/c$]. \
This is clearly
depicted in Fig.1.4 of \cite{Livro}.
%
%

\h On substituting Eq.(\ref{eg}) into Eq.(\ref{sg}) and adopting
the paraxial approximation [which is known to be just valid if $\Delta k_\rho \equiv 1/a << \omega_0/c$], one meets the gaussian beam
\bb \psi_{\rm gauss}(\rho,z,t) \ug \frac{\dis{a^2\,{\rm
exp}\left(\dis{\frac{-\rho^2}{4(a^2 +
i\,z/2k_0)}}\right)}}{\dis{(a^2 +
i\,z/2k_0)}}\,\,\dis{e^{ik_0(z-ct)}}\;\; , \label{fg}\ee

where $k_0=\om_0/c$.  We can verify that the magniture $|\psi|$ of such a beam, which
suffers transverse diffraction, doubles its initial width
$\Delta\rho_0 = 2a$ \ after having traveled the distance $z_{{\rm
dif}} \ug \sqrt{3}\,k_0 \Delta\rho_0^2 /2$, called diffraction
length. The more concentrated a gaussian beam happens to be, the
more rapidly it gets spoiled.

\

{\em The Gaussian Pulse} --- The most common (non-localized) {\em
pulse} is the gaussian pulse, which is got from Eq.(\ref{sg}) by
using the spectrum\cite{chirped}
\bb S(k_\rr,\om) \ug
\frac{2ba^2}{\sqrt{\pi}}e^{-a^2k_\rr^2}e^{-b^2(\om-\om_0)^2}
\label{epg} \ee

where $a$ and $b$ are positive constants.  Indeed, such a pulse is
a superposition of gaussian beams of different frequency.

\h Now, on substituting Eq.(\ref{epg}) into Eq.(\ref{sg}), and
adopting once more the paraxial approximation, one gets the
gaussian pulse:
\bb \psi(\rho,z,t) \ug \frac{a^2\,{\rm
exp}\left(\dis{\frac{-\rho^2}{4(a^2+iz/2k_0)} }\right){\rm
exp}\left(\dis{\frac{-(z-ct)^2}{4c^2b^2}}
\right)}{a^2+iz/2k_0}\;\;, \label{pg} \ee

endowed with speed $c$ and temporal width $\Delta t = 2b$, and
suffering a progressing enlargement of its transverse width, so
that its initial value gets doubled already at position \ $z_{\rm
dif} \ug \sqrt{3}\,k_0 \Delta\rho_0^2 / 2\;$, \ with $\Delta\rho_0 =
2 a$. \ Let us remind that the paraxial approximation is really valid
in the pulse case only if there hold the two conditions \
$\Delta k_\rho \equiv 1/a << \omega_0/c$, \ and \  $\Delta \omega = 1/b << \omega_0$,
imposing a slow variation of the envelope.

\subsection{Further examples: The Non-diffracting Solutions}

Let us finally go on to the construction of the two most renowned
localized waves\cite{Tesi2}: the Bessel beam, and the ordinary X-shaped
pulse. \ First of all, let us recall that, when
superposing (axially symmetric) solutions of the wave equation
in the vacuum, three spectral parameters, $(\om, \ k_\rr, \ k_z)$,
come into the play, which have however to satisfy the constraint
(\ref{c1}), deriving from the wave equation itself. Consequently,
only two of them are independent: and we choose here $\om$ and $k_\rr$.  \ Such a
possibility of choosing $\om$ and $k_\rr$ was already apparent in the
spectral functions generating gaussian beams and pulses, which
consisted in the product of two functions, one depending only on
$\om$ and the other on $k_\rr$.

\h We are going to see that further particular relations between $\om$ and
$k_\rr$ [or, analogously, between $\om$ and $k_z$] can be moreover
enforced, in order to get interesting and unexpected results, such
as the {\em non-diffracting waves}.

\

{\em The Bessel beam} --- Let us start by imposing a {\em linear}
coupling between $\om$ and $k_\rr$ (it could be actually
shown\cite{Durnin2} that it is the unique coupling leading to
NDW solutions).

\h Namely, let us consider the spectral function
\bb S(k_\rr,\om) \ug \frac{\delta(k_\rr -
\dis{\frac{\om}{c}}\sin\theta)}{k_\rr}\,\,\delta(\om - \om_0)
\;\;, \label{eb} \ee

which implies that $k_\rr = (\om\sin\theta)/c$, \ with $0 \leq
\theta \leq \pi/2$: \ A relation that can be regarded as a
space-time coupling. Let us add that this linear constraint
between $\om$ and $k_\rr$, together with relation (\ref{c1}),
yields \ $k_z = (\om\cos\theta)/c$. This is an important fact, since
an {\em ideal} non-diffracting wave {\em must} contain\cite{Tesi2,MRH} a coupling
of the type $\om=V
k_z + b$, where $V$ and $b$ are arbitrary constants. \
The integral function (\ref{sg}), this time with spectrum (\ref{eb}), can be interpreted
as a superposition of plane waves too: But this time the
axially-symmetric Bessel beam appears as the result of the
{\em superposition of plane waves whose wave vectors lay on the surface
of a cone having the propagation line as its symmetry axis and an
opening angle equal to} $\theta$; such $\theta$ being called the
{\em axicon angle\/}. This is clearly shown in Fig.1.5 of \cite{Livro}.
%
%

\h By inserting Eq.(\ref{eb}) into Eq.(\ref{sg}), one gets the
mathematical expression of the so-called {\em Bessel beam:}
\bb \psi(\rho,z,t) \ug J_0\left(\frac{\om_0}{c}\sin\theta\,\,\rho
\right)\,{\rm
exp}\left[i\,\,\frac{\om_0}{c}\cos\theta\,\left(z-\frac{c}{\cos\theta}t\right)\right] \; .
\label{fb} \ee

This beam possesses phase-velocity $v_{\rm ph}=c/\cos\theta$,
and field transverse shape represented by a Bessel function
$J_0(.)$ so that its field in concentrated in the surroundings of
the propagation axis $z$. Moreover, Eq.(\ref{fb}) tells us that
the Bessel beam keeps its transverse shape (which is therefore
invariant) while propagating, with central ``spot'' $\Delta\rho =
2.405 c /(\om\sin\theta)$.

\h The ideal Bessel beam, however, is not square-integrable in the
transverse direction, and is therefore associated with an infinite
power flux: \ i.e., it
cannot be experimentally produced. \ One must have recourse to truncated Bessel beams, generated
by finite apertures: In this case the (truncated) Bessel beams are
still able to travel a long distance while maintaining their
transfer shape, as well as their speed, approximately
unchanged\cite{Durnin1,Durnin2,Durnin3,Overfelt}. For instance, the field-depth
of a Bessel beam generated by a circular finite aperture
with radius $R$ is given [if $R >> \Delta \rho_0 = 2.4 / k_\rho$]  by
\bb Z_{\rm max} \ug \frac{R}{\tan\theta} \; , \ee

where $\theta$ is the beam axicon angle. In the finite aperture
case, the Bessel beam cannot be represented any longer by
Eq.(\ref{fb}), and one must calculate it by the scalar
diffraction theory: by using, for example, Kirchhoff's or
Rayleigh-Sommerfeld's diffraction integrals. But till the
distance $Z_{\rm max}$ one may still use Eq.(\ref{fb}) for
approximately describing the beam, at least in the vicinity of the
axis $\rho=0$, that is, for $\rho << R$. To realize how much a
truncated Bessel beam succeeds in resisting diffraction, one can
consider also a gaussian beam, with the same frequency and central
``spot", and {\em compare} their field-depths. In particular, let
both the beams have $\lambda = 0.63\;\mu$m and initial central
``spot'' size $\Delta\rho_0 = 60\;\mu$m. The Bessel beam will
possess axicon angle $\theta=\arcsin[2.405
c/(\om\Delta\rho_0)]=0.004\;$rad. \ In the case, e.g., of a circular aperture
with radius $3.5\;$mm, it is then easy to verify that the gaussian beam
doubles its initial transverse width already after $3\;$cm, while
after $6\;$cm its intensity has become an order of magnitude
smaller. By contrast, the truncated Bessel beam keeps its
transverse shape until the distance $Z_{\rm max}=R/\tan\theta=
85\;$cm. Afterwards, the Bessel beam rapidly decays, as a
consequence of the sharp cut performed on its aperture (such cut
being responsible also for some intensity oscillations suffered by
the beam along its propagation axis, and for the fact that
the feeding waves, coming from the aperture, at a
certain point get eventually faint). \ All this is clearly depicted in Fig.1.6
of \cite{Livro}.

%
%

\h It may be useful to repeat that a Bessel beam is characterized by an
``extended focus" along its propagation axis, so that its energy cannot be
concentrated inside a small region in the transverse plane: It needs, on
the contrary, to be continuously reconstructed by the energy associated
with the ``lateral rings" (evolving along closing conical surfaces)
which constitute its transverse structure. \ This is quite different from the
case of a Gaussian beam, which possesses a point-like focus: That is, it is
constructed so as to concentrate its energy within a spot that becomes very small
at a certain point of its propagation axis, and afterwards rapidly diffracts.

\h The zeroth-order (axially symmetric) Bessel beam is nothing but
one example of localized beam. Further examples are the higher
order (not cylindrically symmetric) Bessel beams, described by
Eq.(1.13) of \cite{Livro},
%
%
or the Mathieu beams\cite{Mathieu}, and so on.

\

{\em The Ordinary X-shaped Pulse} --- Following the same procedure
adopted in the previous subsection, let us construct pulses by
using spectral functions of the type
\bb S(k_\rr,\om) \ug \frac{\delta(k_\rr -
\dis{\frac{\om}{c}}\sin\theta)}{k_\rr}\,\,F(\om) \; , \label{ex} \ee

where this time the Dirac delta function furnishes the spectral
space-time coupling $k_\rr = (\om\sin\theta)/c$. \ Function $F(\om)$
is, of course, the frequency spectrum; it is left for the moment
undetermined. \ On using Eq.(\ref{ex}) into Eq.(\ref{sg}), one obtains
\bb \psi(\rho,z,t) \ug \int_{-\infty}^{\infty}\,F(\om)\,
J_0\left(\frac{\om}{c}\sin\theta\,\,\rho \right)\,{\rm
exp}\left(\frac{\om}{c}\cos\theta\,\left(z-\frac{c}
{\cos\theta}t\right)\right) \, \drm\om \ . \label{px} \ee

It is easy to see that $\psi$ will be a pulse of the type
\bb \psi \ug \psi(\rho,z-Vt) \ee
with a speed $V=c/\cos\theta$ independent of the frequency
spectrum $F(\om)$.

\h Such solutions are known as X-shaped pulses, and are ({\em non-diffracting\/}) waves
in the sense that they do
obviously maintain their spatial shape during propagation
(see., e.g., Refs.\cite{Lu1,PhysicaA,MRH} and refs. therein;
as well as the following). Their peak velocity is well-known to be
superluminal (cf.also Refs.\cite{RepToSeshadri,Cherenkov,Folman} and refs therein). \
Some relevant, useful comments have been added, for instance, at pages 12-13 of \cite{Livro}.

\h Now, let us for instance consider in Eq.(\ref{px})
the particular frequency spectrum $F(\om)$ given by
\bb F(\om) \ug H(\om)\,\frac{a}{V}\,\,\,{\rm
exp}\left(-\frac{a}{V}\,\om\right) \; , \label{fx} \ee

where $H(\om)$ is the Heaviside step-function and $a$ a positive
constant. Then, Eq.(\ref{px}) yields
\bb \psi(\rho,\zeta) \, \equiv \, X \ug \frac{a}{\sqrt{(a -
i\zeta)^2 + \left(\frac{V^2}{c^2}-1\right)\rho^2}} \;, \label{ox}
\ee

still with $\zeta \equiv z - Vt$.  This solution (\ref{ox}) is the well-known
ordinary, or ``classic", X-wave, which constitutes a simple
example of a superluminal [supersonic, in the case of Acoustics] X-shaped pulse.\cite{Lu1,PhysicaA} \
Notice that function (\ref{fx}) contains mainly low frequencies, so that the
classic X-wave is suitable for low frequencies only.

\h Actually, Lu et al. first introduced them for Acoustic. Soon after
having mathematically and experimentally constructed
their ``classic" (acoustic)  X-wave, they started applying
them to ultrasonic scanning, directly obtaining
very high quality 3D images. \ Subsequently, in 1996 there were mathematically
constructed (see, e.g., Ref.\cite{PhysicaA} and refs. therein) the analogous
X-shaped solutions to the Maxwell equations, by working out scalar superluminal non-diffracting solutions
for each component of the Hertz potential. In reality, Ziolkowski et
al.\cite{Ziolk4} had already found in electromagnetism similar solutions for the simple scalar case,
called by them {\em slingshot} pulses; but their pioneering
solutions had gone almost unnoticed at that time (1993). \ In 1997 Saari et
al.\cite{PSaari} announced the first production of an X-shaped wave
in the optical realm, thus proving experimentally the existence of
superluminal non-diffracting electromagnetic pulses. \ Let us add that X-shaped waves have been
easily produced also in nonlinear media\cite{Conti}.

\h Figure \ref{fig8} depicts (the real part of) an ordinary
X-wave with $V=1.1\,c$ and $a=3\;$m.

\begin{figure}[!h]
\begin{center}
 \scalebox{0.7}{\includegraphics{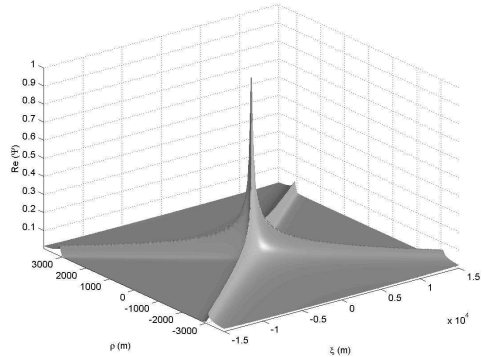}}  
\end{center}
\caption{Plot of the real part of the ordinary X-wave, evaluated
for $V=1.1\,c$ with $a=3\;$m .} \label{fig8}
\end{figure}

\h Solutions (\ref{px}), and in particular the pulse (\ref{ox}),
have got an infinite field-depth, and an infinite energy as well.
Therefore, as it was mentioned in the Bessel beam case, one should
pass to truncated pulses, originating from a finite aperture.
Afterwards, our truncated pulses will keep their spatial shape
(and their speed) all along the depth of field
\bb Z \ug \frac{R}{\tan\theta} \; , \ee

where, as before, $R$ is the aperture radius and $\theta$ the
axicon angle (and $R$ is assumed to be much larger than the X-pulse
spot).

\begin{figure}[!h]
\begin{center}
 \scalebox{1.8}{\includegraphics{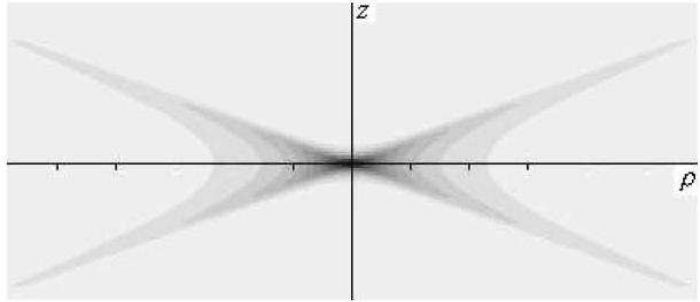}}  
\end{center}
\caption{All the X-waves (truncated or not) must have a leading cone in addition
to the rear cone, such a leading cone
having a role even for the peak stability\cite{Lu1}.  Long ago, this was also predicted,
in a sense, by (non-restricted\cite{IEEE,PhysicaA,Livro}) special relativity: one should not
forget, in fact, that {\em all} wave-equations, and not only Maxwell's, have an intrinsic
relativistic structure... By contrast, the fact that X-waves have a conic shape induced some authors
to look for (untenable) links ---let us now confine ourselves to electromagnetism--- between them and
the Cherenkov radiation, so as to to deny the existence of the leading cone: But X-shaped waves have nothing
to do with Cherenkov!, as it has been thoroughly demonstrated in Refs.\cite{Cherenkov,RepToSeshadri,Folman}.
In practice, when wishing to produce concretely a finite conic non-diffracting
wave, truncated both in space and in time, one is supposed to have recourse {\em in the simplest case} to a
dynamic antenna emitting a radiation cylindrically symmetric in
space and symmetric in time).\cite{Livro} } \label{fig7}
\end{figure}

\h At this point, it is worthwhile to add Fig.\ref{fig7},
with its detailed caption.

\h For further X-type solutions, with less and
less energy distributed along their ``arms", let us refer the reader
to papers like \cite{MRH,MRH2} and references therein, as well as to \cite{Livro}. For
example, it was therein addressed the explicit construction of
infinite families of generalizations of the classic X-shaped wave, with energy
more and more concentrated around their vertex: Cf., e.g., Figs.1.9 in \cite{Livro}).
Elsewhere, the techniques have been found for building up new series of non-diffracting
superluminal solutions to the Maxwell equations suitable for arbitrary frequencies
and bandwidths; and so on.

\section{Eliminating any Backward Components:\\ Totally Forward NDW Pulses}

As we mentioned, Eq.(\ref{S2geral2}), with its
$A_{nm}(k_z,\om)$ given by (\ref{Anm}), even if representing ideal
solutions, is difficult to be used for obtaining analytical solutions
with elimination of the ``non-causal" components; a difficulty which becomes
worse in the case of finite-energy NDWs. As promised, let us come
back to these problems putting forth a method\cite{MRH2} for getting exact
NDW solutions {\em totally free of backward components}.

\h Let us start with Eqs.(\ref{S2geral1},\ref{S2sepc1}), which
describe a general free-space solution (without evanescent waves)
of the homogeneous wave equation, and consider in
Eq.(\ref{S2sepc1}) a spectrum $A_n(k_z,\om)$ of the type
\bb A_n(k_z,\om) \ug \delta_{n\,0} H(\om)H(k_z)A(k_z,\om) \label{spec1}\ee

where $\delta_{n\,0}$ is the Kronecker delta function, $H(\cdot)$
the Heaviside function and $\delta(\cdot)$ the Dirac delta
function, quantity $A(k_z,\om)$ being an arbitrary function.
Spectra of the type (\ref{spec1}) restrict the solutions to the
axially symmetric case, with only positive values to the angular
frequencies and longitudinal wave numbers. With this, the
solutions proposed by us get the integral form
\bb \psi(\rho,z,t) \ug \int_{0}^{\infty}\drm\om\,\int_{0}^{\om/c}
\drm k_z \, A(k_z,\om) \, J_0(\rho\sqrt{\om^2/c^2 -
k_z^2})e^{ik_z}e^{-i\om t} \; , \label{superpo1} \ee

that is, they result to be general superpositions of zero-order Bessel
beams propagating in the positive $z$ direction only. Therefore,
any solution obtained from (\ref{superpo1}), be they
non-diffracting or not, are \emph{completely free} from backward
components.

\h At this point, we can introduce the {\em unidirectional} decomposition

\bb \left\{\begin{array}{clcr} \zeta \ug z-Vt\\
\\
\eta \ug z-ct\end{array}\right. \label{ze} \ee

with $V>c$.

A decomposition of this type has been used till now in the context of
paraxial approximation only\cite{paraxial,ContiPorras}; but we are going to show
that it can be much more effective, giving important results
in the context of exact solutions, and in situations that cannot be analyzed in
the paraxial approach.

\h With (\ref{ze}), we can write the integral solution
(\ref{superpo1}) as
\bb \psi(\rho,\zeta,\eta) \ug
(V-c)\int_{0}^{\infty}\drm\sigma\,\int_{-\infty}^{\sigma} \drm
\alpha \, A(\alpha ,\sigma) J_0(\rho\sqrt{\gamma^{-2}\sigma^2
-2(\beta-1)\sigma\alpha}\,\,)\,e^{-i\alpha\eta}e^{i\sigma\zeta}
\label{sup2} \ee

where $\gamma = (\beta^2 -1)^{-1/2}$, $\beta = V/c$ and where

\bb \left\{\begin{array}{clcr} \alpha \ug \dis{\frac{1}{V-c}}\,\,(\om - Vk_z)   \\
\\
\sigma \ug \dis{\frac{1}{V-c}}\,\,(\om - ck_z)
\end{array}\right. \label{as} \ee

are the new spectral parameters.

\h It should be stressed that superposition (\ref{sup2}) is not
restricted to NDWs: It is the choice of the spectrum
$A(\alpha,\sigma)$ that will determine the resulting NDWs.

\subsection{Totally forward ideal superluminal NDW pulses}

{\em The X-type waves}. \ The most trivial NDW solutions are the X-type
waves. We have seen that they are constructed by frequency
superpositions of Bessel beams with the same phase velocity $V>c$
and \emph{till now} constitute the only known ideal NDW pulses free
of backward components. It is not necessary, therefore, to use the
present method to obtain such X-type waves, since they
can be obtained by using directly the integral representation in
the parameters $(k_z,\om)$, i.e., by using Eq.(\ref{superpo1}).
Even so, let us use our new approach to
construct the ordinary X wave.

\h Consider the spectral function $A(\alpha,\sigma)$ given by
\bb A(\alpha,\sigma) \ug \frac{1}{V-c}\delta(\alpha)e^{-s\sigma}
\label{specx}\ee

One can notice that the delta function in (\ref{specx}) implies
that $\alpha = 0$ $\rightarrow$ $\om = Vk_z$, which is just the
spectral characteristic of the X-type waves. In this way, the
exponential function ${\rm exp}(-s\sigma)$ represents a frequency
spectrum starting at $\om=0$, with an exponential decay and
frequency bandwidth $\Delta\om=V/s$.

\h Using (\ref{specx}) in (\ref{sup2}), we get
\bb \psi(\rho,\zeta) \ug \frac{1}{\sqrt{(s-i\zeta)^2 +
\gamma^{-2}\rho^2}} \; \equiv X \,\,\, , \label{x} \ee

which is the well known ordinary X wave.

\

{\em{Totally forward Superluminal Focus Wave Modes}}. \ Focus wave modes
(FWM) \cite{Ziolk2,PIER98,MRH} are ideal
non-diffracting pulses possessing spectra with a constraint of the
type $\om = \mathcal{V}k_z + b$ (with $b\neq 0$), which links the
angular frequency with the longitudinal wave number, and are known
for their strong field concentrations.

\h Till now, all the known FWM solutions possessed, however,
backward spectral components, a fact that, as we know, forces one
to consider large frequency bandwidths to minimize their
contribution. However we are going to obtain solutions of this
type completely free of backward components, and able to possess
also very narrow frequency bandwidths.

\begin{figure}[!h]
\begin{center}
 \scalebox{2}{\includegraphics{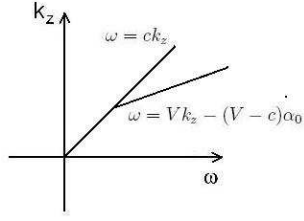}}
\end{center}
\caption{The Dirac delta function in (\ref{specfwm}) confines the
spectral parameters $\om,k_z$ of the Bessel beams to the straight
line $\om = Vk_z - (V-c)\alpha_0$, with $\alpha_0>0$.}
\label{fig1eps}
\end{figure}

\h Let us choose a spectral function $A(\alpha,\sigma)$ like
\bb A(\alpha,\sigma) \ug \frac{1}{V-c}\delta(\alpha +
\alpha_0)e^{-s\sigma} \label{specfwm}\ee

with $\alpha_0>0$ a constant. This choice confines the spectral
parameters $\om,k_z$ of the Bessel beams to the straight line $\om
= Vk_z - (V-c)\alpha_0$, as it is shown in figure \ref{fig1eps} below

\h Substituting (\ref{specfwm}) in (\ref{sup2}), we have
\bb \psi(\rho,\zeta,\eta) \ug
\int_{0}^{\infty}\drm\sigma\,\int_{-\infty}^{\sigma} \drm \alpha
\, \delta(\alpha + \alpha_0)e^{-s\sigma}
 J_0(\rho\sqrt{\gamma^{-2}\sigma^2
-2(\beta-1)\sigma\alpha})e^{-i\alpha\eta}e^{i\sigma\zeta} \,\,\, ,
\label{intfwm} \ee

which, on using identity $6.616$ in Ref.\cite{grad}, results in
\bb \psi(\rho,\zeta,\eta) \ug X\,e^{i\alpha_0\eta}\,{\rm
exp}\left[\frac{\alpha_0}{\beta+1}\left(s-i\zeta - X^{-1}
\right)\right] \label{fwm}\ee

where $X$ is the ordinary X-wave given by Eq.(\ref{x}).

\h Solution (\ref{fwm}) represents an ideal superluminal NDW of the
type FWM, but totally free from backward components.

\h As we already said, the Bessel beams constituting this solution
have their spectral parameters linked by the relation $\om = Vk_z
- (V-c)\alpha_0$; thus, by using (\ref{specfwm}) and (\ref{as}),
it is easy to see that the frequency spectrum of those Bessel
beams starts at $\om_{\rm min} = c\alpha_0$ with an exponential
decay ${\rm exp}(-s\om/V)$, and so possesses the bandwidth
$\Delta\om=V/s$. It is clear that $\om_{\rm min}$ and $\Delta\om$
can assume any values, so that the resulting FWM, Eq.(\ref{fwm}),
can range from a quasi-monochromatic to an ultrashort pulse. This
is a great advantage with respect to the old FWM solutions.

\begin{figure}[!h]
\begin{center}
 \scalebox{2}{\includegraphics{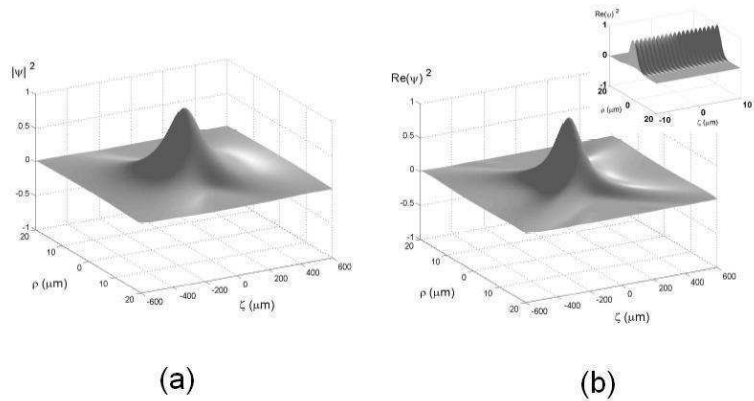}}
\end{center}
\caption{\textbf{(a)} and \textbf{(b)} show, respectively, the
intensity of the complex and real part of a quasi-monochromatic,
totally ``forward", superluminal FWM optical pulse, with
$V=1.5\,c$, $\alpha_0=1.256\times 10^7\,{\rm m}^{-1}$ and $s=
1.194\times 10^{-4}\,{\rm m}$, which correspond to $\om_{\rm
min}=3.77 \times 10^{15}$Hz, and $\Delta\om=3.77 \times
10^{12}$Hz, i.e., to a picosecond pulse with $\lambda_0=0.5\mu$m.}
\label{fig2eps}
\end{figure}

\h As an example, we plot in figure \ref{fig2eps} one case related with the NDW pulse
given by Eq.(\ref{fwm}).

\h In Figs.(\ref{fig2eps}) we have a quasi-monochromatic
optical FWM pulse, with $V=1.5\,c$, $\alpha_0=1.256\times
10^7\,{\rm m}^{-1}$ and $s= 1.194\times 10^{-4}\,{\rm m}$, which
correspond to $\om_{\rm min}=3.77 \times 10^{15}$Hz, and
$\Delta\om=3.77 \times 10^{12}$Hz, i.e., to a picosecond pulse
with $\lambda_0=0.5\mu$m. Figure (\ref{fig2eps}a) shows the intensity
of the complex NDW field, while Fig.\ref{fig2eps}b shows the
intensity of its real part. Moreover, in Fig.\ref{fig2eps}b, in the
right upper corner, it is shown a zoom of this NDW, on the $z$ axis
and around the pulse's peak, where the carrier wave of this
quasi-monochromatic pulse shows up.

\h Now, we apply our method to obtain totally forward
{\em finite-energy} NDW pulses.

\section{Totally Forward, {\em Finite-Energy} NDW Pulses}

Finite-energy NDW pulses are {\em almost} non-diffracting, in the sense
that they can retain their spatial forms, resisting to the
diffraction effects, for long (but not infinite) distances.

\h There exist many analytical solutions representing
finite-energy NDWs\cite{Ziolk2,PIER98,MRH}, but, once more, all the
known solutions suffer from the presence of backward components.
We can overcome this limitation.

\h We are going on skipping here the subluminal NDWs, which
will be addressed below in Section 5, where
also the particularly interesting case of the NDSs ``at rest"
(Frozen Waves) will be briefly considered.

\h Superluminal finite-energy NDW pulses, with peak velocity $V>c$,
can be got by choosing spectral functions in (\ref{superpo1})
which are concentrated in the vicinity of the straight line $\om =
Vk_z + b$ instead of lying on it. Similarly, in the case of
luminal finite-energy NDW pulses the spectral functions in
(\ref{superpo1}) have to be concentrated in the vicinity of the
straight line $\om = ck_z + b$ (note that in the luminal case, one
must have $b\geq 0$).

\h Indeed, from Eq.(\ref{as}) it is easy to see that, by our
approach, finite-energy superluminal NDWs can be actually obtained
by concentrating the spectral function $A(\alpha,\sigma)$ entering
in (\ref{sup2}), in the vicinity of $\alpha = -\alpha_0$, with
$\alpha_0$ a \emph{positive} constant. And, analogously, the
finite-energy luminal case can be obtained with a spectrum
$A(\alpha,\sigma)$ concentrated in the vicinity of $\sigma =
\sigma_0$, with $\sigma_0 \geq 0$.

\h To see this, let us consider the spectrum
\bb A(\alpha,\sigma) \ug \frac{1}{V-c}H(-\alpha -
\alpha_0)e^{a\alpha}e^{-s\sigma} \label{specfe}\ee

where $\alpha_0>0$, $a>0$ and $s>0$ are constants, and $H(\cdot)$
is the Heaviside function.

\h Due to the presence of the Heaviside function, the spectrum
(\ref{specfe}), when written in terms of the spectral parameters
$\om$ and $k_z$, has its domain in the region shown in Fig.\ref{fig4eps}
below.

\begin{figure}[!h]
\begin{center}
 \scalebox{3}{\includegraphics{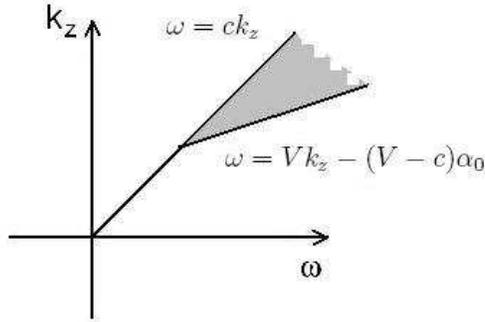}}
\end{center}
\caption{The spectrum (\ref{specfe}), when written in terms of the
spectral parameters $\om$ and $k_z$, has its domain indicated by
the shaded region.} \label{fig4eps}
\end{figure}

\h We can see that the spectrum $A(\alpha,\sigma)$ given by
Eq.(\ref{specfe}) is more concentrated on the line
$\alpha=\alpha_0$, i.e, around $\om = Vk_z - (V-c)\alpha_0$, or on
$\sigma=0$ (i.e., around $\om = ck_z$), depending on the values of
$a$ and $s$: More specifically, the resulting solution will be a
superluminal finite-energy NDW pulse, with peak velocity $V>c$, if
$a>>s$; or a luminal finite-energy NDW pulse if $s>>a$.

\h Inserting the spectrum (\ref{specfe}) into (\ref{sup2}), we
have
\bb \psi(\rho,\zeta,\eta) \ug
\int_{0}^{\infty}\drm\sigma\,\int_{-\infty}^{-\alpha_0} \drm
\alpha \, e^{a\alpha}e^{-s\sigma}
 J_0(\rho\sqrt{\gamma^{-2}\sigma^2
-2(\beta-1)\sigma\alpha}\,\,)e^{-i\alpha\eta}e^{i\sigma\zeta}
\,\,\, , \label{intfwm} \ee

and, by using identity $6.616$ in Ref.\cite{grad}, we get
an expression\cite{MRH2} that can be directly integrated to furnish
\bb \psi(\rho,\zeta,\eta) \ug \frac{\,\,X\,{\rm
exp}\dis{\left\{-\alpha_0\left[(a-i\eta)-\frac{1}{\beta+1}\left(s-i\zeta
- X^{-1} \right) \right] \right\}}}{\dis{(a-i\eta)
-\frac{1}{\beta+1}\left(s-i\zeta - X^{-1}\right)}}\,\,\, ,
\label{felw}\ee

\h As far as we know, the new solution (\ref{felw}) is the first
one to represent finite-energy NDWs completely free of backward
components\cite{MRH2}.

\

{\em{Totally forward, finite-energy superluminal NDW
pulses}}. \ The finite-energy NDW (\ref{felw}) can be superluminal
(peak-velocity $V>c$) or luminal (peak-velocity $c$) depending on
the relative values of the constants $a$ and $s$. To see this in a
rigorous way, in connection with solution (\ref{felw}), in Ref.\cite{MRH2}
it has been calculated how its global maximum of intensity (i.e, its peak),
which is located on $\rho=0$, develops in time. One then obtained the
peak's motion by considering the field intensity of (\ref{felw})
on the $z$ axis, i.e., $|\psi(0,\zeta,\eta)|^2$, at a given time
$t$, and finding out the value of $z$ at which the pulse presents
a global maximum. It was called  $z_{\rm p}(t)$ (the peak's
position) this value of $z$; and the peak velocity evaluated as
$\drm z_{\rm p}(t)/\drm t$.

\h As shown in Ref.\cite{MRH2}, superluminal finite-energy NDW pulses
may be obtained from (\ref{felw}) by putting $a>>s$. In this case,
the spectrum $A(\alpha,\sigma)$ is well concentrated around the
line $\alpha=\alpha_0$, and therefore in the plane $(k_z,\om)$
this spectrum starts at $\om_{\rm min} \approx c\alpha_0$ with an
exponential decay, and the bandwidth $\Delta\om \approx V/s$.

\h Defining the field
depth $Z$ as the distance over which the pulse's peak intensity
remains at least $25\%$ of its initial value\footnote{We can
expect that, while the pulse peak intensity is maintained, the
same happens for its spatial form.}, one obtains\cite{MRH2}
the depth of field
$$ Z \ug \frac{\sqrt{3}\,a}{1-\beta^{-1}} \; , $$

which depends on $a$ and $\beta=V/c$: Thus, the pulse can get
large field depths by suitably adjusting the value of parameter
$a$.

\begin{figure}[!h]
\begin{center}
\scalebox{2}{\includegraphics{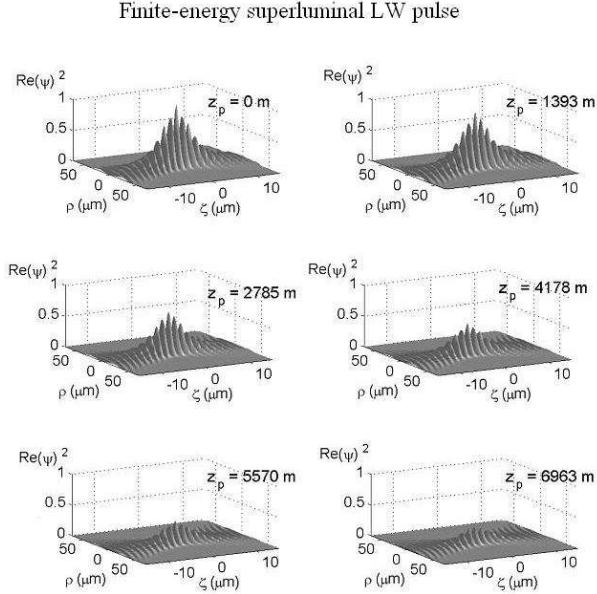}}
\end{center}
\caption{The space-time evolution, from the pulse's peak at $z_{\rm p}=0$ to $z_{\rm p}=Z$,
of a totally ``forward", finite-energy, superluminal NDW optical pulse represented by
Eq.(\ref{felw}), with the following parameter values: $a=20\,$m, $s=3.99\times 10^{-6}\,$m
(note that $a>>s$), $V=1.005\,c$ and $\alpha_0=1.26\times 10^{7}\,{\rm m}^{-1}$.}
\label{MRH2Fig5}
\end{figure}

\h Figure \ref{MRH2Fig5} shows the space-time evolution, from the
pulse's peak at $z_{\rm p}=0$ to $z_{\rm p}=Z$, of a
finite-energy superluminal NDW pulse represented by Eq.(\ref{felw})
with the following parameter values: $a=20\,$m, $s=3.99\times
10^{-6}\,$m (note that $a>>s$), $V=1.005\,c$ and
$\alpha_0=1.26\times 10^{7}\,{\rm m}^{-1}$. For such a pulse, we
have a frequency spectrum starting at $\om_{\rm min} \approx
3.77\times 10^{15}$Hz (with an exponential decay) and the
bandwidth $\Delta\om \approx 7.54 \times 10^{13}$Hz. From these
values and since $\Delta\om/\om_{\rm min}=0.02$, it is a optical
pulse with $\lambda_0 = 0.5\,\mu$m and time width of $13\,$fs. At
the distance given by the field depth
$Z=\sqrt{3}\,a/(1-\beta^{-1})=6.96\,$km the peak intensity is a
fourth of its initial value. Moreover, it is interesting to note
that, in spite of the intensity decrease, the pulse's spot size
$\Delta\rho_0 = 7.5\,\mu$m remains constant during the
propagation.

\

{\em{Totally ``forward", finite-energy luminal NDW
pulses}}. \ Luminal finite energy NDW pulses can be obtained from Eq.(\ref{felw}) by
making $s>>a$ (more rigorously, for $s^2c>>a^2V$).
In this case, the spectrum $A(\alpha,\sigma)$ is well concentrated
around the line $\sigma=0$, and therefore in the plane $(k_z,\om)$
it starts at $\om_{\rm min} \approx c\alpha_0$ with an exponential
decay and the bandwidth $\Delta\om \approx c/a$.

\h On defining the field depth $Z$ as
the distance over which the pulse's peak intensity remains at
least $25\%$ of its initial value, one obtains\cite{MRH2} the
depth of field
\bb Z \ug \frac{\sqrt{3}\,s}{\beta-1} \ee

which depends on $s$ and $\beta=V/c$.

\begin{figure}[!h]
\begin{center}
\scalebox{1.5}{\includegraphics{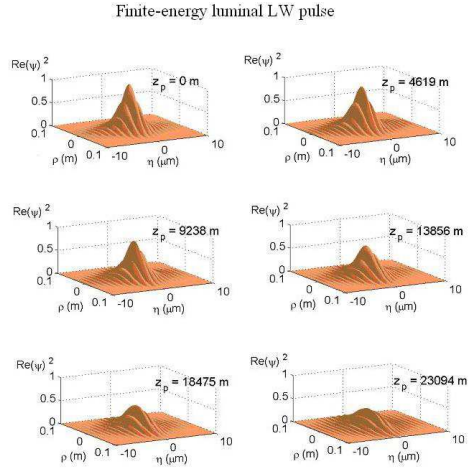}}
\end{center}
\caption{The space-time evolution, from the pulse's peak at
$z_{\rm p}=0$ to $z_{\rm p}=Z$,  of a totally ``forward",
finite-energy, luminal NDW optical pulse represented by
Eq.(\ref{felw}), with $a=1.59\times 10^{-6}$m, $s=1\times 10^{4}$m
(note that $s>>a$), $V=1.5\,c$, $\alpha_0=1.26\times 10^{7}\,{\rm
m}^{-1}$.} \label{fig6eps}
\end{figure}

\h Here, we confine ourselves to depict in Figure \ref{fig6eps} the space-time evolution
of such a pulse, from $z_{\rm p}=0$ to
$z_{\rm p}=Z$. At the distance
given by the field depth $Z=\sqrt{3}\,s/(\beta-1)=23.1\,$km the
peak intensity is just a fourth of its initial value.

\h We can see from the two examples above, and it can also be shown in
a rigorous way, that the superluminal NDW pulses obtained from
solution (\ref{felw}) are superior than the luminal ones obtained
from the same solution, in the sense that the former can possess
large field depths and, at the same time, be endowed with strong transverse
field concentrations. To obtain more interesting and efficient
luminal NDW pulses we should use\cite{MRH,MRH2} spectra concentrated around the
line $\sigma=\sigma_0>0$.

\subsection{A general functional expression for whatever totally-forward NDW pulses}

In the literature concerning the NDWs\cite{BandS} some
interesting approaches appear, yielding functional
expressions which describe NDWs in closed form. Although
interesting, even the NDWs obtained from those approaches
possess backward components in their spectral structure.

\h A general functional expression, capable of furnishing whatever totally-forward NDW pulses,
is, hovever\cite{MRH2}:
\bb \psi(\rho,\phi,\zeta,\eta) \ug e^{i\nu\phi}
\left(\frac{\gamma^{-1}\rho}{s-i\zeta +
X^{-1}}\right)^{\nu}X\,F\left( S \right)\,\, , \label{glw} \ee

with $F(.)$ an arbitrary function, and $X$ the ordinary X-wave (\ref{x}),
while $S$ is
$$ S \ug -i\eta - \frac{1}{\beta +1}(s-i\zeta - X^{-1}) \; . $$

Equation (\ref{glw}) is indeed an exact solution to the
wave equation that can yield both ideal and finite-energy NDW
pulses, with superluminal or luminal peak velocities. And
the NDW solutions obtained from Eq.(\ref{glw}) are totally free of
backward components under the only condition that the chosen
function $F(S)$ be regular and free of singularities at all
space-time points $(\rho,\phi,z,t)$.

\section{Method for the {\em Analytic} Description of {\em Truncated} Beams}

If we are allowed to set forth some more formal material, we like to present now an {\em analytic} method
for describing important beams, {\em truncated} by finite
apertures, in the Fresnel regime. The method works in Electromagnetism (Optics, Microwaves,...), as well as in Acoustics,
etcetera.  But we shall here confine ourselves to Optics, for conciseness' sake.

\h Our method\cite{MRB}, rigorous and effective but rather simple, is based on appropriate superpositions of Bessel-Gauss
beams, and in the Fresnel regime is able to describe in analytic form the 3D evolution of important waves,
like Bessel beams, plane waves, gaussian beams and Bessel-Gauss beams, when truncated by finite apertures. One of the
advantages of our mathematical method is that one can get in few seconds, or minutes, high-precision results which normally
require quite long times of numerical simulation. Indeed, also the coefficients of the Bessel-Gauss beam superpositions
result to be obtainable in a direct way, without any need of numerical evaluations or optimizations.

\subsection{The Method}

We shall leave understood in all solutions the harmonic time-dependence term \ $\exp (-i\om t)$. \ In the \emph{paraxial
approximation}, an axially symmetric monochromatic wave field can be evaluated, knowing its shape on the
$z=0$ plane, through the Fresnel diffraction integral in cylindrical coordinates:
\bb \Psi(\rho,z) =  \frac{-i k}{z}\, e^{i( k z + \frac{k\rho^2}{2z})} \dis{\int_{0}^{\infi}}\,\Psi(\rho',0)\,e^{ik\frac{\rho'^2}{2z}} \,J_0\left(k\frac{\rho\rho'}{z}\right)\rho' d\rho'
\ , \label{fresnel} \ee

where, as usual, $k=2\pi / \lambda$ is the wavenumber, and $\lambda$ the wavelength. In this equation,
$\rho'$ reminds us that the integration is being performed on the plane $z=0$; thus, $\Psi(\rho',0)$
does simply indicate the field value on $z=0$. \ An important solution is obtained by considering on the $z=0$ plane the
``excitation" given by
\bb \Psi(\rho',0) \ug \Psi_{BG}(\rho',0) \ug A J_0(\kr \rho')\exp (-q\rho'^{\,2}) \ , \label{bgz0} \ee

which\cite{Goodman} produces the so-called Bessel-Gauss beam\cite{Gori}:
\bb \Psi_{BG}(\rho,z) \ug -\frac{i k A}{2 z Q}\,e^{i k(z + \frac{\rho^2}{2z})} \, J_0\left(\frac{i k \kr \rho}{2zQ}\right)
e^{-\frac{1}{4Q}(\kr^2 + \frac{k^2\rho^2}{z^2})} \ ,
\label{bg} \ee

which is a Bessel beam transversally modulated by the Gaussian function.  \ Quantity $Q=q - ik/2z$, and $\kr$ is a constant
(namely, the transverse wavenumber associated with the modulated Bessel beam).  \ When $q=0$, the Bessel-Gauss beam
results in the well-known Gaussian beam. \ The Gaussian beam, and Bessel-Gauss', Eq.(\ref{bg}), are among the few
solutions to the Fresnel diffraction integral that can be obtained analytically. \ The situation gets
much more complicated, however, when facing beams truncated in space by finite circular apertures: For instance, a Gaussian
beam, or a Bessel beam, or a Bessel-Gauss beam, truncated
via an aperture with radius $R$. In this case, the upper limit of the integral in Eq.(\ref{fresnel}) becomes the aperture
radius, and the analytic integration becomes very difficult, requiring recourse to lengthy numerical calculations.

\h Let us now go on to our method for the description of truncated beams, characterized by simplicity and, in most cases,
total analyticity. \ Let us start with the Bessel-Gauss beam, Eq.(\ref{bg}), and consider the solution given by the
following superposition of such beams:
\bb \Psi(\rho,z) \ug - \frac{i k}{2 \, z}\,\,e^{i k(z + \frac{\rho^2}{2 \, z})} \dis{\sum_{n=-N}^{N}}\,
\frac{A_n}{Q_n}\, J_0\left(\frac{i \, k \,\kr \,\rho}{2 \, z \, Q_n}
\right)\,e^{-\frac{1}{4 Q_n}(\kr^2 + \frac{k^2 \rho^2}{z^2})} \; , \label{geral} \ee

quantities $A_n$ being constants, and $Q_n$ given by
\bb Q_n = q_n - \frac{ik}{2z} \label{Qn} \; , \ee

where the $q_n$ are constants that can assume {\em complex values.}  Notice that in this superposition all
beams possess the same value of $\kr$. \ We want the solution (\ref{geral}) to be able to represent beams truncated by
circular apertures, in the case, as we know, of Bessel beams, gaussian beams, Bessel-Gauss beams, and plane waves.

\h Given one of such beams, truncated at $z=0$ by an aperture with radius $R$, we have to determine the coefficients
$A_n$ and $q_n$ in such a way that Eq.(\ref{geral}) represents with
fidelity the resulting beam:  If the truncated beam on the $z=0$ plane is given by $V(\rho)$, we have to obtain
$\Psi(\rho,0) = V(\rho)$; that is to say
\bb V(\rho) \ug   J_0(\kr \rho)\,\dis{\sum_{n=-N}^{N}} A_n e^{-q_n\rho^2} \; . \label{geralz0} \ee

The r.h.s. of this equation is indeed nothing but a superposition of Bessel-Gauss beams, all with the same value of $\kr$,
at $z=0$ [namely, each one of such beams is written at $z=0$ according to Eq.(\ref{bgz0})].

\h One has to get the values of the $A_n$ and $q_n$, as well as of $N$, from Eq.(\ref{geralz0}).  Once these values
have been obtained, the field emanated by the finite circular
aperture located at $z=0$ will be given by Eq.(\ref{geral}). \ Remembering that the $q_n$ can be complex, let us make
the following choices:
\bb q_n \ug q_R + i q_{In} \;,\;\; {\rm with}\;\;\; q_{In} \ug - \frac{2 \pi}{L}\, n  \label{qvalor} \; , \ee

where $q_R>0$ is the real part of $q_n$, having the {\em same value} for
every $n$; $q_{In}$ is the imaginary part of $q_n$; and $L$ is a constant with the dimensions of a square length.

\h With such choices, and assuming $N \rightarrow \infi$, Eq.(\ref{geralz0}) gets written as
\bb V(\rho) \ug   J_0(\kr \rho)\,\exp \left(-q_{R} \rho^2 \right)\dis{\sum_{n=-\infi}^{\infi}} A_n \,
\exp \left(i\frac{2 \pi n}{L}\rho^2\right) \; , \label{geralz02} \ee

which has then to be exploited for obtaining the values of $A_n$, \ $\kr$, \ $q_R$ and $L$.

\h In the cases of a truncated Bessel beam (TB) or of a truncated Bessel-Gauss beam (TBG), it it natural to choose
quantity $\kr$ in Eq.(\ref{geralz02}) equal to the
corresponding beam transverse wavenumber. \  In the case of a truncated Gaussian beam (TG) or of a truncated plane wave (TP),
by contrast, in Eq.(\ref{geralz02}) it is natural to choose $\kr = 0$.  \ In all cases, the product
\bb \exp \left(-q_{R} \rho^2 \right)\dis{\sum_{n=-\infi}^{\infi}} A_n \, \exp \left(i\frac{2 \pi n}{L}\rho^2\right) \; ,
\label{expr} \ee

in Eq.(\ref{geralz02}) has to represent:

(i) a function $\circrm (\rho/R)$, in the TB or TP cases;

(ii) a function $\exp \left(-q \, \rho^2 \right)\,\circrm (\rho/R)$, that is, a $\circrm$ function multiplied by a Gaussian
function, in the TBG or TG cases. Of course (i) is a
particular case of (ii) with $q=0$. \ It may be useful to recall that the $\circrm $-function is the step-function
in the cylindrically symmetric case. Quantity $R$ is still the
aperture radius, and $\circrm (\rho/R) = 1$ when $0 \leq \rho \leq R$, and equals $0$ elsewhere.

\h Let us now show how expression (\ref{expr}) can approximately represent the above functions, given in (i) and (ii). \
To such an aim, let us consider a function $G(r)$ defined on an interval $|r| \leq L/2$ and endowed with the Fourier expansion
\bb G(r) \ug \sum_{n=-\infi}^{\infi} A_n \exp (i 2 \pi n r / L) \;\;\;  {\rm for} \;\;\;  |r| \leq L/2  \; , \label{G} \ee

where $r$ and $L$, having the dimensions of a square length, are expressed in square meters ($m^2$). \ Suppose now the
function $G(r)$ to be given by
\bb
 G(r) \ug \left\{\begin{array}{clr}
& \exp \left(q_{R} \, r \right)\,\exp \left(-q \, r \right) \;\;\; {\rm for} \;\;\; |r| \leq R^2  \\

\\

&0 \;\;\; {\rm for}\;\;\; R^2 < |r| < L/2  \ ,
\end{array} \right.  \label{G1}
 \ee

where $q$ is a given constant. 
Then, the coefficients $A_n$ in the Fourier expansion of $G(r)$ will be given by:
\bb A_n =  \dis{\frac{1}{L \, (q_R-q) - i 2\pi n}} \left(e^{(q_R-q - i\frac{2\pi}{L} n)R^2} -
\dis{e^{-(q_R-q - i\frac{2\pi}{L} n)R^2}}\right) \ . \label{An1} \ee

Writing now
\bb r = \rho^2  \label{subs} \; , \ee

we get that quantity (\ref{expr}) in Eqs.(\ref{G},\ref{G1}) can be written as
\bb e^{-q_{R} \rho^2}\,\sum_{n=-\infi}^{\infi} A_n e^{i 2 \pi n \rho^2 / L} = \left\{\begin{array}{clr}
e^{-q \, \rho^2} \;\; & {\rm for}\;\; 0 \leq \rho \leq R  \\

\\

0 \;\; & {\rm for}\;\; R < \rho \leq \sqrt{L/2} \\

\\

e^{-q_{R} \rho^2}\,f(\rho) \approx 0 \;\; & {\rm for}\;\; \rho > \sqrt{L/2}  \ ,
\end{array} \right.\label{16}
 \ee

where the coefficients $A_n$ are still given by Eq.(\ref{An1}), and $f(\rho)$ is a function existing on shorter and shorter
space intervals, assuming as it maximum value $\exp[(q_R-q)R^2]$, when $q_R > q$,  or 1, when
$q_R \leq q$. \  Since $\sqrt{L/2}>R$, for suitable choices of $q_R$ and $L$, we shall have that
$\exp(-q_{R} \rho^2)\,f(\rho)\approx 0$ for $\rho \geq \sqrt{L/2}$.

\h Therefore, we obtain that
\bb e^{-q_{R} \rho^2}\,\sum_{n=-\infi}^{\infi} A_n e^{i 2 \pi n \rho^2 / L} \approx \, e^{-q \, \rho^2}\,\circrm (\rho/R)
\label{expr2} \;\; , \ee

which corresponds to case (i), when $q=0$, and to case (ii). \ Let us recall once more that the $A_n$ are given
by Eqs.(\ref{An1}).

\h On the basis of what was shown before, we have now in our hands a rather efficient method for describing important
beams, truncated by finite apertures: Namely, the TB (truncated Bessel), TG (truncated Gauss), TBG (truncated Besse-Gauss), and
TP (truncated Plane wave) beams.\ Indeed, it is enough to choose the desired
field, truncated by a circular aperture with radius $R$, and describe it at $z=0$ by our Eq.(\ref{geralz02}).

Precisely:
\begin{itemize}
    \item{In the TBG case: the value of $\kr$ in Eq.(\ref{geralz02}) is the transverse wavenumber of the Bessel
      beam modulated by the Gaussian function; \ $A_n$ is given in Eq.(\ref{An1}); \ $q$ is related to the Gaussian function
      width at $z=0$. \ The values $L$ and $q_R$, and the number $N$ of terms in the series (\ref{geralz02}),
      are chosen so as to guarantee a faithful description of the beam at $z=0$ when truncated by a circular aperture
      with radius $R$.}
    \item{The TB, TG and TP are special cases of TBG: in TB, $q=0$; in TG, $\kr = 0$; and in TP, $\kr = 0$ and $q=0$}
\end{itemize}
In conclusion,  {\em once we know the chosen beam on the truncation plane ($z=0$), the beam emanated by the finite aperture
will then be given by solution (\ref{geral})}. Further details can be found in \cite{MRB}.

\h Let us go on to an important example.

\subsection{Application of the method to a truncated Bessel beam}

For the sake of brevity, we apply our method only to the analytic description of the truncated Bessel beam.

\h Let us consider a Bessel beam, with wavelength of 632.8 nm, truncated at $z=0$ by a circular aperture with radius $R$;
that is to say, let us start from $\Psi_{TB}(\rho,0) \ug J_0(\kr \rho)\,\circrm (\rho/R)$. \  Let us choose $R=3.5\;$mm,
and the transverse wavenumber $\kr = 4.07\cdot 10^4 \; {\rm m}^{-1}$, which corresponds
to a beam spot with radius approximatively equal to $\Delta\rho = 59\;\mu$m.

\h At $z=0$ the field is described by Eq.(\ref{geralz02}), where the $A_n$ are given by Eq.(\ref{An1}) and where $q=0$.
In this case, a quite good result can be obtained by the choice $L
=3R^2$, \ $q_R = 6/L$  and  $N=23$. Let us stress that, since such a choice is not unique, very many alternative sets
of values $L$ and $q_R$ exist, which yield excellent results as well.

\begin{figure}[!h]
\begin{center}
 \scalebox{0.7}{\includegraphics{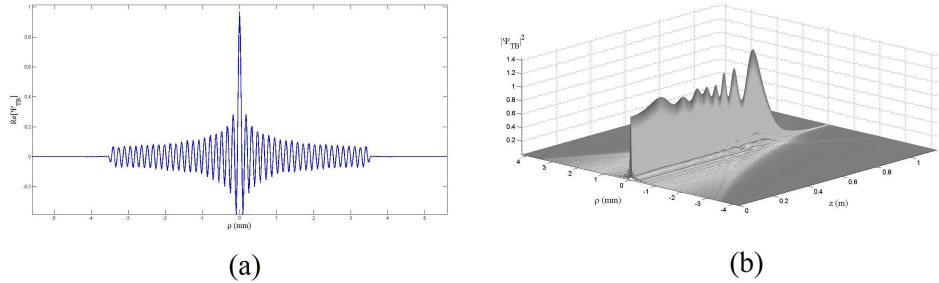}}
\end{center}
\caption{ (Color online)\textbf{(a)}  Field given by Eq.(\ref{geralz02}), representing a Bessel beam at $z=0$, with $\kr = 4.07\cdot 10^4 \; {\rm m}^{-1}$ and truncated by a finite
circular aperture with radius $R=3.5\;$mm. The coefficients $A_n$ are given by Eq.(\ref{An1}), with $q=0$, \ $L = 3R^2$, \ $q_R = 6/L$ and $N=23$. \textbf{(b)} Intensity of a Bessel
beam truncated by a finite aperture, as given by solution (\ref{geral}).} \label{MRBFig1ab}
\end{figure}

\h In the figures (\ref{MRBFig1ab}), image (a) depicts the field given by Eq.(\ref{geralz02}): it represents with high fidelity the Bessel beam truncated
at $z=0$. The resulting field, emanated by the aperture, is given by
solution (\ref{geral}), and its intensity is shown in image (b). One can see that the result really corresponds to a
Bessel beam truncated by a finite aperture.

\section{Subluminal Non-diffracting Waves (or Bullets)}

Let us now obtain in a simple way Non-diffracting
{\em subluminal} Pulses, always as exact analytic solutions to the wave
equations.\cite{sub} \  We shall adopt is this Section a less formal language (perhaps more
intuitive, or more physical), and we shall confine to ideal solutions, but such solutions will
be constructed for arbitrarily chosen frequencies and
bandwidths, and once more avoiding any recourse to the non-causal
(backward moving) components. \ Also the new solutions can be suitable
superpositions of ---zeroth-order, in general--- Bessel beams,
which can be performed by integrating either w.r.t. the angular
frequency $\om$, or w.r.t. the longitudinal wavenumber $k_z\,$: \ Both approaches are
treated below. \ The first one is powerful enough; \ we sketch
the second approach as well, however, since it allows also dealing
---from a new starting point--- with the limiting
case of {\em zero-speed} solutions: Namely, it furnishes a new way, in
terms of {\em continuous} spectra, for obtaining such (``frozen") waves\cite{FWart2,MichelOE2,MichelOE1,Prego}
so promising also from the point of view of the applications. \ Some
attention is successively paid to the known role of Special
Relativity, and to the fact that the NDWs are expected
to be transformed one into the other by suitable Lorentz
Transformations. \ We are moreover going to mention the case of non
axially-symmetric solutions, in terms of higher order Bessel
beams. \ We keep
fixing our attention especially on electromagnetism and optics:
But let us repeat that results of the same kind are valid
whenever an essential role is played by a wave-equation [like in
acoustics, seismology, geophysics, elementary particle physics (as
we shall explicitly see in the slightly different case of the
Schroedinger equation), and also gravitation (for which we have
recently got stimulating new results), and so on].

\h Subluminal NDWs can be obtained too by suitable superpositions of
Bessel beams,\cite{sub}, as in the other cases, but have been rather
neglected for the mathematical difficulties in
getting analytic expressions for them, since the superposition integral runs over a finite
interval. \ Therefore, almost all the few papers devoted to the
subluminal NDWs had recourse to the paraxial\cite{Molone}
approximation\cite{paraxial}, or to numerical
simulations\cite{Salo3}. Only {\em one} analytic solution
was known\cite{9,10,11,Donnelly,MRH2}, biased by the inconvenience
that its frequency spectrum is very large, that it does not
possess a well-defined central frequency, and that
backward-travelling\cite{Ziolk2,PIER98} components were needed for constructing it. \ In this Section
we construct, on the contrary, non-diffracting exact solutions with any spectra,
in any frequency bands and for any bandwidths; and without employing\cite{MRH,Introd} backward-traveling
components. \ One can arrive at such (analytic) solutions, both in the case of integration
over the Bessel beams' angular frequency $\om$, and of
integration over their longitudinal wavenumber $k_z$.

\subsection{A first method for constructing physically \\
acceptable, subluminal Non-diffracting Pulses}

Axially-symmetric solutions to the scalar wave equation are known
to be superpositions of zero-order Bessel beams over the angular
frequency $\om$ and the longitudinal wavenumber $k_z$: \ That is, in
cylindrical co-ordinates,
\bb \Psi(\rho,z,t) \ug
\int_{0}^{\infty}\,\drm\om\,\int_{-\om/c}^{\om/c}\,\drm k_z\,
\overline{S}(\om,k_z) J_0\left(\rho\sqrt{\frac{\om^2}{c^2} -
k_z^2}\right)e^{ik_z z}e^{-i\om t} \; , \label{Eq.(1)} \ee      

where, as usual, \ $k_\rho^2 \equiv \om^2/c^2 - k_z^2$ \ is the transverse
wavenumber; and quantity $k_\rho^2$ has to be positive since
evanescent waves are here excluded. \ We already know that the condition characterizing a non-diffracting wave is the
existence\cite{PIER98,OptComm} of a linear
relation between longitudinal wavenumber $k_z$ and frequency $\om$ for
all the Bessel beams entering the superposition; that is to say, in the chosen
spectrum for each Bessel beam if has to be\cite{MRH,Tesi2}
\bb \om \ug v\,k_z + b \; \, \label{Eq.(2)} \ee          

with $b \geq 0$. [More generally, as shown in Ref.\cite{MRH}, in the plane $\om, k_z$
the chosen spectrum has to call into the play, if
not exactly such a line, at least a region in the proximity of a
straight-line of that type. In the latter case
one obtains solutions endowed with finite energy, and therefore a finite
``depth of field"].

\h The requirement (\ref{Eq.(2)}) is a specific space-time coupling, implied by the chosen spectrum
$\overline{S}$. Let us recall that Eq.(\ref{Eq.(2)}) has to be obeyed by the spectra of
any one of the three possible types (subluminal, luminal or
Superluminal) of non-diffracting pulses: Indeed, with the choice (\ref{Eq.(2)}), the pulse re-gains
its initial shape after the space-interval ${\Delta z}_1 = 2\pi
v/b$. \ [But the more general case can be also
considered\cite{MRH,Capitulo} when $b$ assumes any values
$b_m=m\,b$ (with $m$ an integer), and the periodicity
space-interval becomes ${\Delta z}_m = {\Delta z}_1 / m\,$. \
We are referring ourselves, now, to the real (or imaginary)
part of the pulse, since its magnitude is endowed with
rigid motion].

\h Let us first derive in the subluminal case the only exact solution
known till recent time, the Mackinnon's\cite{9} one, represented by Eq.(\ref{Eq.(10)}) below.
Since the
transverse wavenumber $k_\rho$ of each Bessel beam entering
Eq.(\ref{Eq.(1)}) has to be real, it can be easily shown (as first noticed in
\cite{Salo3})
that in the subluminal case $b$ cannot vanish, but it must be $b>0$. \ Then,
on using conditions (\ref{Eq.(2)}) and $b>0$, the subluminal
localized pulses can be expressed as integrals over the frequency only:
\bb \Psi(\rho,z,t) \ug  \exp{[-ib {z \over v}]} \,
\int_{\om_-}^{\om_+}\,\drm\om \; S(\om) \, J_0(\rho k_\rho)\,
\exp{[i\om {\zeta \over v}]} \; , \label{Eq.(3)} \ee           

where now
\bb k_\rho \ug {1 \over v} \, \sqrt{2b\om - b^2
- (1-v^2/c^2) \om^2} \; \, \label{Eq.(4)} \ee                   

with
\bb \zeta \, \equiv \,  z - v t \; \, \label{Eq.(5)} \ee    

and with
\bb
\left\{ \begin{array}{clr}
\om_{-} \ug \dis{{b \over {1+v/c}}} \\
\\
\om_{+} \ug \dis{{b \over {1-v/c}}}
\end{array}   \right. \; . \label{Eq.(6)}
\ee                                        

As anticipated, the Bessel beam superposition in the subluminal
case is an integration over a finite interval of $\om$,
which also does show that the backward-travelling
components correspond to the interval $\om_- < \om < b$. \ [It
could be noticed that Eq.(\ref{Eq.(3)})  does not represent the most general
exact solution, which on the contrary is {\em a
sum\/}\cite{Capitulo} of such solutions for the various possible
values of $b$ mentioned above: That is, for the values $b_m=m\,b$
with spatial periodicity ${\Delta z}_m = {\Delta z}_1 / m\,$. \
But we can confine ourselves to solution (\ref{Eq.(3)})  without any real
loss of generality, since the actual problem is evaluating in
analytic form the integral entering Eq.(\ref{Eq.(3)}). For any mathematical
and physical details, see Ref.\cite{Capitulo}].

\h Now, if one adopts the change of variable
\bb \om \, \equiv \, {b \over {1-v^2/c^2}}\;(1 + {v \over c} s) \; \,
\label{Eq.(7)} \ee    

equation (\ref{Eq.(3)}) becomes\cite{Salo3}
\begin{eqnarray}
\lefteqn{\Psi(\rho,z,t) \ug {b \over c} \,{v \over {1-v^2/c^2}} \, \exp{[-i
{b \over v} z]} \;
\exp{\left[ i{b \over v} \, {1 \over {1-v^2/c^2}} \, \zeta \right]}} \nonumber
\\
& & {} \times \int_{-1}^{1}\,\drm s \; S(s) \, J_0\left( {b \over
c} \, {\rho \over {\sqrt{1-v^2/c^2}}} {\sqrt{1-s^2}} \right) \;
\exp{\left[ i {b \over c} {1 \over {1-v^2/c^2}} \zeta s \right]}
\; . \label{Eq.(8)}
\end{eqnarray}                                               

In the following we shall adhere ---as it is an old habit of ours--- to some symbols
standard in  Special Relativity, since the whole topic of subluminal, luminal
and Superluminal NDWs is strictly connected\cite{IEEE,PhysicaA,SupCh}
with the principles and structure of SR [cf.\cite{BarutMR,Review} and refs. therein],
as we shall mention also in the specific remarks which follow below). Namely, we put
 \ $\beta \equiv v/c \;$ \ and \ $\gamma \equiv 1 / {\sqrt{1-\beta^2}} \;$.                   

\h Equation (\ref{Eq.(8)}) has till now yielded {\em one}
analytic solution, for $S(s) \, = \;$constant: the {\em Mackinnon
solution\/}\cite{9,Donnelly,11,JosaMlast}
\begin{eqnarray}\lefteqn{ \Psi(\rho,\zeta,\eta) \ug 2{b \over c} v \, \gamma^2\,
{\exp{\left[ i{b \over c} \, \beta \gamma^2 \; \eta \right]}}}
\nonumber
 \\
& & {} \times \sinc \ {\sqrt{ {b^2 \over c^2}\, \gamma^2 \left(
\rho^2 + \gamma^2 \; \zeta^2 \right)
}}  \; , \label{Eq.(10)}                                        
\end{eqnarray}

which however, for its above-mentioned drawbacks, is endowed with
little physical and practical interest. \ In Eq.(\ref{Eq.(10)}) the $\sinc$
function has the ordinary definition \ $\sinc \ x \, \equiv \, (\sin \; x)/x \,$, and
\bb \eta \, \equiv \, z -Vt, \ \ \ \ \ \ \ {\rm with} \ \ V \equiv
{c^2 \over v}
 \; , \label{Eq.(11)} \ee  \  

where $V$ and $v$ are related by the {\em de Broglie relation.} \ Notice that
$\Psi$ in Eq.(\ref{Eq.(10)}), and in the following ones, is eventually a function
(besides of $\rho$) of $z,t$ only via quantities $\zeta$ and $\eta$.

\h However, we can construct further subluminal pulses, corresponding to whatever spectrum
 and devoid of backward-moving components, just by exploiting the fact that in our equation (\ref{Eq.(8)}) the integration
interval is finite: That it, by transforming it into a good, instead
of a harm. Let us first observe that Eq.(\ref{Eq.(8)}) will also yield exact, analytic solutions
for {\em any} exponential spectra of the type
\bb
S(\om) \ug \exp{[{i2n\pi \om \over \Omega}]}
 \; , \label{Eq.(12)}
\ee  

with $n$ any integer number: Which means for any spectra of this type it holds \
$S(s)= \exp{[in\pi / \beta]} \,
\exp{[in\pi s]}$, \ as can be easily checked. \ In Eq.(\ref{Eq.(12)}) we have set \
$ \Omega \, \equiv \, {\om_+} - {\om_-} \; . $ \ In this more general case, the solution writes
\begin{eqnarray}
\lefteqn{ \Psi(\rho,\zeta,\eta) \ug 2b\beta\, \gamma^2 \,
\exp{\left[ i{b \over c} \, \beta \, \gamma^2 \, \eta \right]}}
\nonumber
 \\
& & {} \times \exp{[in {\pi \over \beta}]}
 \ \sinc \ {\sqrt{{b^2 \over c^2}\, \gamma^2 \, \rho^2
+ \left( {b \over c} \, \gamma^2 \, \zeta + n \pi
\right)^2}}
 \; . \label{Eq.(13)}
\end{eqnarray}     

Notice that also in Eq.(\ref{Eq.(13)}) quantity $\eta$ is
defined as in Eqs.(\ref{Eq.(11)}) above, where $V$ and $v$ obey the de
Broglie relation $vV=c^2$, the subluminal quantity $v$ being the
velocity of the pulse envelope, and $V$ playing the role (in the
envelope's interior) of a superluminal phase velocity.

\h We now take {\em advantage}
of the finiteness of the integration limits for
expanding any arbitrary spectra $S(\om)$ in a Fourier series in the interval
$\om_{-} \leq \om \leq \om_{+}\;$; that is:
\bb S(\om) \ug \sum_{n=-\infty}^{\infty} \, A_n \,
\exp{[+in {2\pi \over \Omega} \om]} \; , \label{Eq.(14)}
\ee                                                           

where (we went back, now, from the $s$ to the $\om$ variable):
\bb A_n \ug {1 \over \Omega} \, \int_{\om_-}^{\om_+} \drm \om \, S(\om) \,
\exp{[-in {2 \pi \over \Omega} \om]}  \; \,
\label{Eq.(15)} \ee                                               

quantity $\Omega$ being defined above.

\h Then, on remembering the special, ``Mackinnon-type" solution (\ref{Eq.(13)}),
we can infer from expansion
(\ref{Eq.(12)}) that, for any arbitrary spectral function $S(\om)$, one can work out
a rather general axially-symmetric analytic solution for the subluminal case:
\begin{eqnarray}
\lefteqn{ \Psi(\rho,\zeta,\eta) \ug 2b\beta\, \gamma^2 \,
{\exp{\left[ i{b \over c} \, \beta \, \gamma^2 \; \eta \right]}}}
\nonumber
 \\
& & {} \times \sum_{n=-\infty}^{\infty} \, A_n \, \exp{[in{\pi
\over \beta}]} \ \sinc \ {\sqrt{ {b^2 \over c^2}\, \gamma^2 \rho^2
+ \left( {b \over c} \gamma^2 \; \zeta + n \pi \right)^2}}
\; , \label{Eq.(16)}     
\end{eqnarray}

coefficients $A_n$ being still given by Eq.(\ref{Eq.(15)}).

\h The present approach presents several advantages. We can easily choose
spectra localized within the prefixed frequency interval (optical waves, microwaves,
etc.) and endowed with the desired bandwidth.   Moreover, we have seen that
spectra can now be chosen
such that they have zero value in the region $\om_{-} \leq \om \leq b$,
which is responsible for the backward-traveling components of the
subluminal pulse. \ Even when the adopted spectrum $S(\om)$
does not possess a known Fourier series (so that the coefficients
$A_n$ cannot be exactly evaluated via Eq.(\ref{Eq.(15)}), one can calculate
approximately such coefficients without meeting any problem, since
our general solutions (\ref{Eq.(16)}) will still be exact solutions.

\h Let us set forth some examples.

\subsection{Examples}

In general, optical pulses generated in the lab possess a
spectrum centered at some frequency value, $\om_0$, called the
carrier frequency. \ The pulses can be, for instance, ultra-short,
when $\Delta\om/\om_0 \geq 1$; or quasi-monochromatic, when
$\Delta\om/\om_0 << 1$, where $\Delta\om$ is the spectrum
bandwidth.

\h These kinds of spectra can be mathematically represented by a
gaussian function, or by functions with similar behaviour. \ One can find various
examples In Refs.\cite{AIEP,sub}.

\

{\em First example} --- Let us consider, e.g., a gaussian spectrum:
\bb S(\om) \ug \frac{a}{\sqrt{\pi}}\exp{[-a^2(\om-\om_0)^2]} \label{Eq.(17)} \ee   

whose values are negligible outside the frequency interval $\om_-
< \om < \om_+$ over which the Bessel beams superposition in Eq.(\ref{Eq.(3)})
is made, it being  $\om_- = b/(1+\beta)$ and $\om_+ =
b/(1-\beta)$. [Let us stress that, once $v$ and $b$ have been fixed, the values of
$a$ and $\om_0$ can afterwards be selected in order to kill the
backward-travelling components, that correspond, as we know, to $\om < b\;$].
The Fourier expansion in Eq.(\ref{Eq.(14)}), which yields, with the above
spectral function (\ref{Eq.(17)}), the coefficients
\bb A_n \, \simeq \, \dis{{{1} \over {W}}} \, \exp{[-in {2\pi
\over \Omega} \om_0]} \;
\exp{[{{-n^2 \pi^2} \over {a^2 W^2}}]} \; , \label{Eq.(18)} \ee   

constitutes an excellent representation of the gaussian spectrum (\ref{Eq.(17)}) in
the interval $\om_- < \om < \om_+$ (provided that, as we requested, our
gaussian spectrum does get {\em negligible} values outside the frequency interval
$\om_- < \om < \om_+$). \ In other words, a subluminal pulse with frequency
spectrum (\ref{Eq.(17)}) can be written as Eq.(\ref{Eq.(16)}), with the coefficients
$A_n$ given by Eq.(\ref{Eq.(18)}): the evaluation of such coefficients $A_n$
being rather simple. \ Let us repeat that, even if the values of
the $A_n$ are obtained via a (rather good, by the way)
approximation, we based ourselves on the {\em exact} solution
Eq.(\ref{Eq.(16)}).

\begin{figure}[!h]
\begin{center}
 \scalebox{.5}{\includegraphics{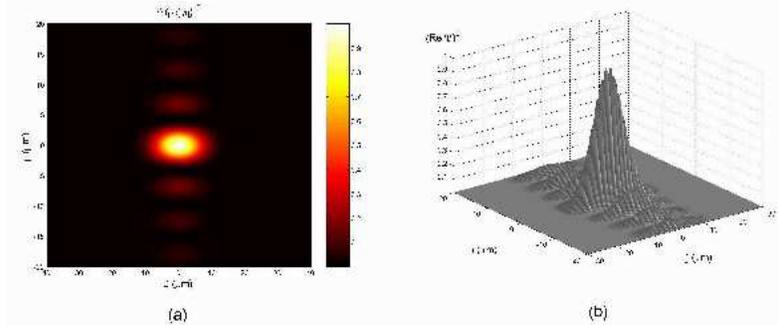}}
\end{center}
\caption{\textbf{(a)} The intensity orthogonal projection for a pulse
corresponding to Eqs.(70,71) in the case of an optical frequency: Namely,
for a subluminal pulse with velocity $v=0.99\;c$, carrier angular
frequency $\om_0= 2.4\times 10^{15}\;$Hz (that is,
$\la_0=0.785\;\mu$m) and bandwidth (FWHM) $\Delta\om=\om_0
/20=1.2\times 10^{14}\;$Hz; which results to be an optical pulse of $24$ fs. One has also to specify
a value for the the frequency: let it be
$b=3\times 10^{13}\;$Hz; as a consequence one has
$\om_{-}=1.507\times 10^{13}\;$Hz and $\om_{+}=3\times
10^{15}\;$Hz. \ [This is exactly a case in which the
pulse has no backward-traveling components,
since the chosen spectrum possesses totally negligible values for
$\om < b$]; \ \textbf{(b)} The three-dimensional intensity
pattern of the {\em real part} of the same pulse, which reveals the
carrier wave oscillations.} \label{fig36}
\end{figure}

\h One can, for instance, obtain exact solutions representing
subluminal pulses for optical frequencies: Cf. Figs.\ref{fig36}.  \ The construction of the
considered pulse results already satisfactory when considering about 51 terms ($-25 \leq n \leq
25$) in the series entering Eq.(\ref{Eq.(16)}).

\h Figures \ref{fig36} show that pulse, evaluated  just by summing the mentioned
fifty-one terms: Fig.(a) depicts the orthogonal projection
of the pulse intensity; \ Fig.(b) shows the three-dimensional
intensity pattern of the \emph{real part} of the pulse, which
reveals the carrier wave oscillations.

\h Let us stress (see below) that the ball-like shape for the field intensity is typically associated with
the subluminal NDWs, while the typical Superluminal ones are
known to be X-shaped\cite{Lu1,PhysicaA,SupCh}, as predicted
since long by special relativity in its ``non-restricted"
version: See Refs.\cite{BarutMR,Review,PhysicaA,IEEE,RepToSeshadri} and refs
therein. \ It can be indeed noted
that each term of the series in Eq.(\ref{Eq.(16)}) corresponds to an
ellipsoid or, more specifically, to a spheroid, for each velocity
$v$.

\

{\em A second example} --- Let us consider now the very simple
case when, within the integration limits $\om_{-}$, $\om_{+}$,
the complex exponential spectrum (\ref{Eq.(12)}) is replaced by the real
function (still linear in $\om$)
\bb S(\om) \ug \frac{a}{1-\exp{[-a(\om_+ - \om_-)]}}\,\exp{[a(\om
- \om_{+}]}
 \; , \label{Eq.(20)}
\ee  

with $a$ a positive number [for $a=0$ one goes back to the
Mackinnon case]. Spectrum (\ref{Eq.(20)}) is exponentially concentrated in
the proximity of $\om_{+}$, where it reaches its maximum value,
and (on the left of $\om_{+}$) becomes more and more concentrated
as the arbitrarily chosen value of $a$ increases; its
frequency bandwidth being $\Delta\om=1/a$.

\h On performing the integration as in the case of spectrum (\ref{Eq.(12)}), instead
of solution (\ref{Eq.(13)}) in the present case one eventually gets the solution
\begin{eqnarray}
\lefteqn{ \Psi(\rho,\zeta,\eta) \ug \frac{2ab\beta\gamma^2 \,
\exp{[ab\gamma^2]} \, \exp{[-a\om_{+}]}}{1-\exp{[-a(\om_+ -
\om_-)]}}}
 \nonumber
 \\
 \nonumber
 \\
 & & {} \times
 \exp{\left[ i{b \over c} \, \beta \, \gamma^2 \, \eta \right]}
 \ \sinc{\left[ {b \over c} \, \gamma^2 \; {\sqrt{\gamma^{-2} \, \rho^2
- (av+i\zeta)^2}} \right]} \; . \label{Eq.(21)}
\end{eqnarray}     

\h This Eq.(\ref{Eq.(21)}) appears to be the simplest
closed-form solution, after Mackinnon's, since both of them do not need any recourse
to series expansions. In a sense, our solution (\ref{Eq.(21)}) may be
regarded as the subluminal analogue of the (Superluminal) X-wave
solution; a difference being that the standard X-shaped solution
has a spectrum starting with 0, where it assumes its maximum
value, while in the present case the spectrum starts at $\om_{-}$
and gets increasing afterwards, till $\om_+$. \ More important is
to observe that the gaussian spectrum has a priori two advantages
w.r.t. Eq.(\ref{Eq.(20)}): It may be more easily centered around any value
$\om_{0}$ of $\om$, and, when increasing its concentration in the
surroundings of $\om_{0}$, the spot transverse width does not
increase indefinitely, but tends to the spot-width of a Bessel
beam with $\om=\om_0$ and $k_z=(\om_0 - b)/V$, at variance with
what happens for spectrum (\ref{Eq.(20)}). \ Anyway, solution (\ref{Eq.(21)}) is
noticeable, since it is really {\em the simplest} one. \ An example is
consituted by Figure 37 in \cite{AIEP}, referring to an optical pulse of 0.2 ps.

\subsection{A second method for constructing\\
subluminal Non-diffracting Pulses}

The previous method appears to be very efficient for finding out analytic
subluminal NDWs, but it looses its validity in the limiting case $v
\rightarrow 0$, since for $v=0$ it is $\om_{-} \equiv \om_{+}$ and the
integral in Eq.(\ref{Eq.(3)}) degenerates, furnishing a null value. \ By
contrast, we are interested also in the $v=0$ case, since it corresponds,
as we said, to some of the most interesting, and potentially useful, NDWs: \
That is, to the ``stationary" solutions to the wave equations endowed with
a {\em static} envelope, and that we call Frozen Waves. \ Before going on, let us recall
that the theory of Frozen Waves was
initially developed in Refs.\cite{MichelOE1,FWart2}, by having recourse to discrete
superpositions in such a way to bypass the need of numerical simulations. \
[In the case of continuous superpositions, some numerical
simulations had been performed in Refs.\cite{Dartora12}]. \ However,
the method presented in this subsection does allow finding out analytic
exact solutions, without any need of numerical simulations,
also for Frozen Waves consisting in {\em continuous superpositions}.

\h Actually,
we are going to see that the present method works regardless of the chosen
field-intensity shape, and in regions with size of the order
of the wavelength. \ It is possible to get such results by starting again from Eq.(\ref{Eq.(1)}),
with constraint (56), but going on ---this time--- to integrals over
$k_z$, instead of over $\om$. \ It is enough to write relation (\ref{Eq.(2)})
in the form \ $k_z \ug (\om - b) / v$, \ for expressing the exact solutions (\ref{Eq.(1)}) as
\bb \Psi(\rho,z,t) \ug  \exp{[-ibt]} \;
\int_{k_z{\rm min}}^{k_z{\rm max}}\,\drm k_z \; S(k_z) \, J_0(\rho k_\rho)\,
\exp{[i \zeta k_z]} \; , \label{Eq.(22)} \ee     

with
\bb
k_z{\rm min} \ug \dis{ {-b \over c} \, {1 \over {1+\beta}}} \; ; \ \ \ \ \ \ \ \ \ \
k_z{\rm max} \ug \dis{ {b \over c} \, {1 \over {1-\beta}} }  \label{Eq.(23)}
\ee        

and with
\bb {k_\rho}^2 \ug -{k_z^2 \over \gamma^2} + 2 {b \over c} \beta k_z +
{b^2 \over c^2} \; , \label{Eq.(24)} \ee    

where quantity $\zeta$ is still defined according to Eq.(\ref{Eq.(5)}), always with
$v<c$.

\h One can show that the unique exact solution previously
known\cite{9} may be rewritten in form (75) {\em with}
$S(k_z) = \;$constant. \ Then, on following the same procedure exploited
in our first method,
%
%
%
%
%
%
one can again observe \cite{AIEP} that any spectra $S(k_z)$ can be expanded,
on the interval $k_z{\rm min} < k_z < k_z{\rm max}$, into the Fourier
series:
\bb S(k_z) \ug \sum_{n=-\infty}^{\infty} \, A_n \,
\exp{[+in {2\pi \over K} k_z]} \; , \label{Eq.(28)}
\ee    

with coefficients given now by
\bb A_n \ug {1 \over K} \, \int_{k_z{\rm min}}^{k_z{\rm max}} \drm k_z \,
S(k_z) \, \exp{[-in {2 \pi \over K} k_z]} \; \,
\label{Eq.(29)} \ee    

where \ $ K \equiv k_z{\rm max} - {k_z{\rm min}}$.

\h At the end of the whole procedure\cite{AIEP}, the general exact solution
representing a subluminal NDW, for any spectra $S(k_z)$, can be
eventually written
\begin{eqnarray}
\lefteqn{ \Psi(\rho,\zeta,\eta) \ug 2\,{b \over c} \, \gamma^2 \,
\exp{[i{b \over c} \, \beta \, \gamma^2 \, \eta]}}  \nonumber
 \\
& & {} \times \sum_{n=-\infty}^{\infty} \, A_n \, \exp{[in \pi \beta]}
 \ \sinc \ {\sqrt{{b^2 \over c^2}\, \gamma^2 \, \rho^2 +
\left( {b \over c} \, \gamma^2 \, \zeta + n \pi \right)^2}}
 \; , \label{Eq.(30)}
\end{eqnarray}     

whose coefficients are expressed in Eq.(\ref{Eq.(29)}), and where quantity $\eta$
is defined as above, in Eq.(\ref{Eq.(11)}).

\h Interesting examples could be easily worked out.

\section{\bf ``Stationary" solutions with zero-speed envelopes: {\em Frozen Waves}}  

Here, we shall refer ourselves to the (second) method, expounded
above. \ Our solution (\ref{Eq.(30)}), for the case of
envelopes {\em at rest}, that is, in the case $v=0$ [which implies
$\zeta = z$], becomes

\hfill{$
\Psi(\rho,z,t) \ug \dis{ 2\,{b \over c} \; \exp{[-i b t]} \; \sum_{n=-\infty}^{\infty} \, A_n \
\sinc \ {\sqrt{{b^2 \over c^2}\, \rho^2 +
\left( {b \over c} \, z + n \pi \right)^2}}} \; ,
$\hfill} (85)        

with coefficients $A_n$ given by Eq.(\ref{Eq.(29)}) with $v=0$, so that its integration
limits simplify into $-b/c$ and $b/c$, respectively. Thus, one gets

\hfill{$ A_n \ug \dis{{c \over {2b}} \, \int_{-b/c}^{b/c} \drm k_z \,
S(k_z) \, \exp{[-in {c \pi \over b} k_z]}} \; .
$\hfill} (83')        

\h Equation (85) is a new exact solution, corresponding to
``stationary" beams with a {\em static} intensity envelope. \ Let us
observe, however, that even in this case one has energy
propagation, as it can be easily verified from the power flux
$\Sbf_{\rm s} = -\nabf\Psi_{\cal R} \ \pa\Psi_{\cal R}/\pa t$
(scalar case) or from the Poynting vector $\Sbf_{\rm v} = (\Ebf
\wedge \Hbf)$ (vectorial case: the condition being that
$\Psi_{\cal R}$ be a single component, $A_z$, of the vector
potential $\Abf$).\cite{PhysicaA} \ We have indicated by
$\Psi_{\cal R}$ the real part of $\Psi$. \ For $v=0$, Eq.(\ref{Eq.(2)})
becomes \ $ \om \ug b \, \equiv \, \om_0$, \ so that the particular subluminal
waves endowed with null velocity are actually monochromatic beams.  [Let us explicitly observe that
for smaller and smaller speeds the subluminal NDWs become {\em pulses} more and more localized
in space: However, for $v=0$ the frequency bandwidth becomes zero, and we end up no
longer with pulses but with {\em beams\/}].

\h We size the present opportunity for presenting here two simple figures, which
recall in an intuitive way some of the geometrical characteristics of our {\em Frozen Waves:}

\begin{figure}[!h]
\begin{center}
 \scalebox{4.0}{\includegraphics{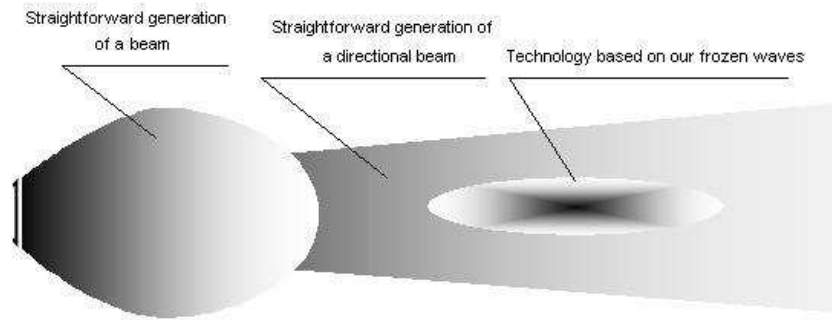}}
\end{center}
\caption{See the explications contained in the figures themselves,
which apparently refer to: \ (a) ordinary (electromagnetic or
acoustical) transmission; \ (b) standard directional transmission;
\ and \ (c) well-localized transmission allowed by our Frozen Wave
techniques. [Courtesy of Andrei Utkin]} \label{CNPqfig1}
\end{figure}

\newpage

\begin{figure}[!h]
\begin{center}
 \scalebox{3.5}{\includegraphics{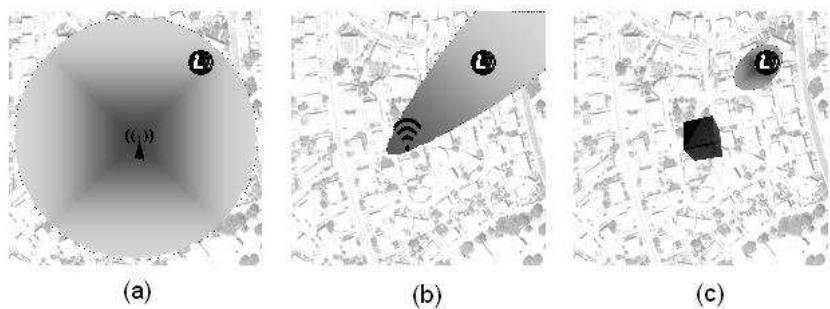}}
\end{center}
\caption{Areas covered by the electromagnetic (or acoustic)
signals in the case, once more, of: \ (a) omnidirectional
trasnsmission; \ (b) standard directional transmission; \ and \
(c) spot-to-spot signal transmission permitted by our Frozen Wave
tecniques. [Courtesy of Andrei Utkin]} \label{CNPqfig2}
\end{figure}

\h It may be stressed that the present (second) method, without any need of the paraxial approximation,
does yield {\em exact} expressions for (well localized) beams with sizes {\em of the
order} of their wavelength. \ It may be noticed, moreover, that the
already-known exact solutions ---for instance, the Bessel beams---
are nothing but particular cases of solution (85).

{\em An example:} \ On choosing (with $0 \leq q_- < q_+ \leq 1$)
the spectral double-step function

\

\hfill{$
S(k_z) \ug \left\{ \begin{array}{clr}
\dis{\frac{c}{\om_0(q_+ - q_-)}} \ \ \ \ \ \ \ \ \ \ & {\rm for} \ q_-\om_0/c \leq k_z \leq q_+\om_0/c \\
\\
0 \ \ \ \ \ \ \ \ \ \ & {\rm elsewhere} \; ,
\end{array}   \right.
$\hfill} (86)        

\

the coefficients of Eq.(85) become

\

\hfill{$ A_n \ug \dis{\frac{ic}{2\pi n\om_0(q_+ - q_-)} \; \left[
e^{-iq_+\pi n} - e^{-iq_-\pi n} \right]}
 \; .
$\hfill} (87)        

\

\h The double-step spectrum (86) corresponds, with regard to the longitudinal wave number,
to the mean value $\overline{k}_z = \om_0(q_++q_-)/2c$ \ and to
the width $\Delta k_z=\om_0(q_+-q_-)/c$. From such relations, it follows
that $\Delta k_z/\overline{k}_z = 2(q_+-q_-)/(q_++q_-)$.

\h For values of $q_-$ and $q_+$ that do not satisfy the inequality $\Delta
k_z / \overline{k}_z << 1$, the resulting solution will be a {\em non-paraxial}
beam.

\h An exact solution can be found in Figure 38 of \cite{AIEP}, which describes a beam with a spot
diameter of $0.6\;\mu$m (for $\lambda_0 = 1 \; \mu$m) and, moreover, with a rather good
longitudinal localization. \ In the case considered therein, about 21
terms in the sum entering Eq.(83) resulted to be quite
enough for a good evaluation of the series. Such a beam was
highly non-paraxial (having $\Delta
k_z/\overline{k}_z = 1$ ), and therefore could not have been
obtained by ordinary gaussian beam solutions, which are valid in
the paraxial regime only. Notice that, for simplicity, we are referring ourselves
to scalar wave fields only; but, in the case of non-paraxial optical beams, the
vector character of the field has to be taken into account.
%
%

\subsection{A new approach to the {\em Frozen Waves}}

A noticeable property of our present
method is that it allows a spatial modeling even of monochromatic
fields (that correspond to envelopes {\em at rest\/}; so that, in the
electromagnetic cases, one can speak, e.g., of the modeling of
``light-fields at rest"). \ Let us repeat  that such a modeling
---rather interesting, especially for applications\cite{brevetto}---
was already performed in
Refs.\cite{MichelOE1,FWart2,MichelOE2}, in terms of discrete superpositions
of Bessel beams.

\h But the method presented in the last Section allows us to make use of
{\em continuous} superpositions, in order to get a predetermined longitudinal
(on-axis) intensity pattern, inside a desired space interval
$0<z<L$. \ Such continuous superposition writes\cite{AIEP,FWart2}

\

\hfill{$ \Psi(\rho,z,t) \ug \dis{ e^{-i\om_0 t} \;
\int_{-\om_0/c}^{\om_0/c} \drm k_z \ S(k_z) \ J_0(\rho k_\rho) \
e^{iz k_z} } \; ,
$\hfill} (88)    

\

which is nothing but the previous Eq.(\ref{Eq.(20)}) with $v=0$
(and therefore $\zeta = z$). \ In other words, Eq.(88) does just
represent a {\em null-speed} subluminal wave. \ The FWs were expressed in the past as discrete
superpositions, because it was not known at that time how to treat
analytically a continuous superposition like (88). \ We are now able, however,
to deal also with
integrals: without numerical simulations, as we said, but in terms once more
of analytic solutions.

\h Indeed, the exact solution of Eq.(88) is given by Eq.(85), with
coefficients (83');  and one can choose the spectral function
$S(k_z)$ in such a way that $\Psi$ assumes the on-axis pre-chosen
static intensity pattern $|F(z)|^2$. \ Namely, the equation to be
satisfied by $S(k_z)$, to such an aim, is derived by associating
Eq.(88) with the requirement $|\Psi(\rho=0,z,t)|^2 = |F(z)|^2$,
which entails the integral relation

\hfill{$  \dis{\int_{-\om_0/c}^{\om_0/c} \drm k_z \; S(k_z) \;
e^{i z k_z}} \ug F(z) \; .
$\hfill} (89)    

Equation (89) would be trivially solvable in the case of an
integration between $-\infty$ and $+\infty$, since it would merely
be a Fourier transformation; but obviously this is not the case,
because its integration limits are finite. Actually, there are
functions $F(z)$ for which Eq.(89) is not solvable at all, in the sense
that no spectra $S(k_z)$ exist obeying the last equation. \
For instance, if we consider the {\em Fourier} expansion

\hfill{$ F(z) \ug \dis{\int_{-\infty}^{\infty}} \drm k_z \;
\widetilde{S}(k_z) \; e^{iz k_z} \; ,
$\hfill}

when $\widetilde{S}(k_z)$ does assume non-negligible values outside
the interval $-\om_0/c < k_z < \om_0/c$, then in Eq.(89) {\em no} $S(k_z)$
can forward that particular $F(z)$ as a result.

\h {\em However}, some procedures can be devised, such that one can
nevertheless find out a function $S(k_z)$ that approximately
(but satisfactorily) complies with Eq.(89).

\h {\em The first procedure} consists of writing $S(k_z)$ in the form

\hfill{$ S(k_z) \ug \dis{ {1 \over K} \; \sum_{n=-\infty}^{\infty}
\, F \left( {{2n\pi} \over K} \right) \; e^{-i2n\pi k_z / K} } \; ,
$\hfill} (90)    

where, as before, $K=2\om_0/c$. Then, Eq.(90) can be easily verify
as guaranteeing that the integral in Eq.(89) yields the
values of the desired $F(z)$ {\em at the discrete points} \ $z =
2n\pi / K$. \ Indeed, the Fourier expansion (90) is already of the
same type as Eq.(82), so that in this case the coefficients $A_n$
of our solution (85), appearing in Eq.(83'), do simply become

\hfill{$
A_n \ug \dis{{1 \over K} \; F(-{{2n\pi} \over K})} \; .
$\hfill} (91)    

\h This is a powerful way for obtaining a desired longitudinal (on-axis)
intensity pattern, especially for tiny spatial
regions, because it is not necessary to solve any integral to find
out the coefficients $A_n$, which by contrast are given directly by Eq.(91).

\begin{figure}[!h]
\begin{center}
 \scalebox{.5}{\includegraphics{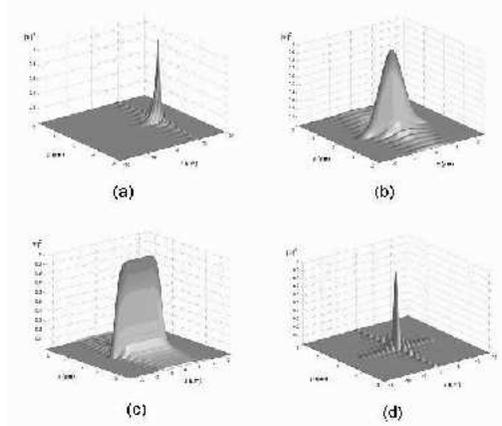}}
\end{center}
\caption{Frozen Waves with the on-axis longitudinal field pattern
chosen as: \textbf{(a)} Exponential; \ \textbf{(b)} Gaussian;  \
\textbf{(c)} Super-gaussian;  \ \textbf{(d)} Zero order \ Bessel
function} \label{fig39}
\end{figure}

\h Figures \ref{fig39} depict some interesting applications of this method.
A few  desired longitudinal intensity patterns $|F(z)|^2$ have been chosen,
and the corresponding Frozen Waves calculated by using Eq.(85) with the
coefficients $A_n$ given in Eq.(91). The desired patterns are enforced
to exist within very small spatial intervals only, in order to show the
capability of the method to model\cite{sub} the field intensity shape
also under such strict requirements.

\h In the four examples below, we considered a wavelength
$\lambda=0.6\;\mu$m, which corresponds to $\om_0=b=3.14\times
10^{15}\;$Hz. \ Details can be found in \cite{AIEP}

The first longitudinal (on-axis) pattern considered by us is
 \ $F(z) = \exp [a(z-Z)]$ for $0 \leq z \leq Z$, \ and zero elsewhere;
that is a pattern with an exponential increase, starting from $z=0$
untill $Z=10\;\mu$m and with $a=3/Z$. \ The intensity of the corresponding Frozen Wave is
shown in Fig.\ref{fig39}a.

The second longitudinal pattern (on-axis) taken into consideration
is the gaussian one, given by \ $F(z) = \exp [-q (z/Z)^2]$ \ for $-Z \leq z \leq Z$,
and zero elsewhere, with $q=2$ and $Z=1.6\;\mu$m. \ The intensity of the corresponding
Frozen Wave is shown in Fig.\ref{fig39}b.

In the third example, the desired longitudinal pattern is supposed to
be a super-gaussian, \ $F(z) = \exp [-q(z/Z)^{2m}]$ \ for $-Z \leq z \leq Z$,
and zero elsewhere, where $m$ controls the edge sharpness. We choose
$q=2$, $m=4$ and $Z=2 \; \mu$m. \ The intensity of the Frozen Wave
obtained in this case is shown in Fig.\ref{fig39}c.

Finally, in the fourth example, let us choose the longitudinal
pattern as being the zero-order Bessel function \ $F(z) = J_0(q\,z)$
for $-Z \leq z \leq Z$, and zero elsewhere,
with $q=1.6\times 10^{6}\;\rm{m}^{-1}$ and $Z=15\;\mu$m. \ The intensity
of the corresponding Frozen Wave is shown in Fig.\ref{fig39}d.

\h Any static envelopes of this type can
be easily transformed into propagating pulses by the mere application
of Lorentz transformations (LT).

\

\h {\em Another procedure} exists for evaluating $S(k_z)$, based on the assumption
that \ $S(k_z) \, \simeq \, \widetilde{S}(k_z)$, which constitutes a good approximation
whenever $\widetilde{S}(k_z)$ assumes
{\em negligible} values outside the interval $[-\om_0/c, \ \om_0/c]$. \ In
such a case, one can have recourse to the method associated with Eq.(\ref{Eq.(28)})
and expand $\widetilde{S}(k_z)$ itself in a Fourier series, getting
eventually the relevant coefficients $A_n$ by Eq.(\ref{Eq.(29)}). \ Let us recall
that it is still $K \equiv k_z{\rm max} - k_z{\rm min} = 2\om_0/c$.

\h It is worthwhile to call attention to the circumstance
that, when constructing FWs in terms of a sum of discrete
superpositions of Bessel beams (as it has been done in
Refs.\cite{MichelOE1,FWart2,MichelOE2,MRH,brevetto}), it was easy to obtain
extended envelopes like, e.g., ``cigars": Where ``easy" means by
using only a few few terms of the sum. By contrast, when we construct
FWs ---following this Section--- as continuous superpositions,
then it is easy to get highly {\em localized} (concentrated) envelopes.
\ Let us explicitly mention, moreover, that the method presented
in this Section furnishes FWs that are no longer periodic along
the $z$-axis (a situation that, with our old
method\cite{MichelOE1,FWart2,MichelOE2,MRH}, was obtainable only when the
periodicity interval tended to infinity).

\subsection{Frozen Waves in absorbing media}

Let us mention that it is possible to obtain even in {\it absorbing media} non-diffracting {\em ``stationary"} wave fields
capable to assume, approximately, any desired longitudinal
intensity pattern within a chosen interval $0\leq z \leq L$ of the
propagation axis $z$. These new solutions are more easily realizable in practice, at the extent to be
more indicated for the various applications already mentioned by us.

\h We know that, when propagating in a non-absorbing medium, the
NDWs\cite{MichelOE1,FWart2} maintain their spatial shape
for long distances. The situation is not the same when
dealing with absorbing media: In such cases, both the ordinary
and the non-diffracting beams (and pulses) will be
exponentially attenuated along the propagation axis.

\h It can be however shown that, through suitable superpositions
of equal-frequency Bessel beams, it is possible to obtain
non-diffracting beams {\it in absorbing media}, whose longitudinal
intensity pattern can assume any desired shape within a chosen
interval $ 0 \leq z \leq L$ of the propagation axis $z$. For example,
one can obtain non-diffracting beams capable to
resist the loss effects, maintaining amplitude and spot size of
their central core for long distances.

\h The corresponding method, with some interesting examples, is expounded in
Ref.\cite{MichelOE2} and in Chapter 2 of \cite{Livro}

\subsection{Experimental production of the Frozen Waves}

Frozen Waves have been eventually produced, in recent times\cite{bbi28},
in Optics, as reported also in another Chapter of this Volume; while we
expect their production also in Acoustics, even if till the present
moment only simulated experiments have been performed\cite{Prego}.

\

\section{On The role of Special Relativity, and of Lorentz Transformations}

Strict connections exist between, on one hand, the principles and
structure of Special Relativity and, on the other hand, the whole
subject of subluminal, luminal, Superluminal Localized Waves, and
it is expected since long time that a priori they
are transformable one into the other via suitable Lorentz
transformations (cf. Refs.\cite{BarutMR,Review,JWeber,NorthHoll,chmonPLA,Review0,NCim,Saari2004}).

Let us first confine ourselves to the cases faced in this last
Sections. Our subluminal localized pulses, that may be called ``wave
bullets", behave as {\em particles\/}: Indeed, our subluminal
pulses [as well as the luminal and Superluminal (X-shaped) ones,
that have been so amply investigated in the past literature] do exist
as solutions of any wave equations, ranging from electromagnetism
and acoustics or geophysics, to elementary particle physics (and
even, as we discovered recently, to gravitation physics). \ From
the kinematical point of view, the velocity composition
relativistic law holds also for them. \ The same is true, more in
general, for any localized waves (pulses or beams).

\h Let us start for simplicity by considering, in an initial reference-frame
O, just a ($\nu$-order) Bessel beam \ $\Psi(\rho,\phi,z,t) \ug J_\nu(\rho k_\rho) \;
e^{i\nu\phi} \; e^{iz k_z} \; e^{-i\om t}$. \ In a second reference-frame O', moving
with respect to (w.r.t.) O with speed $u$ ---along the positive z-axis and in the positive
direction, for simplicity's sake---, it will be observed\cite{Saari2004} the
new Bessel beam
\bb
\Psi(\rho ',\phi ',z ',t ') \ug J_\nu(\rho ' {k'}_{\rho'}) \;
e^{i\nu\phi '} \; e^{iz' {k'}_{z'}} \; e^{-i\om ' t'} \; ,
\label{Eq.(42)} \ee    

obtained by applying the
appropriate Lorentz transformation (a Lorentz
``boost") with \ $\gamma = [\sqrt{1-u^2/c^2}]^{-1}$, and \ ${k'}_{\rho'} = k_\rho; \ \ {k'}_{z'} =
\ga (k_z-u\om/c^2); \ \ \om ' = \ga (\om - uk_z)$;  this can be easily seen, e.g., by putting
 \ $\rho = \rho '; \ \ \  z = \ga (z'+ut'); \ \ \  t = \ga (t'+ uz'/c^2)$ \ directly into
Eq.(\ref{Eq.(42)}).

\h Let us now pass to subluminal {\em pulses}. We can investigate
the action of a Lorentz transformation (LT), by expressing them
either via the first method, or via the second one, of our Section 5.
 \ Let us consider for instance, in the frame O, a
$v$-speed (subluminal) pulse\cite{AIEP} given in our Section 5.
\ When we go on to a second observer O' moving {\em with the same
speed} $v$ \ w.r.t. frame O, and, still for the sake of
simplicity, passing through the origin $O$ of the initial frame at
time $t=0$, the new observer O' will see the
pulse\cite{Saari2004}
\bb \Psi(\rho ',z ',t ') \ug e^{-i t' {\om'}_0} \,
\int_{\om_{-}}^{\om_{+}} \drm \om \; S(\om) \; J_0(\rho '
{k'}_{\rho'}) \; e^{i z' {k'}_{z'}}
\; , \label{Eq.(45)} \ee     

with \ ${k'}_{z'} = {\ga}^{-1} \om/v - \ga b/v; \ \ \om' = \ga b = \om'_0; \ \
{k'}_{\rho'} = {\om'}_0/c^2 - {{k'}_{z'}}^2$, \ as one gets from the mentioned Lorentz
boost\cite{AIEP},
with $u = v$ \ (and $\gamma$ defined as usual\cite{AIEP}). \ Notice that ${k'}_{z'}$
is a function of $\om$; and that here $\om'$ is a constant.

\h If we explicitly insert into Eq.(\ref{Eq.(45)})  the relation \ $ \om =
\ga (v{{k'}_{z'}} + \ga b)$, \ which is nothing but a re-writing of the first
one of the relations following Eq.(\ref{Eq.(45)}) above, then Eq.(\ref{Eq.(45)}) becomes\cite{Saari2004}

\bb \Psi(\rho ',z ',t ') \ug \ga v \; e^{-i t' \om_0} \;
\int_{-\om_0/c}^{\om_0/c} \drm {k'}_{z'} \;
\overline{S}({k'}_{z'}) \; J_0(\rho ' {k'}_{\rho'}) \; e^{i z'
{k'}_{z'}} \; ,
\label{Eq.(47)} \ee     

where $\overline{S}$ is expressed in terms of the previous function $S(\om)$,
entering Eq.(\ref{Eq.(45)}), as follows: \ $\overline{S}({k'}_{z'}) \ug S(\ga v {k'}_{z'} + \ga^2 b)$.
Equation (\ref{Eq.(47)}) describes monochromatic beams with axial symmetry
(and does coincide also with what derived within our second
method, in Section 5, when posing $v=0$).

\h The conclusion is that a subluminal pulse, given by
our Eq.(\ref{Eq.(3)}), which appears as a $v$-speed {\em pulse} in a frame
O, will appear\cite{Saari2004} in another frame O' (traveling
w.r.t. observer O with the same speed $v$ in the same direction
$z$) just as the {\em monochromatic beam} in Eq.(\ref{Eq.(47)}) endowed with
angular frequency ${\om'}_0 = \ga b$, whatever be the pulse
spectral function in the initial frame O: \ even if the kind of
monochromatic beam, one arrives to, does of course depend
on the chosen $S(\om)$. The vice-versa is also true, in general.
 \ [Notice, incidentally, that one gets in particular a Bessel-type beam when $S$
is a Dirac's delta-function: $S(\om)= \delta(\om-\om_0)$; \ moreover, let us notice that,
on applying a LT to a Bessel beam, one obtains another Bessel beam, with a different
axicon-angle]. \ Let us set forth explicitly an observation that up to now has been
noticed only in Ref.\cite{sub}.  Namely, let us mention
that, when starting not from Eq.(\ref{Eq.(3)}) but from the most general
solutions which
---as we have already seen--- are {\em sums} of solutions (\ref{Eq.(3)}) over
the various values $b_m$ of $b$, then a Lorentz transformation
will lead us to {\em a sum} of monochromatic beams: actually, of
harmonics (rather than to a single monochromatic beam). \ In
particular, if one wants to obtain a sum of harmonic beams, one
has to apply a LT to more general subluminal pulses.

\h Let us add that {\em also} the various Superluminal localized
pulses get transformed\cite{Saari2004} one into the other by the
mere application of ordinary LTs; while it may be expected that
the subluminal and the Superluminal NDWs are to be linked (apart
from some known technical difficulties, that require a particular
caution\cite{RepToSeshadri}) by the Superluminal Lorentz ``transformations" expounded
long ago, e.g., in Refs.\cite{Review,JWeber,NCim,BarutMR} and refs.
therein.

\begin{figure}[!h]
\begin{center}
 \scalebox{1.2}{\includegraphics{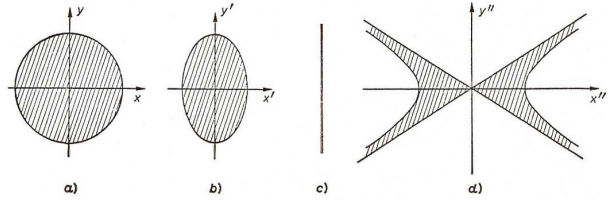}}
\end{center}
\caption{From {\em Non-restricted special Relativity\/}, also called ``Extended special
Relativity"\cite{RepToSeshadri,Review} one can
recall the following.  An intrinsically spherical (or pointlike, at the limit)
object appears in the vacuum as an ellipsoid contracted along the
motion direction when endowed with a speed $v<c$. \ By contrast,
if endowed with a speed $V>c$ (even if the $c$-speed barrier
cannot be crossed, neither from the left nor from the right), it
would appear\cite{BarutMR,Review,RepToSeshadri} no longer as a particle, but as
occupying the region delimited by a double cone and a two-sheeted
hyperboloid ---or as a double cone, at the limit--, and moving
with
Superluminal speed $V$ [the cotangent square of the cone
semi-angle, with $c=1$, being $V^2-1$. For simplicity, a space axis is
skipped. \ This figure is taken from our Refs.\cite{BarutMR,Review}. \ It is
remarkable that the shape of the localized (subluminal and
Superluminal) pulses, solutions to the {\em wave equations}, appears to
follow the same behaviour; this can have a role for a better
comprehension even of the corpuscle/wave duality, that is, of de Broglie and Schroedinger
wave-mechanics. \ See also Fig.\ref{fig40}.}
\label{LRFig5}
\end{figure}
\begin{figure}[!h]
\begin{center}
 \scalebox{.4}{\includegraphics{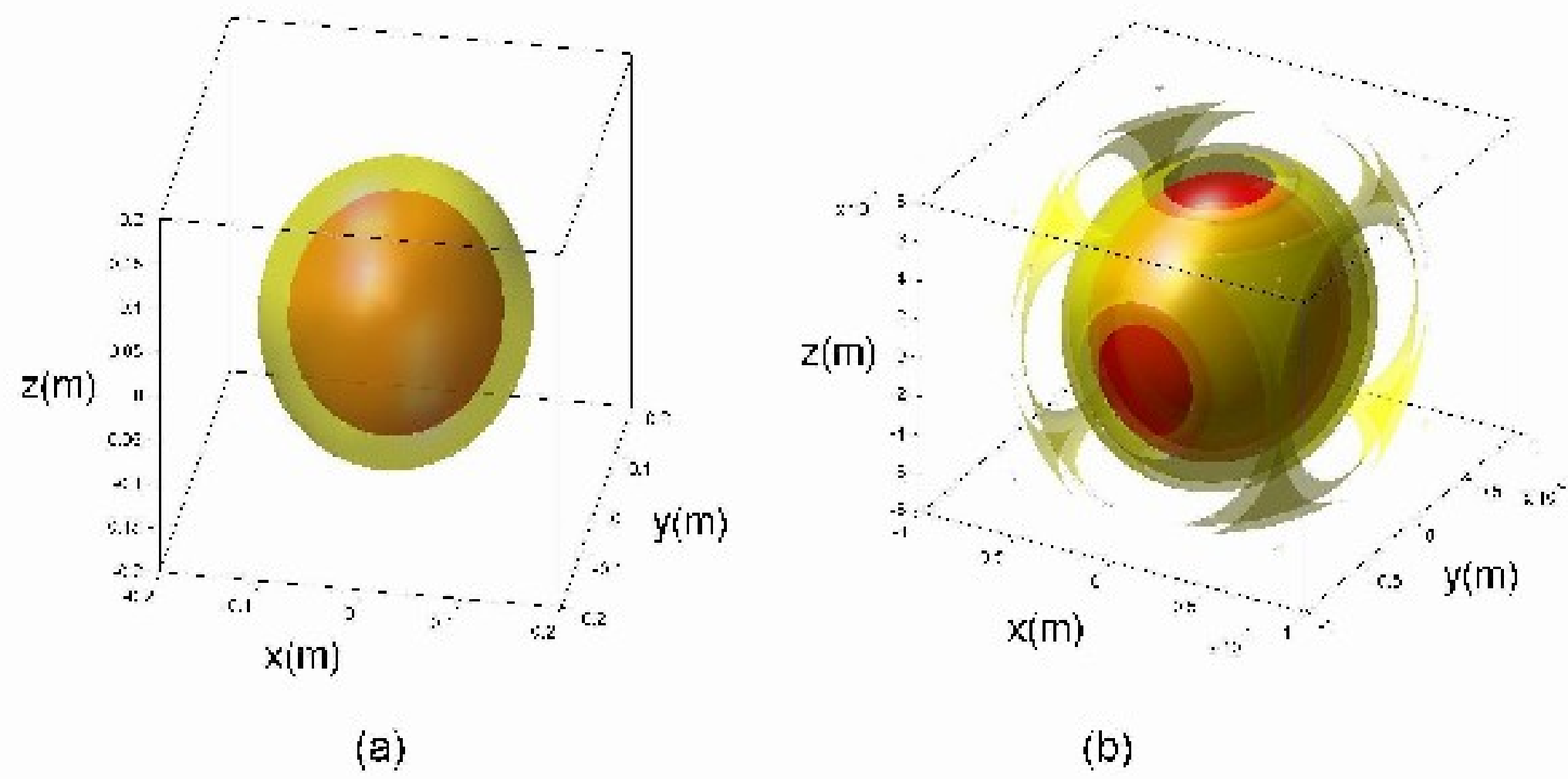}}
\end{center}
\caption{In the previos figure we have seen how SR, in its
non-restricted version (NRR), predicted\cite{BarutMR,Review} that,
while the {\em simplest} subluminal object is obviously a sphere (or, in
the limit, a space point), the simplest Superluminal object is on
the contrary an X-shaped pulse (or, in the limit, a double cone).
\ The circumstance that the Localized Solutions to the
{\em wave equations } do follow the same pattern is rather interesting,
and is expected to be useful ---in the case, e.g., of elementary
particles and quantum physics--- for a deeper comprehension of de
Broglie's and Schroedinger's wave mechanics. With regard to the
fact that the {\em simplest} subluminal NDWs, solutions to the wave
equations, are ``ball-like", let us depict by these figures, in
the ordinary 3D space, the general shape of the Mackinnon's
solutions as expressed by Eq.(124), numerically evaluated for
$v<<c$.  In figures (a) and (b) we graphically represent the field
iso-intensity surfaces, which in the considered case result to be
(as expected) just spherical.} \label{fig40}
\end{figure}


\ Let us recall at this point that, in the years 1980-82, special
relativity, in its non-restricted version, predicted that, while
the {\em simplest} subluminal {\em object} is obviously a sphere (or, in the
limit, a space point), the simplest Superluminal object is on the
contrary an X-shaped pulse (or, in the limit, a double cone): This is
shown in Fig.\ref{LRFig5}. \ The
circumstance that the localized solutions to the
{\em wave equations} follow indeed the same pattern is rather interesting,
and might be of help ---in the case, e.g., of elementary
particles and quantum physics--- for a deeper comprehension of de
Broglie's and Schroedinger's wave mechanics, and of the corpuscle/wave
duality. \  With regard to the
fact that the {\em simplest} subluminal NDWs, solutions to the wave
equation, are ``ball-like", let us present in Figs.\ref{fig40}, in
ordinary 3D space, the general shape of the simple Mackinnon's
solutions, as expressed by Eq.(\ref{Eq.(10)}) for $v<<c$: In such figures we
graphically depict the field iso-intensity surfaces, which (as expected) result
to be just spherical in the considered case.

\h We have also seen, among the others, that, even if our first method
(subsection 5.1) cannot {\em directly} yield zero-speed envelopes, such envelopes
``at rest", Eq.(85), can be however obtained by applying a $v$-speed LT
to Eq.(130). In this way, one starts from many frequencies [Eq.(\ref{Eq.(16)})] and
ends up with one frequency only [Eq.(85)], since $b$ gets transformed into
{\em the} frequency of the monochromatic beam. \ Let us add a warning: The topic of
Superluminal LTs is a
delicate one\cite{Review,JWeber,NCim,BarutMR}, at the extent that
the majority of the recent attempts to re-address this question
and its applications (cf., e.g., Ref.\cite{RepToSeshadri} and references therein)
risk to be defective: in some cases they did not
even respect the necessary covariance of the wave equation itself.

\h Further details on these topics can be found in Refs.\cite{AIEP,Livro,IEEE,PhysicaA},
where, in connection with the fact that the X-shaped pulses are
endowed with Superluminal peak-velocities, an overview was presented of the various
experimental sectors of physics in which superluminal motions do seem
to appear. Namely, also a bird's-eye view was given therein of the
{\em experiments} till now performed with evanescent waves (and/or
tunneling photons), and with the NDW solutions to the wave equations.

\

\section{Non-axially symmetric solutions: The case of higher-order
Bessel beams} 

Let us stress that till now we paid attention to exact
solutions representing axially-symmetric (subluminal) pulses only:
that is to say, to pulses obtained by suitable superpositions of
zero-order Bessel beams.

\h It is however interesting to look also for analytic solutions
representing {\em non}-axially symmetric subluminal pulses, which
can be constructed in terms of superpositions of $\nu$-order
Bessel beams, with $\nu$ a positive integer ($\nu>0$). \ This can be
attempted both in the case of subsection 5.1 (first method), and in
the case of Sect.5.3 (second method). \ For brevity's sake, let us take only
the first method (subsection 5.1) into consideration.

\h One is immediately confronted with the difficulty
that {\em no} exact solutions are known for the integral in Eq.(\ref{Eq.(8)})
when $J_0(.)$ is replaced with $J_\nu (.)$. \ One can overcome this difficulty
by following a simple method,
which allows obtaining ``higher-order" subluminal waves in
terms of the axially-symmetric ones. \ Indeed, it is well-known
that, if $\Psi(x,y,z,t)$ is an exact solution to the ordinary wave
equation, then \ $\partial \Psi / \partial x$ \ and \
$\partial \Psi / \partial y$ \ are also exact
solutions [incidentally, even \ $\partial^n \Psi /
\partial z^n$ \ and \ $\partial^n \Psi / \partial t^n$ will be
exact solutions]. \ By contrast, when
working in cylindrical co-ordinates, if $\Psi(\rho,\phi,z,t)$ is a
solution to the wave equation, quantities $\partial \Psi / \partial \rho$ \
and \ $\partial \Psi /
\partial \phi$ are {\em not} solutions, in general. \ Nevertheless, it
is not difficult at all to reach the noticeable conclusion that,
once \ $\Psi(\rho,\phi,z,t)$ \ is a solution, then also
\bb \overline{\Psi}(\rho,\phi,z,t) \ug e^{i\phi}\left(
\frac{\partial \Psi}{\partial \rho} +
\frac{i}{\rho}\frac{\partial\Psi}{\partial \phi}\right) \label{ho}
\ee  

is an exact solution! For instance, for an axially-symmetric solution
of the type $\Psi=J_0(k_{\rho}\rho)\,\exp[ik_z]\,\exp[-i\om t]$, equation
(\ref{ho}) yields $\overline{\Psi}=-k_{\rho}
\,J_1(k_{\rho}\rho)\,\exp[i\phi]\,\exp[ik_z]\,\exp[-i\om t]$,
which is actually one more analytic solution. \ In other words, it is enough to
start for simplicity from a
zero-order Bessel beam, and to apply Eq.(\ref{ho}), successively,
$\nu$ times, in order to get as a new solution \ $\overline{\Psi}
= (-k_{\rho})^\nu \, J_\nu(k_{\rho}\rho)\,\exp[i \nu
\phi]\,\exp[ik_z]\,\exp[-i\om t]$, \ which is a $\nu$-order Bessel
beam. \

\h In such a way, when applying $\nu$ times Eq.(\ref{ho}) to the
(axially-symmetric)
subluminal solution $\Psi(\rho,z,t)$ in Eqs.(\ref{Eq.(16)},\ref{Eq.(15)},\ref{Eq.(14)}) \
[obtained from Eq.(\ref{Eq.(3)}) with spectral function $S(\om)$], we
get the subluminal non-axially symmetric pulses \
$\Psi_{\nu}(\rho,\phi,z,t)$ \ as new analytic solutions,
consisting as expected in superpositions of $\nu$-order Bessel
beams:
\bb  \Psi_{n}(\rho,\phi,z,t) \ug \int_{\om_-}^{\om_+} \drm \om \;
S'(\om)\;J_\nu(k_{\rho}\rho) \; e^{i \nu \phi} \; e^{ik_z z} \; e^{-i \om t} \; ,
\label{supho} \ee   

where $k_{\rho} (\om)$ is given by Eq.(\ref{Eq.(4)}), and quantities $S'(\om) =
(-k_{\rho}(\om))^\nu S(\om)$ are the spectra of the new pulses.
\ If $S(\om)$ is centered at a certain carrier frequency (it is a
gaussian spectrum, for instance), then $S'(\om)$ too will
approximately result to be of the same type.

\h Now, if we wish the new solution $\Psi_{\nu}(\rho,\phi,z,t)$ to
possess a pre-defined spectrum $S'(\om) = F(\om)$, we can first
take Eq.(\ref{Eq.(3)}) and put \ $S(\om) = F(\om) / (-k_{\rho}(\om))^\nu$ \
in its solution (\ref{Eq.(16)}), and afterwards apply to it, $\nu$ times,
the operator \ $U \equiv \exp[i\phi] \; [\partial /
\partial\rho + (i / \rho) \partial /
\partial\phi)] \, $: \ As a result, we will obtain the desired
pulse, $\Psi_{\nu}(\rho,\phi,z,t)$, endowed with $S'(\om) =
F(\om)$.

{\bf An example} --- On starting from the subluminal axially-symmetric pulse
$\Psi(\rho,z,t)$, given by Eq.(\ref{Eq.(16)}) with the {\em gaussian}
spectrum (\ref{Eq.(17)}), we can get the subluminal, non-axially symmetric,
exact solution $\Psi_{1}(\rho,\phi,z,t)$ by simply calculating
\bb \Psi_{1}(\rho,\phi,z,t) \ug {{\partial \Psi} \over
{\partial \rho}} \ \; e^{i\phi} \; , \label{ho1} \ee  

which actually yields the ``first-order" pulse $\Psi_{1}(\rho,\phi,z,t)$,
which can be more compactly written in the form:
\bb \Psi_{1}(\rho,\phi,\eta,\zeta) \ug 2 \, {b \over c} \, v \,
\gamma^2 \; {\exp{\left[ i{b \over c} \, \beta \, \gamma^2 \; \eta
\right]}} \sum_{n=-\infty}^{\infty} \; A_n \; \exp{[in{\pi \over
\beta}]} \ \; \psi_{1n}
\label{Eq.(52)} \ee   

with
\bb \psi_{1n}(\rho,\phi,\eta,\zeta) \equiv \dis{ {b^2 \over c^2}
\, \gamma^2 \rho  \ \; Z^{-3} \ [ Z \, \cos Z - \sin Z ] \ \;
e^{i\phi} }
 \; , \label{Eq.(53)} \ee  

where
\bb Z \equiv \dis{ \sqrt{{b^2 \over c^2} \, \gamma^2 \rho^2 +
\left({b \over c} \, \gamma^2 \zeta + n \pi  \right)^2 }} \; .
\label{Eq.(54)}
\ee  

\begin{figure}[!h]
\begin{center}
 \scalebox{.35}{\includegraphics{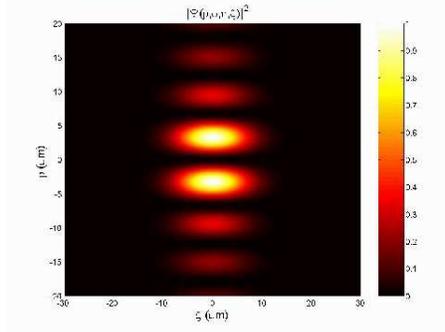}}
\end{center}
\caption{Orthogonal projection of the field intensity
corresponding to the higher order subluminal {\em pulse} represented by
the exact solution Eq.(\ref{ho1}), quantity $\Psi$ being given by Eq.(\ref{Eq.(16)}) with
the gaussian spectrum (\ref{Eq.(17)}). The pulse intensity happens to have this
time a ``donut"-like shape.} \label{fig41}
\end{figure}

This exact solution, let us repeat, corresponds to
superposition (\ref{supho}), \ with \ $S'(\om) = k_{\rho}(\om)
S(\om)$, \ quantity $S(\om)$ being given by Eq.(\ref{Eq.(17)}). \
It is represented in Figure \ref{fig41}. The {\em pulse} intensity has
a ``donut-like" shape.

\h Let us take the liberty of recalling that in Chap.2 of \cite{Livro},
in connection with the {\em frozen waves}, we argued about the possibility of
increasing even more our control on their transverse shape also by using higher-order
Bessel beams in the FWs fundamental superposition Eq.(2.74) in \cite{Livro}. \
That new approach can be understood and accepted on the basis of
simple and intuitive arguments, which
can be found in Ref.\cite{MRH}.

\h In the mentioned Chap.2 of \cite{Livro} we showed for example how to obtain a
{\em cylindrical surface of ``static" light}, in correspondence with a chosen
space interval $0 \leq z \leq L$ (for instance, with $L=238\,\mu$m).

\begin{figure}[!h]
\begin{center}
\scalebox{.6}{\includegraphics{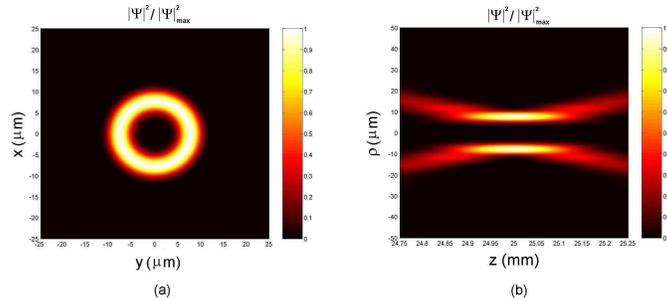}}
\end{center}
\caption{\textbf{(a)} Transverse section at $z=L/2$ of the
considered higher-order Frozen Wave. \ \textbf{(b)} Orthogonal projection
of the three-dimensional intensity pattern of the same
higher-order FW. \ The resulting field resembles
indeed a cylindrical surface of ``static" light with the chosen radius $7.75\,\mu$m
(and the chosen length $238\,\mu$m).}
\label{S5MHRFig3sec5}
\end{figure}

\h Figure \ref{S5MHRFig3sec5} depicts the longitudinal intensity pattern
as it was approximately obtained, shifted from $\rho=0$ to a different
value of $\rho$ (in this case, $\rho = 7.75\,\mu$m). The resulting field resembles
indeed a cylindrical surface of ``static" light with radius $7.75\,\mu$m
and length $238\,\mu$m.

\

\section{An Application to Biomedical Optics: NDWs and the ``GLMT" (Generalized Lorenz-Mie Theory)}

We mentioned above in several places the possible applications of NDWs, quoting
even a patent of ours\cite{brevetto} regarding the FWs.  Let us exploit here at least the
theoretical aspects of an application in biomedical optics.

\h As we know, NDWs  have become a hot topic nowadays in a variety of fields. \ Let us
recall in particular that their use, replacing laser beams for achieving multiple traps, has found many potential
applications in medicine and biomedicine\cite{Arlt,Herman,Garces,ref4,ref5}. Even though their
multi-ringed structure is not suitable for an effective three-dimensional trap when single
beam setups are employed, nevertheless, with today techniques for their generation and real-time control,
non-diffracting beams have become (better then focused Gaussian beams or others), indispensable
``laser-type" beams for biological studies by means of optical tweezing and micromanipulation techniques.

\h The theory involved in optical trapping and micro manipulation (for a review see, e.g.,
Ref.\cite{bib6}) is strongly dependent upon the relative size and electromagnetic parameters
of the scatterer, which is in general assumed to have some symmetric shape (sphere,
cylinder, ellipsis,...). If we take the electromagnetic properties of the particle and of the
surrounding medium to be of the same order (as it usually happens for
biological particles immersed in water or oil), two situations are of particular theoretical
interest, related with the possibility of avoiding, or just eliminating, too large an amount of algebra
or numerical calculations.

\h The first one is met when the size parameter $s$ of the scatterer is
much larger than the wavelength $\lambda$ of the wave ($s >> \lambda$), so that
geometrical optics considerations become the fastest and most convenient way to find out the
physical properties of interest\cite{ref5,bib7,bib8,bib9}.

\h The second one, on the contrary,
concerns very small particles: that is, scatterers whose overall dimension may be considered a
small fraction of the wavelength ($s << \lambda$), so that the Rayleigh theory becomes the
most suitable theoretical approach for solving the associated scattering problem
\cite{bib10}. Indeed, both the ray optics method and the Rayleigh theory are extremely accurate
within their range of validity and remain valid for any incident wave (as long as it is
adequately modeled).

\h However, for $s$ close to $1$, it results to be difficult to formulate analytic closed-form expressions for
the physical properties of interest. In this particular situation, indeed, none of the two aforementioned
approaches is of any help, and one is forced to adopt alternative approaches or
techniques, such as the so-called Lorenz-Mie theory (for plane waves and spherical particles)
or its generalized version, the GLMT (Generalized Lorenz-Mie Theory)\cite{re11,re12,re13,re14}
for arbitrary wave fields. We adopt the GLMT in this Section mainly because it seems to be
the most established numerical/theoretical formalism for
arbitrary-size particles in scattering problems (for further methods see, for instance,
Ref.\cite{re15} and references therein).

\

\begin{figure}[!h]
\begin{center}
 \scalebox{3.0}{\includegraphics{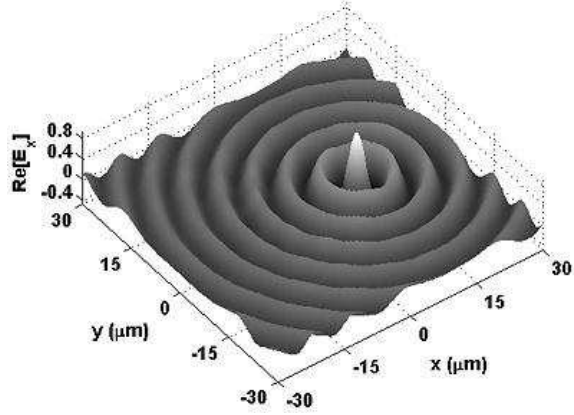}}
 \end{center}
\caption{The $E_x$ component generated by the ``GLMT" for an ordinary Bessel beam (BB) with $\lambda =
532 \mu$m. The axicon angle $\theta_a$ was chosen to be $5\,^{\circ}$, a limiting number
for which the paraxial approximation may still be considered valid.}
\label{HLfig1}
\end{figure}

\

\h In the framework of the GLMT and for spherical scatterers, a $v$-th order paraxial Bessel beam
\bb \psi = J_v\left(k_\rho\rho\right){\rm exp}(i\omega t){\rm exp}(iv\phi)\label{eq1}  \ee

must be described in terms of the beam-shape coefficients (BSCs) $g_{n,TM}^m$ and $g_{n,TE}^m$
($n$, $m$ being integers), because of the mathematical structure commonly used for the incident electromagnetic field
(which is based on power series expansions in terms of vector spherical harmonics\cite{re15}). The BSCs are,
thus, the coefficients of such expansion, and are responsible for an adequate
description of the spatial intensity profile of the wave.

\h Much effort has been devoted during the last years in order to get reliable and
useful descriptions of scalar Bessel beams, envisioning optical trapping and micro
manipulation, particle sizing applications and so on. In fact, if the radial component of the
electric field, $E_r$, is given, or known, then the BSCs $g_{n,TM}^m$ will read\cite{re15},
in a spherical coordinate system whose origin coincides with the center of the particle, as:
\bb
\begin{array}{clr}
g_{n,TM}^m & \ug \dis{\frac{(2n+1)^2}{2\pi^2n(n+1)c_n^{pw}}\frac{(n-|m|)!}{(n+|m|)!}\int_{0}^{2\pi}
\int_{0}^{\pi}\int_{0}^{\infty}\frac{E_r(r,\theta,\phi)}{E_0}r\Psi_n^{(1)}(kr)}\times\\
\\
& \times\dis{P_n^{|m|}({\rm cos}\theta){\rm exp}(-im\phi){\rm sin}\theta\,d(kr)\,d\theta\,d\phi} \; ,
\end{array}   \label{eq2}  \ee

or
\bb
\begin{array}{clr}
g_{n,TM}^m & \ug \dis{\frac{2n+1}{4\pi n(n+1)c_n^{pw}}\frac{(n-|m|)!}{(n+|m|)!}\frac{a}
{\Psi_n^{(1)}(ka)}\int_{0}^{2\pi}\int_{0}^{\pi}\frac{E_r(r=a,\theta,\phi)}{E_0}}\times\\
\\
& \times\dis{P_n^{|m|}({\rm cos}\theta){\rm exp}(-im\phi){\rm sin}\theta\,d\theta\,d\phi} \; ,
\end{array}  \; , \label{eq3}  \ee

where Eq.(\ref{eq3}) follows from a suitable choice of the spatial parameter $a$. In the above
expressions, $c_n^{pw} =(-i)^{n+1}(2n+1)/(kn(n+1))$, \ while $k$ is the wave number in the external
medium, and the $\Psi_n^{(1)}$ are spherical Bessel functions; at last, quantities \ $P_n^{|m|}({\rm cos}\theta)$ \ are the
associated Legendre polynomials, and $E_0$ the electric field strength. The coefficients
$g_{n,TE}^m$ follow from similar considerations.

\h Unless Eq.(\ref{eq2}) or Eq.(\ref{eq3}) are numerically evaluated, they a priori give us no
direct insight on the behavior of the beam shape coefficients $g_{n,TM}^m$ and $g_{n,TE}^m$, which may be, or
may not be, written in terms of any of the following parameters, or values:  $n$, $m$, the size-parameter $s$, the {\it spot}
$\Delta\rho$ of the impinging Bessel beam, and the perpendicular distance $\rho_0$ between the optical axis of the beam
and the center of the particle. \ Several researchers have devoted time to the derivation of
numerically efficient and fast computing techniques and formulae, instead of simply
implementing recursive algorithms for computing triple and double integrations as given by
(\ref{eq2}) and (\ref{eq3}), respectively\cite{bi16,bi17,bi18,bi19,bi20}.

\h We have recently shown that, in spherical coordinates, a scalar ordinary Bessel beam (BB) can
be accurately described by means of what has been called\cite{rre21} the Integral Localized
Approximation (ILA), a method that considerably revolutionized the numerical
aspects of the generalized Lorenz-Mie theory, by making it possible to obtain, in a numerically-efficient way,
closed-form expressions\cite{bi16,bi17,bi18,bi19,bi20,rre21,rre22} for the BSCs.  \ For example, a zero order Bessel
beam (BB) propagating along his axiz $z$ and polarized along $x$, when displaced along the $x$ direction of
a distance $\rho_0 = x_0$, has its BSCs, $g_{n,TM}^m$ and $g_{n,TE}^m$, given by the
simple expressions\cite{rre21}:
\bb g_{n,TM}^0 = i\frac{2n(n+1)}{2n+1}J_1(\varpi)J_1(\xi){\rm exp}(ik_zz_0)\label{eq4}  \ee

\bb
\begin{array}{clr}
g_{n,TM}^{m \neq 0} & \ug \dis{\frac{1}{2}\left(\frac{-2i}{2n+1}\right)^{|m|-1}}\times \\
\\
& \times\dis{\left[J_{|m|-1}(\varpi)J_{|m|-1}(\xi)+J_{|m|+1}(\varpi)J_{|m|+1}(\xi)\right]{\rm exp}(ik_zz_0)} \; ,
\end{array} \label{eq5}  \ee

\bb g_{n,TE}^0 = 0 \label{eq6}  \ee

\bb
\begin{array}{clr}
 g_{n,TE}^{\pm |m| \neq 0} & \ug \dis{\frac{\mp i}{2}\left(\frac{-2i}{2n+1}\right)^{|m|-1}}\times \\
\\
& \times\dis{\left[J_{|m|-1}(\varpi)J_{|m|-1}(\xi)-J_{|m|+1}(\varpi)J_{|m|+1}(\xi)\right]{\rm exp}(ik_zz_0)} \; ,
\end{array} \label{eq7}  \ee

quantity $k_z$ being the longitudinal wave number, \ $z_0$ a constant which accounts for the correct
phase of the wave at some observation point, \ $\varpi = \left(n+1/2\right){\rm
sin}\theta_a$,  and $\xi = x_0k {\rm sin}\theta_a$ \ ($\theta_a$ being the axicon angle). Once the BSCs
have been found, all the EM field components can be readily obtained by using double summation
expressions\cite{re15}. For instance, $E_r$ reads
\bb E_r\left(r,\theta,\phi\right) = -iE_0\sum_{n=1}^{\infty}\left(-i\right)^n\left(2n+1\right)
\frac{\Psi_n^{(1)}(kr)}{kr}\sum_{m=-n}^{n}g_{n,TM}^{m}\pi_n^{|m|}(\theta){\rm
sin}\theta{\rm exp}(im\phi)\ \label{eq8}  \ee

whose original  value $E_x$ is given by (\ref{eq8}) when imposing $E_x = E_r(r=|x_0|,  \theta=\pi/2,\phi=0)$
for $x > 0$ \ and \ $E_x = E_r(r=|x_0|, \theta=\pi/2,\phi=\pi)$ \ for $x < 0$, \ as we have depicted
in Fig.\ref{HLfig1}. \ Unfortunately, the higher the radial displacement $\rho_0$ of the beam
relative to the particle, the higher the number of BSCs that come into play in
Eqs.({\ref{eq4}}-{\ref{eq8}}), or, more generally, in the evaluation of all the physical properties of
interest (radiation pressure cross-sections, torques, spatial intensity distribution, and so
on). \ Nevertheless, the set of Eqs.({\ref{eq4}}-{\ref{eq7}}) can speed-up numerical calculations by a
factor of 100, or even 1,000, with respect to that expected from a direct use of (\ref{eq2}) and (\ref{eq3})
\cite{rre21}. With such a fast computing technique, together with equivalent expressions
for some other specific polarizations, there have been investigated some of the most fundamental trapping properties of
(absorbent or lossless) arbitrary size spheres, simple or stratified, with positive or negative refractive
indexes:  The results being more or less in accordance with
what should be expected in a real experiment\cite{re15,rre21,rre23,rre24}. By ``more or less"
we mean that the ILA does not predict the changes in the intensity profile of
the beam, after its passage through the lens system and the objective of the microscope... (a good
theoretical approach to this case has been recently demonstrated for a focused Gaussian beam
\cite{rre25,bbi26}).

\

\begin{figure}[!h]
\begin{center}
\scalebox{1.5}{\includegraphics{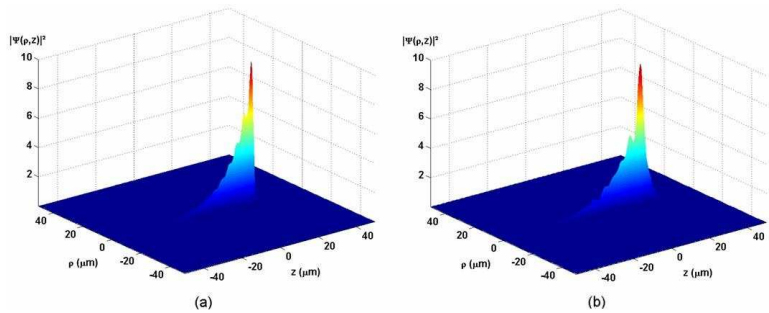}}
\end{center}
\caption{(a) A frozen wave with exponential growth, generated by the method in
Ref.\cite{rfr35} through a superposition of Bessel beams, all with the same frequency $\omega = 6.12
\times 10^{15}$Hz. \ (b) Same as (a), but now using the ILA for computing the BSCs of each Bessel beam.}
\label{HLfig2}
\end{figure}

\

\h The same approach was more recently applied for the BSCs of higher order Bessel beams,
under the paraxial approximation, for studying the optical forces exerted over biological cells
\cite{rre24}. Even though the paraxial restriction may not be adequate in some cases, it
allows one to rapidly evaluate the angular and linear momentum transfer characteristics for
a wide range of spherical-like, simple or stratified structures and biological particles.
Incidentally, if the beam is an authentic Maxwellian wave (which is not the case for a Gaussian
beam), the ILA provides a fast and reliable
alternative for investigating scattering problems within the GLMT. It should be emphasized
that the formulation in Refs.\cite{rre25,bbi26} leads to analytical BSC expressions similar
to those given by Eqs.(\ref{eq4}) and (\ref{eq5}), thus demonstrating how close the ILA outputs are to the
exact quadrature expressions, (\ref{eq2}) and (\ref{eq3}) or, equivalently, to what provided
in Ref.\cite{rre25}.

\begin{figure}[!h]
\begin{center}
\scalebox{1.25}{\includegraphics{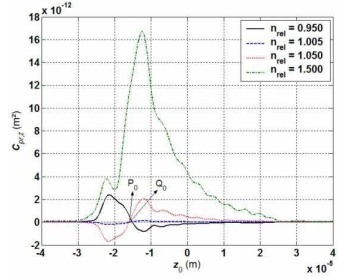}}
\end{center}
\caption{Radiation pressure cross-section exerted on a spherical dielectric particle of radius
$a = 3.75 \mu$m, as a function of its relative refractive index and of the distance $z_0$
from $z = 0$. \ The external medium is assumed to be water. Points of longitudinal stable equilibrium
are denoted by $P_0$ and $Q_0$.}
\label{HLfig3}
\end{figure}

\h One of the particularities of the GLMT is that, when the incident beam is replaced by
another one with different parameter $s$, all subsequent formulae and numerical code remain
unchanged, avoiding redefinition or inclusion of additional lines in the numerical algorithm which
contains the expressions for the physical parameters to be calculated. Further, once the
BSCs for a given BB are given, any impinging wave constructed by means of a suitable
superposition of them can also be easily described and investigated. This is of great interest in
the case of static (zero speed) longitudinal intensity patterns, generated by superposing
$N$ equal-frequency zero-order Bessel beams with different longitudinal wave number ---which is the
interesting case\cite{FWart2} of the Frozen Waves (FWs) [whose experimental production has been recently
realized, let us repeat, for the case of longitudinal intervals of the order\cite{bbi28} of 1 m]. \ Notice
that  the BSCs of paraxial FWs would simply involve a summation of $N$ individual BSCs, each
one adequately weighted in order to model some pre-chosen longitudinal intensity
pattern. \ This simple and direct technique enables the description of FWs for a
large number of potential applications, as already said elsewhere. Figure \ref{HLfig2} reveals, for
example, the equivalent of the longitudinal exponential intensity pattern first introduced
in Ref.\cite{FWart2} for mid-range purposes. It is clearly seen that, indeed, the GLMT is
capable of handling this new class of ``laser beams" and provide pretty good results for their
associated optical properties, such as the longitudinal radiation pressure
cross-section profile of Fig.\ref{HLfig2}, as shown in Fig.\ref{HLfig3}.

\h The transverse intensity control provided by the superposition of higher order BBs could also be
taken into account by using the analytic expressions for the BSCs, provided, for
example, in Ref.\cite{rre24} for single BBs. \ Finally, future theoretical work may allow one to deal
with both scalar and vector FWs, since it is understandable that, once an accurate
description of arbitrary order scalar BBs is given, their equivalent vector fields are somehow
functions of those same Bessel functions that enter into their expressions, and that can therefore
be described by the GLMT in terms of their field components\cite{ref5,bbi29}.

\

\

\h Bessel beams have also been theoretically introduced as one of the first ``laser beam" for
studying the mechanical properties of simple negative refractive-index (NRI) scatterers
\cite{rre23,bbi30}. For such particles, the matching condition (that is, the identity of the impedance
of the external medium with that of the particle) is known to produce
non-zero radial and scattering optical-forces, even if the wave suffers no reflection at the
surface of the sphere, in contrast with the ordinary case of positive refractive index (PRI)
particles\cite{rfr31,rfr32,rfr33}. Using Bessel beams (both in the ray optics and in the GLMT),
it has been possible to show, for example, that a given NRI spherical particle can be
radially either attracted by or repelled from the bright or dark annular intensity disks: This
behavior being strongly affected by how the incident wave distributes itself in
space, i.e., by its spot and relative transverse distance.

\h If the medium inside which the particle is embedded is lossy (or if the scatterer itself
is absorbent), it is also possible to conceive the incorporation of
Diffraction-Attenuation Resistant Beams (DARBs) into some optical tweezers setup\cite{MichelOE2,rfr35},
so that any pre-fixed longitudinal intensity provides the experimentalist with the
expected optical properties. However, the generation and implementation of DARBs for
arbitrary-range applications are still open problems.

\

\section{Soliton-like solutions to the ordinary Schroedinger equation, within standard QM}

As we know, not only nonlinear, but also a large class of linear equations (including the wave equations)
admit of ``soliton-like" solutions, which propagate without distortion in one direction. In the case of the
(linear) wave equations, for such soliton-like solutions we have used the name of Non-diffracting Waves (NDW).
It was soon thought that, since these solutions to the wave equations are non-diffractive and particle-like,
they are a priori suitable, more than gaussian's, for describing elementary particle motion, and may well be
related their wave nature\cite{13,14}. In fact, localized solutions were soon found also for Klein-Gordon and for
Dirac equations\cite{13,14}. \ In this Section we show\cite{schX} that, {\em mutatis mutandis}, non-diffracting solutions exist
even for the ordinary (linear) Schroedinger equation within standard Quantum Mechanics; were we may obtain both
approximate and exact solutions. \ In the ideal case such solutions (even if localized, and "decaying) are not
square-integrable, analogously to plane or spherical waves: one has to show therefore
how to obtain finite-energy solutions. \ The approach can of course be extended also for a particle
moving in the presence of a potential\cite{schX}.

\h Little work\cite{15} was done in the past for the case of the ordinary {\em Schroedinger}
equation: see, e.g., besides \cite{7}, also Refs.\cite{14}. Indeed, the Schroedinger case is different, since
the relation between the energy $E$ and the impulse magnitude $p \equiv |\imp|$ is quadratic [$E=p^2/(2m)\/$]
in non-relativistic cases, like in Schroedinger's, at variance with the relativistic ones. \ We might mention that
many non-diffracting (especially X-shaped) solutions have been constructed for the linear\cite{16} or nonlinear\cite{17}
equations that in Optics bear the name of {\em ``Schroedinger equation"}, even if they are mathematically very
different from the ordinary Schroedinger's. \  Moreover, a special kind of non-diffracting
packet solutions, in terms of Airy functions, were found in the seventies for the case
of the actual 1D Schroedinger equation, and extended later on to the 3D case.  All that
has been recently applied to the case of Optics, originating the discovery of Airy-type waves,
now well-known for their remarkable properties\cite{20,21,22a,22b,23}: Such Airy waves being solutions, once
more, to the so-called (linear) ``Schroedinger equation" of Optics. \ But, as we were saying, the non-diffracting
solutions to the ordinary Schroedinger equation, within standard Q.M.,
are quite apt themselves at describing elementary particles. \ They will result rather different from the ones
found in Optics, both for the mentioned fact that the optical Schroedinger equation is mathematically different
from the ordinary Schroedinger equation, and for the fact that our approach and methods are quite
different from the ones adopted in Optics.

\h Before going on, let us first recall that in the time-independent
realm ---or, rather, when the dependence on time is only harmonic,
i.e., for monochromatic solutions--- the (quantum, non-relativistic) Schroedinger
equation happens to be mathematically identical to the (classical, relativistic) Helmholtz
equation\cite{24,25,26,27}. \ And many trains of localized X-shaped pulses have been found, as
{\em superpositions} of solutions to the Helmholtz equation, which propagate, for
instance, along cylindrical or co-axial waveguides\cite{Coaxial}; \ but we shall skip all the
cases\cite{MRF,MFR} of this type, since we are concerned here with propagation in free space,
even when in the presence of an ordinary potential. \ Let
us also mention that, in the general time-{\em dependent}
case, that is, in the case of pulses, the Schroedinger and the
ordinary wave equation are no longer mathematically identical,
since the time derivative results to be of the fist order in the
former and of the second order in the latter. [It has been shown
that, nevertheless, at least in some cases, they still share
various classes of analogous solutions, differing only in their
spreading properties\cite{25}]. Moreover, the
Schroedinger equation implies the existence of an {\em intrinsic}
dispersion relation even for free particles; this is another difference
to pay attention to: the solutions to the wave equation suffer
only diffraction (and no dispersion) in the vacuum, while those of the
Schroedinger equation suffer also (an intrinsic) dispersion even in the vacuum.

\

\subsection{Bessel {\em beams} as non-diffracting solutions (NDS) to the Schroedinger
equation}

Let us consider the Schroedinger equation for a free particle (an electron,
for example)
\bb
\nablabf^2 \psi + {{2im} \over \hbar} \, {{\pa \psi} \over {\pa t}}  \ug 0 \; .
\label{eq1S}
\ee      

\h If we confine ourselves to solutions of the type

\hfill{$
\psi(\rho,z,\varphi;t) \ug F(\rho,z,\varphi) \; \e^{-iEt/\hbar} \; ,
$\hfill}

their spatial part $F$ is known to obey the reduced equation \ $\nablabf^2 F + k^2 F  \ug 0$, \ with \ $k^2 \equiv p^2/\hbar^2$ \
and \ $p^2 = 2mE$ (quantity $p \equiv |\imp|$ being the particle momentum, and therefore $k \equiv |\kbf|$ the
total wavenumber). \ Such a reduced equation is nothing but the Helmholtz equation, for which various simple localized-beam
solutions are already known: In particular, the so-called Bessel beams, which have been experimentally
produced since long.
 \ Actually, let us look ---as usual--- for factorized solutions (in the simple case of cylindrical symmetry w.r.t. the $z$-axis),
by supposing
the constant longitudinal wavenumber $k_z \equiv p_z/\hbar$. \ [Since the present formalism is used both in quantum mechanics and in electromagnetism, with a difference in the customary
nomenclature, for clarity's sake let us here stress that $k \equiv p/\hbar$; $k_\rho \equiv k_\perp \equiv p_\perp/\hbar$; $\omega
\equiv E/\hbar$; while $k_z \equiv k_\parallel = p_\parallel /\hbar \equiv p_z/\hbar$ is often represented by the (for us) ambiguous symbol
$\beta$]. \ As a consequence, the (transverse) wavefunction obeys a Bessel differential equation, in which it enters the
constant transverse wavenumber $k_\rho \equiv p_\rho /\hbar$ with the {\em condition} \ $k_\rho^2 = k^2 - k_z^2 \equiv 2mE/{\hbar^2} -
k_z^2$. \ To avoid any divergencies, it must be $k_\rho^2 \geq 0$, that is,
$k^2 \geq k_z^2$; namely, it must hold [see Fig.1 in Ref.\cite{schX}] the constraint

$$E \, \geq \, {p_z^2 \over {2m}} \ .$$

A simple solution is therefore [$p \equiv \hbar k$]:
\bb
\psi(\rho,z;t) \ug J_0(\rho p_\rho/\hbar) \; \exp{[i(z p_z - Et)/\hbar]}
\label{eq4S}
\ee      

together with the above condition. \ {\em Equation (\ref{eq4S}) can be regarded as a Bessel beam solution to the Schroedinger
equation} [the other Bessel functions are not acceptable here, because of their divergence at $\rho = 0$ or for $\rho \rightarrow
\infty$], with forward propagation (i.e., positive $z$ direction) for $k_z > 0$. \ This result
is not surprising, since ---once we suppose the whole time variation to be expressed by the function $\exp{[i\omega t]}$---
both the ordinary wave equation and the Schroedinger equation transform into the Helmholtz equation. \ Actually, the only
difference between the Bessel beam solutions to the wave equation and to the Schroedinger equation consists in the different
relationships among frequency, longitudinal, and transverse wavenumber;
in other words (with $E \equiv \omega\hbar$):
\bb
 p_\rho^2 \ug E^2/c^2 - p_z^2 \ \ \ \ \ {\rm for \
the \ wave \ equation};
\label{eq5aS}
\ee      

\bb
p_\rho^2 \ug 2mE - p_z^2 \ \ \ \ \ \ {\rm for \ the \
Schroedinger \ equation}. \
\label{eq5bS}
\ee      

In the case of beams, the experimental production of NDSs to the
Schroedinger equation can be {\em similar} to the one exploited for the
NDSs to the wave equations (e.g., in Optics, or Acoustics):  Cf.,
e.g. Figure 1.2 in the first one of Refs.\cite{Introd},
and refs. therein, where the simple case of a source consisting in an array of circular slits, or
rings, were considered.\footnote{For pulses, however, the
generation technique must deviate from Optics', since in the
Schroedinger equation case the phase velocity of the Bessel beams produced
through an annular slit would depend on the energy.}  \ In the
Table we refer to a Bessel beam of photons, and a Bessel beam of
(e.g.) electrons, respectively. We list therein the relevant
quantities having a role, e.g., in Electromagnetism, and the
corresponding ones for the Schroedinger equation's spatial part
$\hbar^2 \nablabf^2 F + 2mE \, F = 0$, with $F = R(\rho) \;
Z(z)\,$. The second and the fourth lines have been written down
for the so-called simple Durnin et al.'s case, when the Bessel beam is produced
by an annular slit (illuminated by a plane wave) located at the focus
of a lens\cite{18,19,Durnin1,Durnin2}.

\

\begin{tabular}{|c|c|}
  \hline
  WAVE EQUATION & SCHROEDINGER EQUATION \\
  \hline
  $k  \ug {\omega \over c}$ & $p \ug \sqrt{2mE}$ \\
  $k_\rho \, \simeq \, {r \over f} \, k$ & $p_\rho \simeq {r \over f} \, p$ \\
  $k_\rho^2 \ug {\omega^2 \over c^2} - k_z^2$ & $p_\rho^2 \ug 2mE - p_z^2$ \\
  $k_z^2 \ug {\omega^2 \over c^2}(1-{r^2 \over f^2})$ & $p_z^2 \ug 2mE (1-{r^2 \over f^2})$ \\
  \hline
\end{tabular}

\

In this Table, quantity $f$ is the focal distance of the lens (for
instance, an ordinary lens in optics; and a magnetic lens in the case of
Schroedinger charged wavepackets), and $r$ is the radius of the
considered ring. [In connection with the last line of the Table, let us recall
that in the wave equation case the phase-velocity $\omega / k_z$
is almost independent of the frequency (at least for limited frequency
intervals, like in optics), and one gets a constant group-velocity and
an easy way to build up X-shaped waves. By contrast,
in the Schroedinger case, the phase-velocity of each (monochromatic)
Bessel beam depends on the frequency, and this makes it difficult to
generate an ``X-wave" (i.e., a wave depending on $z$ and $t$ only via the
quantity $z-Vt$) by using simple methods, as Durnin et al.'s,
based on Bessel beams superposition. In the case of charged particles,
one should compensate such a velocity variation by suitably modifying
the focal distance $f$ of the Durnin's lens, e.g. on having recourse
to an additional magnetic, or electric, lens.]

\h Before going on, let us stress that one could easily eliminate
the restriction of axial symmetry: In such a case, in fact,
solution (\ref{eq4S}) would become

\

\hfill{$
\psi(\rho,z,\varphi;t) \ug J_n(\rho p_\rho/\hbar) \ \dis{\e^{i z p_z/\hbar} \, \e^{-iEt/\hbar}
\, \e^{i n \varphi}} \; ,
$\hfill}

\

with $n$ an integer. \ The investigation of not
cylindrically-symmetric solutions is interesting especially in the
case of localized {\em pulses}: and we shall deal
with them below.

\

\subsection{Exact non-diffracting solutions to the Schroedinger equation}

Coming to the problem of finding out ``soliton-like" solutions to the {\em ordinary} Schroedinger
equation, let us switch to a more comprehensive formalism. \ Namely, in cylindrical coordinates and
neglecting evanescent waves, a quite general function $\psi$ of $\rho,\phi,z$ and
$t$, expressed in terms of Fourier and Hankel transformations, can be written as:
\bb \Psi(\rho,\phi,z,t) \ug
\sum_{n=-\infty}^{\infty} \left[ \int_{0}^{\infty} d\kr \, \int_{-\infi}^{\infty} dk_z \, \int_{-\infty}^{\infty} d\om \, \kr
A_n^{'}(\kr,k_z,\om) J_n(\kr \rho) e^{ik_z z} e^{-i\om t} e^{i n \phi} \right] \ .
\label{eq2S} \ee

Notice that the last equation is nothing but Eq.(\ref{S2geral1}), by us considered
in subsection 1.3 when having in mind a rather general, ideal solution to linear, homogeneous wave equations in free space
(still disregarding the evanecsent sector).
The essential point, for Eq.(\ref{eq2S}) to represent a (general) solution to the Scroedinger
equation, is imposing now that the $A_n(k_z,\om)$ be given by
\bb A_n^{'}(\kr,k_z,\om) \ug A_n(k_z,\om)\,\delta \left[ \kr^2 -
\left( \frac{2m\omega}{\hbar} - k_z^2 \right) \right] \ .
\label{eq3S}  \ee

\h We request moreover that
\bb
A_n(k_z,\om) \ug \sum_{n=-\infty}^{\infty} \; S_{mn} \; \delta \left[ \omega - (V k_z + b'_m) \right] \; ,
\label{eq4S}  \ee

with
\bb
 b'_m \ug \frac{2m \pi V}{\Delta z_0} \; .
\label{eq5S}  \ee

The last two equations guarantee that the general solution (\ref{eq2S}) to Eq.(\ref{eq1S}) is a NDW,
that is, a wave capable of keeping indefinitely its spatial shape while propagating. Let us recall that such a property,
when assuming propagation in the $z$ direction, may be mathematically expressed as in Eq.(\ref{S2def}) of Subsection 1.3,
where $\Delta z_0$ is a chosen length, and $V$ is the
pulse peak-velocity, with $0\leq V \leq \infty$. \ For the moment, the meaning of the spectral parameters $k_z, \ k_\rho, \omega$
appearing above in not important, since they are dumb integration variables.

\h Notice that in the general solution (\ref{eq2S}), together with Eqs.(\ref{eq3S}-\ref{eq5S}), all Bessel functions $J_n(\kr \rho)$,
with any $n$, appear. \ Just for simplicity, however, we can choose
\bb
S_{mn} \ug S'(\omega) \; \delta_{0n} \; \delta_{ l m} \; ,
\label{eq6S}  \ee

where the $delta$'s are now Kronecker's symbols, and $l$ is a positive integer; so as to reduce ourselves to the
mere case of zeroth-order Bessel functions. \ Since we are now dealing with
quantum mechanics, let us go on to the new notations
$$ k \equiv p/\hbar; \ \ \ k_\rho \equiv p_\rho / \hbar; \ \ \ k_z \equiv p_z / \hbar; \ \ \ \omega \equiv E / \hbar$$

and put \ $b'_l = 2 l \pi V / {\Delta z_0} \equiv b / \hbar$.

\h Since the present formalism is used both in quantum mechanics and in electromagnetism, with a difference in the customary
nomenclature, for clarity's sake let us repeat once more that $k \equiv p/\hbar$; $k_\rho \equiv k_\perp \equiv p_\rho/\hbar$; $\omega
\equiv E/\hbar$; while $k_z \equiv k_\parallel = p_\parallel /\hbar \equiv p_z/\hbar$.

\h We can now integrate Eq.(\ref{eq2S}) in $k_\rho$ and in $k_z$, obtaining non-diffracting solutions
to the Schroedinger equation as the following superpositions (integrations over the frequency, or the energy)
of Bessel-beam solutions [with $b \geq 0$]:
\bb \Psi(\rho,z,\zeta) \ug \dis{\e^{{{-i b} \over \hbar V} z}} \;
\int_{E_{-}}^{E_{+}} \drm E \; J_0(\rho p_\rho/\hbar) \; S(E) \;
\dis{\e^{i{E \over \hbar V} \zeta}} \; ,
\label{eq7S}
\ee

where it is still
\bb
\zeta \; \equiv \; z - V t \; ,
\label{eq8S}
\ee

while
\bb
p_\rho \ug {1 \over V} \, \sqrt {-E^2 + (2mV^2+ 2 b) E - b^2}
\label{eq9S}
\ee

and
\bb
E_{\pm} \ug mV^2 \left( 1 \pm \sqrt{1+{{2b} \over
{mV^2}}} \right) + b \; .
\label{eq10S}
\ee

Notice that the in Eq.(\ref{eq7S}) [as well as in equations below like (\ref{eq17aS})], the solution $\Psi$
depends on $z$, besides via $\zeta$, only via a phase factor; the modulus
$|\Psi|$ of $\Psi$ goes on depending on $z$ (and on $t$) only through the
variable $\zeta \equiv z-Vt$.  This means, as we already know, that the magnitude of each solution
does not change during propagation: that is, the solutions are NDWs and keep their shape while
traveling.

\h The simple integral solution (\ref{eq7S}), which yields non-diffracting solutions with azimutal
symmetry, admits of a simple physical interpretation:  It implies integrating the Bessel beams
\ $J_0(\rho p_\rho/\hbar) \;
\exp[i{p_z \over \hbar} \zeta] \; \exp[i{E \over \hbar} t]$, \ with \ $p_\rho = \sqrt{2mE - p_z^2}$,
in the interval  \ $E_- \leq E \leq E_+$, \ \ along the {\em straight line} \ $E = V p_z + b$: This
is known to eliminate evanescent waves.

\begin{figure}[!h]
\begin{center}
 \scalebox{2.0}{\includegraphics{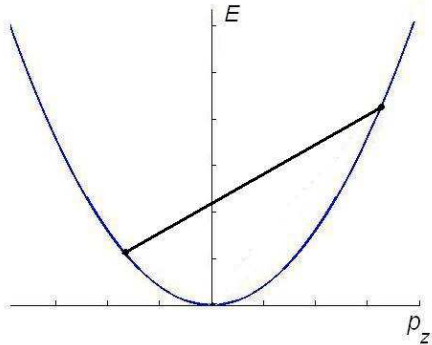}}
\end{center}
\caption{The allowed region is the one internal to the parabola, since (to avoid divergencies) it must be \ $ E \geq p_z^2 / (2m)$. \
In this case, the parabola and the chosen straight-line have equations $E = p_z^2/(2m)$ and $E = Vp_z + b$, respectively.
The two values of the intersections of this straight-line with the parabola are given in Eq.(\ref{eq10S}). \
Inside the parabola $p_\rho^2 \geq 0$.} \label{fig3S}
\end{figure}

\

{\bf Examples \ ---} \ An interesting solution to Eq.(\ref{eq7S}) is for instance obtained when assuming the {\em real}
exponential spectrum
\bb S(E) \ug s_0 \; \dis{ \exp [a (E - E_+)] } \; ,
\label{eq11S}
\ee

$a$ and 
of $E_+$. \ On integrating\cite{schX}, we get [$\Ncal$ being a constant]:
\bb
\Psi(\rho,\eta,\zeta) \ug  \Ncal s_0 2V \sqrt{P} \; \dis{ \exp [i {{mV}
\over {\hbar}} \eta] \ \exp [-a V \sqrt{P}] \ {{\sin Y} \over Y} } \; , \label{eq17aS}
\ee

where
\bb
Y \; \equiv \; {{\sqrt{P}} \over \hbar} \, \sqrt{\rho^2 -
(\hbar aV+i\zeta)^2} \; , \label{eq17bS}
\ee

and $P \equiv m^2 V^2 + 2 m b$, while $\eta \equiv z-vt$ is a function of $b$. \ Notice that for $a=0$,
one ends up with a solution similar to Mackinnon's\cite{9}. \ Equations (\ref{eq17aS},\ref{eq17bS}) are the simplest closed-form
nondiffracting solution to the Schroedinger equation: In Figs.3 of
Ref.\cite{schX} we have depicted his square magnitude, when choosing for simplicity $b=0$ [namely, Fig.3a therein corresponds to
$a=E_+ / 5$, while Fig.3b therein corresponds to $a=5 E_+$].

\begin{figure}[!h]
\begin{center}
 \scalebox{2.1}{\includegraphics{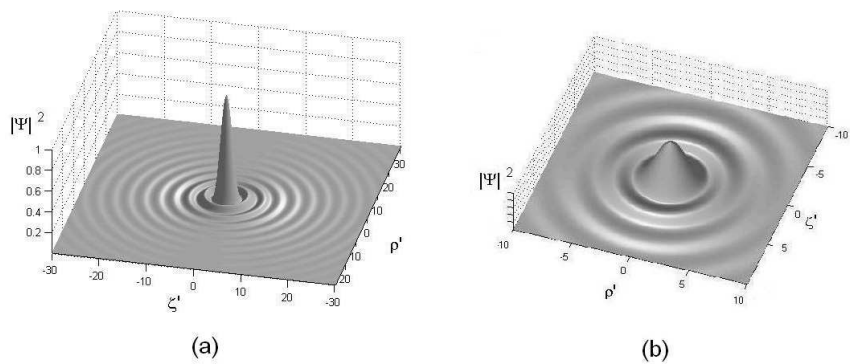}}
\end{center}
\caption{In these, and the following Figure, we depict the square magnitude of elementary solutions
of the type (\ref{eq17aS}), corresponding to the {\em real}
spectrum $S(u) = s_0 \exp[(E - E_+) a]$, as a function of $\rho ' \equiv \rho \sqrt{P}/\hbar$ and of
$\zeta ' \equiv \zeta \sqrt{P}/\hbar$.  Quantity $a$ is a positive number (when $a=0$ one ends up with a solutions similar
to Mackinnon's\cite{9}, while $b$ for simplicity has be chosen equal to zero. \ Figure (a) corresponds to $a=E_+ / 5$. \
For figure (b), normalized with respect to $\rho$ and $\zeta$, we have still assumed  for simplicity $b=0$, so that
$\Aove = maV^2$: More precisely, it refers to $\Aove = 0$ and does clearly show the ``ball-like"
structure one expects in such a case.  It should be however noted that, for the
Schroedinger equation, also an X-shaped structure is always appearing ---more evident here in
figure (a),--- even in the most ball-like solutions.}
\label{figs5}
\end{figure}

\h Some physical (interesting) comments on such results will appear elsewhere. Here, let us only add a few
brief comments, illustrated by some more Figures.  Let us first recall
that the Non-diffracting Solutions to the ordinary wave equations resulted
to be roughly {\em ball-like} when their peak-velocity is
subluminal\cite{sub,BarutMR}, and {\em X-shaped\/}\cite{Lu1,PhysicaA,BarutMR} when superluminal. Now, let us see what happens in the different case of the
Schroedinger equation.  \ Normalizing $\rho$ and $\zeta$, we can write Eq.(\ref{eq17bS}) as

$$Y \ug \sqrt{{\rho'}^2 - (\Aove + i\zeta')^2}$$

with \ $\rho' \equiv \sqrt{P} \rho / \hbar$ \ and \ $\zeta' \equiv
\sqrt{P} \zeta / \hbar$, \ while \ $\Aove
\equiv aA = \sqrt{P} a V$. \ For simplicity, let us stick to the case $b=0$; therefore, the
simple relation will hold: \ $\Aove = maV^2$. \ For the Schroedinger equation, we can observe
that:

(i) If we choose $\Aove = 0$, which can be associated with $V=0$, we get the solutions in
Fig.\ref{figs5}b: that is, a mainly ball-like structure (even if, differently from the
ordinary wave equation cases, an X-shaped structure does timidly start to appear).

\begin{figure}[!h]
\begin{center}
 \scalebox{2.}{\includegraphics{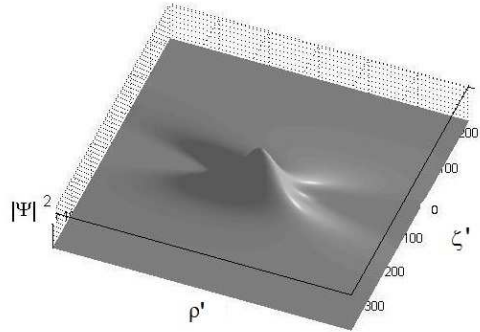}}
\end{center}
\caption{The solution, under all the previous conditions, with an increased value of
$\Aove$, namely with $\Aove = 20$. An X-shaped structure more evidently appears, contributing
in a more clear way to the general form of the solution (see the text).}
\label{fig6}
\end{figure}

(ii) If we increase by contrast the value of $\Aove$, by choosing e.g. $\Aove=20$ (which can be
associated with larger speeds), one notices that now also the X-shaped structure does evidently
contribute: See, e.g., Fig.\ref{fig6}.

(iii) To have a preliminary idea of the ``internal structure" of our soliton-like solutions
to the (ordinary) Schroedinger equation, we have to plot, instead of the square magnitude
of $\Psi$, its real or imaginary part: In Figs.6 of Ref.\cite{schX} we chose the square of its real part. Then, even in
the $\Aove = 0$
case, one can start to see in those figures the appearance of the X shape, which becomes more and more evident as the value of
$\Aove$ increases. \ We confine ourselves here to stress that the (square of the) real part of $\Psi$ does show, in 3D, also
some ``internal oscillations": see, e.g., Fig.\ref{fig8} corresponding to the value $\Aove = 5$. \ We
shall face elsewhere topics like their possible connections with the de Broglie
picture of quantum particles.

\begin{figure}[!h]
\begin{center}
 \scalebox{2.5}{\includegraphics{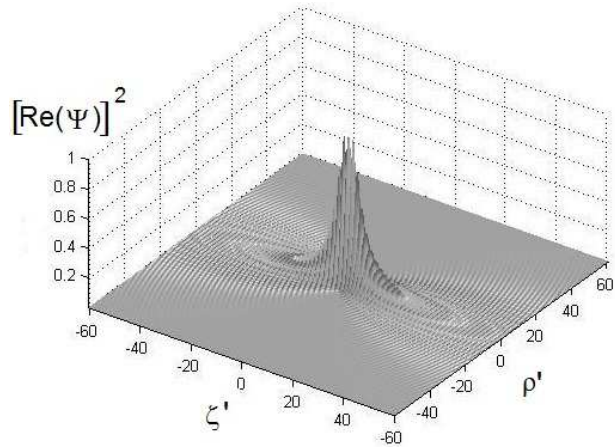}}
\end{center}
\caption{The (square of the) real part of $\Psi$ shows, in 3D, also some ``internal
oscillations": this Figure corresponds, e.g., to the value $\Aove = 5$.}
\label{fig8}
\end{figure}

\subsection{A general exact Localized Solution}

Let us go back to the choice of spectrum $S(E)$.  Since in our
equation (\ref{eq7S}) the integration interval is limited [$E_- < E < E_+$], in such
an interval {\em any} spectral function $S(E)$ can be expanded into
the Fourier series
\bb S(E) \ug \sum_{n=-\infty}^{\infty} \, a_n \, \dis{
\e^{i{{2\pi} \over D} n E} } \; ,
\label{eq18S}
\ee

with
\bb a_n \ug {1 \over D} \dis{ \int_{E_-}^{E_+} \drm E \; S(E) \;
\e^{-i{{2\pi} \over D} n E} } \; ,
\label{eq19S}
\ee

quantity $S(E)$ being an {\em arbitrary} function, and $D$ being still defined as
$D \equiv E_{+} - E_{-}$. \ Inserting Eq.(\ref{eq18S}) into Eq.(\ref{eq19S}), and following\cite{schX}
the same procedure exploited in the previous Subsection, we get the {\em general exact non-diffracting solution}
to the Schroedinger equation in the form
\bb \Psi(\rho,\eta,\zeta) \ug \Ncal \, {2A} \, \dis{ e^{i
{{mV} \over {\hbar}}\eta}}  \
\sum_{n=-\infty}^{\infty} \, a_n \; \dis{ \exp{[{i{{2\pi} \over D}
n B}]} } \; {\sin {Z} \over Z} \; , \label{eq20S}
\ee

where

\bb
Z \, \equiv \, \sqrt{ \left( {A \over {\hbar V}} \zeta + n\pi \right)^2 +
{P \over \hbar^2} \rho^2} \; .
\label{eq15S}
\ee

and \ $A = V \sqrt{P}$; \ $B = mV^2 + b$, \ and  $\Ncal$ is a suitable normalization constant.\
Notice that solution (\ref{eq20S}) yields non-diffracting solutions with azimuthal symmetry for
whatever spectrum $S(E)$ in Eq.(\ref{eq7S}). It is moreover worthwhile to note that, even when truncating the series
in Eq.(20) at a certain value $n=N$, the solutions obtained is
{\em still} an exact non-diffracting solution to the Schroedinger equation.

\h We already mentioned the problem of producing Bessel beams of electrons, instead of optical Bessel beams. \ As to the
possible generating set-ups, an interesting problem from the experimental point of view is that in Optics
one starts usually from a laser source; in the case of quantum
mechanics, one might have recourse to ``laser beams of particles", as the ones under
investigation since more than a decade.

\

\section{A brief mention of further topics}

\

\subsection{Airy and Airy-types waves}

Many non-diffracting (especially X-shaped) solutions have been constructed for the linear\cite{16} or nonlinear\cite{17}
equations that in Optics bear the name of {\em ``Schroedinger equation"}, even if they are mathematically very
different from the ordinary Schroedinger's. \  Moreover, a special kind of non-diffracting
packet solutions, in terms of Airy functions, were found in the seventies for the case
of the actual 1D Schroedinger equation, and extended later on to the 3D case.  All that
has been recently applied to the case of Optics, originating the discovery of Airy-type waves,
now well-known for their remarkable properties\cite{20,21,22a,22b,23}: Such Airy waves being solutions, once
more, to the so-called (linear) ``Schroedinger equation" of Optics.

\h We wish to repeat here this information, for its intrinsic interest and its relevance, and for the fact
that one ---or rather two--- of the following Chapters of this Book will be mainly devoted to
the Airy waves.

\h The results presented above, in Sect.10 of this Chapter, are rather different, however, from the ones found in Optics,
both for the mentioned fact that the optical Schroedinger equation is mathematically different from the ordinary
Schroedinger equation, and for the fact that our approach and methods are quite different from the ones
adopted in Optics.

\

\subsection{``Soliton-like" solutions to the Einstein equations of General Relativity, and Gravitational waves}

Some interesting progress has been performed by one of us [MZR] even in the sector of
the Einstein (non-linearized) equations of General Relastivity, finding out therefore new
possible solutions for gravitational waves. But there is no room here for presenting
details.

\

\subsection{Super-resolution}

Strong super-resolution effects can be attained by suitable superpositions of evanescent
Bessel beams. But this topic too will be reviewed elsewhere, for the tyranny of space.

\

\

{\bf Acknowledgements}\\

For useful discussions the authors are grateful,
among the others, to M.Assis, R.G.Avenda\~{n}o, M.Balma, I.A.Besieris, R.Bonifacio,
D.Campbell, R.Chiao, C.Conti, A.Friberg, D.Faccio, F.Fontana,
P.Hawkes, R.Grunwald, G.Kurizki, M.Mattiuzzi, P.Milonni,
J.L.Prego-Borges, P.Saari, A.Shaarawi, M.Tygel, A.Utkin and
R.Ziolkowski. \ {\em Due to reasons of space, many important references had to be
skipped in this introductory Chapter: We apologize with the relevant Authors.} \
This paper has been written as the introductory chapter (Chap.1) of a book on
Non-Diffracting Waves published (2014) by J.Wiley-VCH, Berlin; and will constitute
a part of a much longer Review (in preparation).

\

\

\

\end{document}